\documentclass{JHEP3} 
\pdfoutput=1

\usepackage{psfrag}
\usepackage{subfigure}
\usepackage{graphicx}

\usepackage{amsmath}
\usepackage{amsfonts}   
\usepackage{amssymb}    
\usepackage[numbers ,square, comma, sort&compress]{natbib} 

\def\tfrac#1#2{{\textstyle\frac{#1}{#2}}}
\def\be{\begin{equation}}
\def\ee{\end{equation}}
\def\ber{\begin{eqnarray}}
\def\eer{\end{eqnarray}}

\newcommand\newc{\newcommand} 
\def\ch2{$\chi^2$}
\newc\neutone{\chi_1^0} \newc\mneutone{m_{\neutone}}
\newc\neuttwo{\chi_2^0} \newc\mneuttwo{m_{\neuttwo}}
\newc\neutthree{\chi_3^0} \newc\mneutthree{m_{\neutthree}}
\newc\neutfour{\chi_4^0} \newc\mneutfour{m_{\neutfour}}
\newcommand{\msl}{\mslepton}
\newcommand{\msq}{\msquark}
\newc\msquark{m_{\squark}}
\newc\msquarkl{m_{\squarkl}} 
\newc\msquarkr{m_{\squarkr}}

\newcommand{\bfm}[1]{{\mbox{\boldmath $#1$}}}

\newcommand{\data}{d}
\newcommand{\nuis}{\psi}
\newcommand{\params}{\theta}
\newcommand{\basis}{\Theta}

\newcommand{\mhalf}{m_{1/2}}      \newcommand{\mzero}{m_0}
\newcommand{\tanb}{\tan\beta}
\newcommand{\azero}{A_0}

\newcommand{\mtpole}{M_t}
\newcommand{\alphas}{\alpha_s(M_Z)^{\overline{MS}}}
\newcommand{\alphaemmz}{\alpha_{\text{em}}(M_Z)^{\overline{MS}}}

\newcommand{\fb}{\rm fb}

\newcommand\squark{\widetilde q} 
\newcommand\squarkl{{\widetilde q}_L} 
\newcommand\squarkr{{\widetilde q}_R}

\newcommand\slepton{\widetilde l} 
\newcommand\mslepton{m_{\slepton}}

\title{A Coverage Study of the CMSSM Based on ATLAS Sensitivity  Using Fast Neural Networks Techniques}

\author{Michael Bridges\\
       Astrophysics Group, Cavendish Laboratory, University of Cambridge \\
       J.J. Thomson Avenue, 	Cambridge, CB3 0HE, UK,
        E-mail: \email{mb435@mrao.cam.ac.uk}}
        
\author{Kyle Cranmer\\
	Center for Cosmology and Particle Physics, New York University \\
	Washington Place, New York, NY 10003, USA, 
	E-mail: \email {cranmer@cern.ch}}
       
\author{Farhan Feroz\\
       Astrophysics Group, Cavendish Laboratory, University of Cambridge \\
       J.J. Thomson Avenue, 	Cambridge, CB3 0HE, UK, 
        E-mail: \email{ff235@mrao.cam.ac.uk}}

\author{Mike Hobson\\
       Astrophysics Group, Cavendish Laboratory, University of Cambridge \\
       J.J. Thomson Avenue, 	Cambridge, CB3 0HE, UK, 
        E-mail: \email{mph@mrao.cam.ac.uk}}

\author{Roberto Ruiz de Austri\\
       Instituto de F\'isica Corpuscular, IFIC-UV/CSIC \\
       Valencia, Spain, 
       E-mail: \email{rruiz@ific.uv.es}}

\author{Roberto Trotta\\
       Astrophysics Group, Imperial College London, Blackett Laboratory \\
       Prince Consort Rd, London SW7 2AZ, UK, 
       E-mail: \email{r.trotta@imperial.ac.uk}}

\abstract{We assess the coverage properties of confidence and credible
  intervals on the CMSSM parameter space inferred from a Bayesian
  posterior and the profile likelihood based on an ATLAS sensitivity
  study.  In order to make those calculations feasible, we introduce a
  new method based on neural networks to approximate the mapping
  between CMSSM parameters and weak-scale particle masses. Our method
  reduces the computational effort needed to sample the CMSSM
  parameter space by a factor of $\sim 10^4$ with respect to
  conventional techniques.  We find that both the Bayesian posterior
  and the profile likelihood intervals can significantly over-cover
  and identify the origin of this effect to physical boundaries in
  the parameter space.  Finally, we point out that the effects intrinsic to the 
  statistical procedure are conflated with simplifications to the likelihood functions from 
  the experiments themselves.}


\preprint{}
\begin{document}

\section{Introduction}\label{sec:intro}

Experiments at the Large Hadron Collider (LHC) will soon start testing
many models of particle physics beyond the Standard Model (SM).
Particular attention will be given to the Minimal Supersymmetric SM
(MSSM) and other scenarios involving softly-broken
supersymmetry (SUSY).

We are interested in the statistical properties of the emerging methodology used to infer the parameters of these models.  Here we consider a restricted class of SUSY models with certain universality 
assumptions regarding the SUSY breaking parameters. This scenario is commonly
called mSUGRA or the
Constrained Minimal Supersymmetric Standard Model (CMSSM)~\cite{kkrw94,sugra-reviews}. The model is defined in terms of five free
parameters: common scalar ($\mzero$), gaugino ($\mhalf$) and
tri--linear ($\azero$) mass parameters (all specified at the GUT
scale) plus the ratio of Higgs vacuum expectation values $\tanb$ and
$\text{sign}(\mu)$, where $\mu$ is the Higgs/higgsino mass parameter
whose square is computed from the conditions of radiative electroweak
symmetry breaking (EWSB).

A common procedure to explore the model's parameter space
consisted of evaluating the likelihood function on a fixed grid, often
encompassing only 2 or 3 dimensions at the time~\cite{grid-cmssm1,
  grid-cmssm2, grid-cmssm3, grid-cmssm4, grid-cmssm5, grid-cmssm6}.
Of course, the number of likelihood evaluations in a grid scan scales
exponentially with the dimensionality of the parameter space, making
it impractical for a full exploration of the CMSSM parameter space.
More recently new approaches based on both Frequentist and Bayesian
statistics coupled with Markov Chain Monte Carlo (MCMC) methodology
have been applied
\cite{Baltz:2004aw,Allanach:2005kz,Allanach:2006jc,deAustri:2006pe,Allanach:2006cc,Roszkowski:2006mi,Allanach:2007qk,Roszkowski:2007fd,Roszkowski:2007va,Allanach:2008tu,Allanach:2008iq,Martinez:2009jh}. The
efficiency of the MCMC techniques allow for a full exploration of
multidimensional models.  However, the likelihood function of these
models is complex, multimodal with many narrow features, making the
exploration task with conventional MCMC methods challenging often with low sampling efficiency.

More recently, a novel scanning algorithm, called MultiNest~\cite{Feroz:2007kg,Feroz:2008xx,Trotta:2008bp}, has been proposed in the
Bayesian context. It is based on
the framework of Nested Sampling, recently invented by
Skilling~\cite{Skilling:2006}. MultiNest has been developed in such a
way as to be an extremely efficient sampler of the posterior
distribution even for likelihood functions defined over a parameter
space of large dimensionality with a very complex structure. It has
been applied for the exploration of several models within the MSSM
\cite{Feroz:2008wr,Trotta:2008bp,Feroz:2009dv,AbdusSalam:2009qd,Trotta:2009gr,AbdusSalam:2009tr,Cabrera:2009dm,Bertone:2010rv,Scott:2009jn}
and its minimal extension, the so--called Next-to-Minimal Supersymmetric SM (NMSSM)
\cite{LopezFogliani:2009np}.  Nested sampling has also been
shown as a powerful technique for model reconstruction based on LHC data
\cite{Roszkowski:2009ye}.

Having implemented sophisticated statistical and scanning methods, 
several groups have turned their attention to
evaluating the sensitivity to the choice of priors
\cite{Allanach:2007qk,Lafaye:2007vs,Trotta:2008bp} and of scanning
algorithms \cite{Akrami:2009hp}.  Those analyses indicate that current constraints are not strong enough to dominate the posterior and that
the choice of prior does influence the resulting inference.  While
confidence intervals derived from the profile likelihood or a
chi-square have no formal dependence on a prior, there is a sampling
artifact when the contours are extracted from samples produced from
MCMC or MultiNest~\cite{Trotta:2008bp}.

Given the sensitivity to priors and the differences between the
intervals obtained from different methods, it is natural to seek out a
quantitative assessment of their performance.  A natural quantity for
this evaluation is \textit{coverage}: the probability that an interval
will contain (cover) the true value of a parameter.  The defining
property of a 95\% confidence interval is that the procedure that
produced it should produce intervals that cover the true value 95\% 
of the time; thus, it is reasonable to check if the
procedures have the properties they claim.  Coverage is a frequentist
concept: one assumes the true parameters take on some specific,
unknown values and considers the outcome of repeated experiments.
Intervals based on Bayesian techniques are meant to contain a given
amount of posterior probability for a single measurement and are
referred to as credible intervals to make clear the distinction.
While Bayesian techniques are not designed with coverage as a goal, it
is still meaningful to investigate their coverage properties.  In
fact, the development of Bayesian methods with desirable frequentist
properties is an active area of research for professional
statisticians~\cite{reid,baines}.  Moreover, the frequentist intervals
obtained from the profile likelihood or chi-square functions are based
on asymptotic approximations and are not guaranteed to have the
claimed coverage properties.

In this work we study the coverage properties of both Bayesian and
Frequentist procedures commonly used to study the CMSSM.  In
particular, we consider the coverage of the ATLAS ``SU3'' CMSSM
benchmark point.  The task requires extensive computational
expenditure, which would be unfeasible with standard
analysis techniques.  Thus in this paper we explore the use of a class
of machine learning devices called Artificial Neural Networks (ANNs)
to approximate the most computationally intensive sections of the
analysis pipeline.  Such techniques have been successfully 
employed before in the cosmological setting to accelerate inference 
massively~\cite{Auld:2006pm}.  Within particle physics, ANNs have been 
used extensively for event classification, but we are not aware of their
use in approximating the mapping between fundamental Lagrangian 
parameters and observable quantities such as a mass spectrum.

This paper will be arranged as follows: in Section \ref{sec:CMSSM} we
will briefly review the Bayesian methodology and the CMSSM framework
that this work operates within while in Section \ref{sec:nn} we will
outline the use of neural networks to approximate SUSY spectrum
calculators. In Section \ref{sec:nn_performance} we will demonstrate
our method, discuss the accuracy of our network predictions against
the dedicated spectrum calculator, and demonstrate computational
performance improvements.  In Section \ref{sec:coverage} we present
the results of a coverage study employing the trained networks. We
conclude with a discussion in Section \ref{sec:conclusions}.

\section{Statistical framework and supersymmetry model}
\label{sec:CMSSM} 

The cornerstone of Bayesian inference is  Bayes' Theorem, which reads
\be \label{eq:bayes}
 p(\basis | \data) = \frac{p(\data |
\basis) p(\basis)}{p(\data)}. \ee
The quantity $p(\basis | \data)$ on the l.h.s. of eq.~\eqref{eq:bayes}
is called the {\em posterior}, while on the r.h.s., the quantity
$p(\data | \basis)$ is the likelihood (when taken as a function
$\basis$ for fixed data, $d$). The quantity $p(\basis)$ is the {\em
  prior} which encodes our state of knowledge about the values of the
parameters $\basis$ before we see the data. The state of knowledge is
then updated to the posterior via the likelihood. Finally, the
quantity in the denominator is called {\em evidence} or {\em model
  likelihood}. If one is interested in constraining the model's
parameters, the evidence is merely a normalization constant,
independent of $\basis$, and can therefore be dropped (for further
details, see e.g.~\cite{Trotta:2008qt}).

We denote the parameter set of the CMSSM introduced above ($\mzero$,
$\mhalf$, $\azero$ and $\tanb$) by $\params$ (we fix sgn($\mu$) to be
positive, motivated by arguments of consistency with the measured
anomalous magnetic moment of the muon), while $\nuis$ denotes the
``nuisance parameters'' whose values we are not interested in
inferring, but which enter the calculation of the observable
quantities.  Here, the relevant nuisance parameters are the SM
quantities
\begin{equation}
\nuis \equiv \{ \mtpole, m_b(m_b)^{\overline{MS}}, \alphas , \alphaemmz \} \, ,
\end{equation}
where $\mtpole$ is the pole top quark mass, $m_b(m_b)^{\overline{MS}}$
is the bottom quark mass at $m_b$, while $\alphaemmz$ and $\alphas$
are the electromagnetic and the strong coupling constants at the $Z$
pole mass $M_Z$, the last three evaluated in the $\overline{MS}$
scheme.  Note that there are no nuisance parameters associated with
experimentally related systematics, which would instead be anticipated to be the case in future
inference problems based on the results of LHC experiments.  We denote
the full 8--dimensional set of parameters by \be \basis =
(\params,\nuis).
\label{basis:eq}
\ee

\begin{figure}  
\begin{center} 
\includegraphics[width=0.45\linewidth, angle = -90]{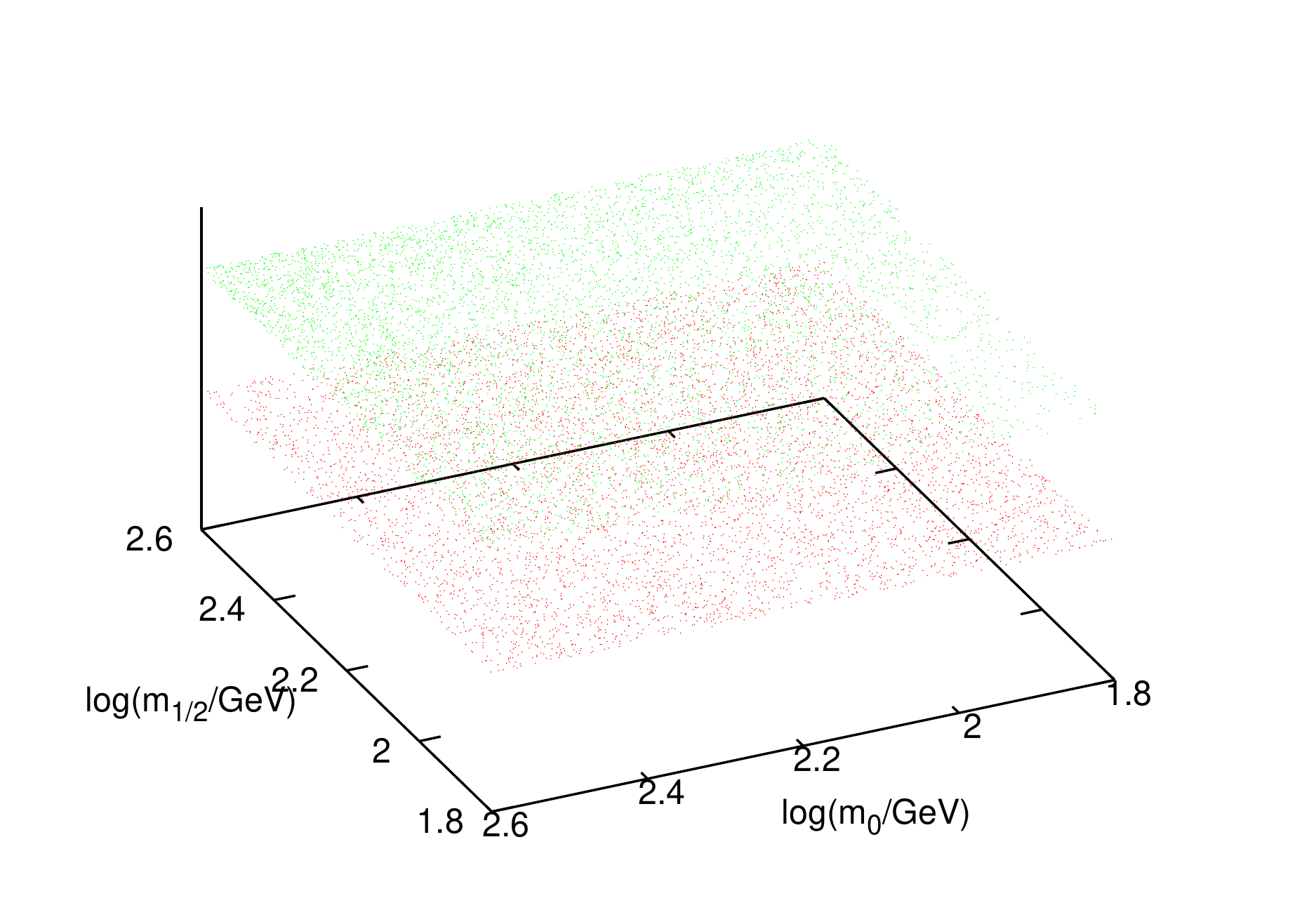}
\end{center} 
\caption{\label{fig:holes} Physical (upper) and unphysical (lower) samples 
in $m_0 - m_{1/2}$ space.   }  
\end{figure}

The likelihood evaluation in a SUSY particle model analysis involves
the computation of the mass spectrum. It implies the use of an
iterative procedure in which the renormalization group equations
(RGEs) are solved. The current public numerical codes reach a
precision at the level of a few percent in computing the
spectrum~\cite{Allanach:2003jw}.  For it, the RGEs are implemented up
two-loop level and the calculation of the physical masses involve the
inclusion, at least, of the full one-loop radiative corrections.  The
typical running time for the spectrum calculator is a few seconds per
model point, which can seriously hinder all Monte-Carlo style
explorations of the parameter space where $10^5-10^7$ parameter points
need to be examined. In addition these parameter spaces have a
unphysical regions which emit tachyonic solutions and/or in which EWSB
is not fulfilled.  Figure~\ref{fig:holes} illustrates that these regions
can be evenly spread across some projections of the $\params$
parameter space.  Testing and discarding these unphysical points leads to large
timing inefficiencies in the scan.

\section{Approximating the spectrum calculators with neural networks}\label{sec:nn}

Any analysis of the CMSSM parameter space must relate the high-scale
parameters $\params$ with observable quantities, such as the sparticle
mass spectrum at the LHC.  As described above, this is achieved by
evolving the RGEs, which is a computationally intensive procedure.  We
use {\sc SoftSusy} \cite{softsusy} for calculating the sparticle mass
spectrum, and denote the predicted weak-scale sparticle masses by
$\bfm{m}$.  One can view the RGEs simply as a mapping from $\Theta \to
\bfm{m}$, and attempt to engineer a computationally efficient
representation of the function.  There is a vast literature for
multivariate function approximation, which includes techniques such as
neural networks~\cite{Rumelhart}, radial basis
functions~\cite{Powell}, support vector machines~\cite{Vapnik}, and
regression trees~\cite{Breiman, Friedman}.

Here we use a multilayer perceptron (MLPs), a type of feed-forward
neural network. We restrict ourselves to three-layer MLPs, which
consist of an input layer ($\Theta_l$), a hidden layer ($h_j$), and an
output layer ($y_i$). In such a network, the value of the nodes in the
hidden and output layers are given by
\begin{eqnarray}
\mbox{hidden layer:} & h_j=g^{(1)}(f_j^{(1)}); &
f_j^{(1)} = \sum_l w^{(1)}_{jl}\Theta_l +
  \xi_j^{(1)}, \\
\mbox{output layer:} & y_i=g^{(2)}(f_i^{(2)}); & f_i^{(2)} =
\sum_j w^{(2)}_{ij}h_j + \xi_i^{(2)},
\end{eqnarray}
where the index $l$ runs over input nodes, $j$ runs over hidden nodes
and $i$ runs over output nodes. The functions $g^{(1)}$ and $g^{(2)}$
are called activation functions and are chosen to be bounded, smooth
and monotonic. The non-linear nature of the former is a key
ingredient in constructing a viable network; we use $g^{(1)}(x)=\tanh x$ and
$g^{(2)}(x)=x$.

The weights $w$ and biases $\xi$ are the quantities we wish to
determine, which we denote collectively by $\bfm{a}$. As these
parameters vary, a very wide range of non-linear mappings between the
inputs and outputs are possible.  In fact, according to a `universal
approximation theorem' \cite{Cybenko,Leshno}, a standard multilayer
feed-forward network with a locally bounded piecewise continuous
activation function can approximate any continuous function to {\it
  any} degree of accuracy if (and only if) the network's activation
function is not a polynomial and the network contains an adequate number of hidden nodes. 
This result applies when activation
functions are chosen a priori and held fixed if $\bfm{a}$ is allowed
to vary. In practice, one often must empirically find a network
architecture with a suitable balance of training requirements,
generalization, and prediction accuracy.
   
\begin{table}
\begin{center}
\caption{Parameter ranges chosen for CMSSM parameters $\theta$ and 
nuisance parameters $\psi$. We have adopted uniform priors on the variables and ranges shown here. }
\begin{tabular}{|c|}
\hline
\textbf{CMSSM parameters}: $\theta$ \\
\hline
 $\log(50)$ $<$ $\log m_0, \log m_{1/2}$ (GeV) $<$ $\log(500)$\\
 $-4$ $<$ $A_0$ (TeV) $<$ $4$\\
 $2$ $<$ $\tan \beta$ $<$ $62$\\
\hline
\textbf{SM nuisance parameters}: $\psi$\\
\hline
 $3.92$ $<$ $M_t$ $<$ $4.48$\\
 $163.7$ $<$ $m_b(m_b)^{\overline{MS}}$ $<$ $178.1$\\
 $0.1096$ $<$ $\alpha_{em}(M_Z)^{\overline{MS}}$ $<$ $0.1256$\\
 $127.846$ $<$ $\alpha_s(M_Z)^{\overline(MS)}$ $<$ $127.99$\\
 \hline
\end{tabular}
\label{table:priors}\\
\end{center}
\end{table}

\subsection{Network training, accuracy, and performance}
\label{sec:nn_performance}

The procedure of adjusting the weights and biases of a network (and
sometimes the architecture of the network itself) as to approximate
the {\sc SoftSusy} mapping is referred to as training, and requires
the use of sophisticated algorithms. The training procedure is usually
cast as an optimization problem with respect to some loss function
$\mathcal{L}$. A training data set $\mathcal{D} = \{{\Theta}^{(k)},\bfm{m}^{(k)}\}$ was
used for this procedure, where $k$ is an index for input-output pairs
generated with {\sc SoftSusy}.  The training dataset was typically
chosen to have a few thousand points for this application.  If the
training sample adequately represents the function over the chosen
region and the network has good generalization capabilities, then the
trained network should approximate the function well on an independent
testing sample. Below we describe the training algorithm and assess
its accuracy with an independent testing sample.

Any training process requires the quantification of errors between the
network outputs $y_i$ and the corresponding values from the training
via some loss function. In this application we chose a simple squared
error objective function in the network parameters $\bfm{a}$:
\be 
\label{eq:loss_function} 
\mathcal{L}(\mathcal{D}|\bfm{a}) = \tfrac{1}{2}\sum_k \sum_i\left[m^{(k)}_i-y_i(\Theta^{(k)};\bfm{a})\right]^2. 
 \ee This
is a highly non-linear, multi-modal function in hundreds or thousands
of dimensions.

The problem of network training is analogous to one of parameter
estimation in $\bfm{a}$ with Equation \eqref{eq:loss_function} playing
the role of the log-likelihood function.  For small networks this
function can be adequately optimised using a form of gradient descent
described as \emph{back propagation} \cite{backprop}. A complete
Bayesian framework for the training of neural networks has been given
by \cite{MacKaythesis}. In this context the log of the posterior
probability of $\bfm{a}$ given the training data $\mathcal{D}$ is: 
\be
\label{eq:net_like}
\ln  P(\bfm{a}|\mathcal{D})  \propto
-\mathcal{L}(\mathcal{D}|\bfm{a}) + \ln \Pi(\bfm{a}) 
\ee 
where $\Pi$ is a prior distribution of the network parameters
$\bfm{a}$. While it is conceptually feasible to use conventional Monte
Carlo sampling based methods such as Metropolis-Hastings or Nested
Sampling to optimise this function and thus recover the optimal set of
$\bfm{a}$, the computational expense of this approach would be
prohibitively high. In this work we have used an optimising suite of
software called {\sc MemSys} based on the principle of Maximum
Entropy.

The {\sc MemSys} algorithm chooses a positive-negative entropy function $S$
as the prior distribution, $\Pi(\bfm{a})$ with a
hyperparameter $\alpha$ which controls the relative balance between prior and likelihood
within the posterior distribution~\cite{Hobson} so that Equation~\eqref{eq:net_like}
becomes: 
\be 
\label{eq:net_maxent_like} 
\ln P(\bfm{a}|\mathcal{D}) \propto -\mathcal{L}(\mathcal{D}|\bfm{a}) + 
\alpha \ln S(\bfm{a}) .
\ee
For each choice of $\alpha$ there is a set of network parameters $\hat{\bfm{a}}$ that
maximises the posterior distribution. As $\alpha$ is varied one obtains a path of
maximum posterior solutions $ \hat{\bfm{a}}(\alpha) $ called the
\emph{maximum-entropy trajectory}. The {\sc MemSys} algorithm begins the exploration of
the joint distribution of $\bfm{a}$ and $\alpha$ by slowly introducing the likelihood
term in the posterior with initially large values of $\alpha$. For each sampled value of
$\alpha$ the algorithm converges to the maximum posterior solution $\hat{\bfm{a}}$ via
conjugate gradient descent. The parameter $\alpha$ is then slowly reduced via a rate parameter such
that the new maximum entropy solution $\hat{\bfm{a}}'$ lies close to the previous
solution $\hat{\bfm{a}}$, thus ensuring a smooth descent to the optimal point in
$\bfm{a}$ and $\alpha$. {\sc MemSys} then returns an estimate of the maximum
posterior network weights and biases. For
more complete details of the {\sc MemSys} algorithm we refer the reader
to~\cite{memsys}.

\begin{figure}
        \subfigure[$m_{\nu}^1$]{
          \includegraphics[width=.22\columnwidth, angle=-90]{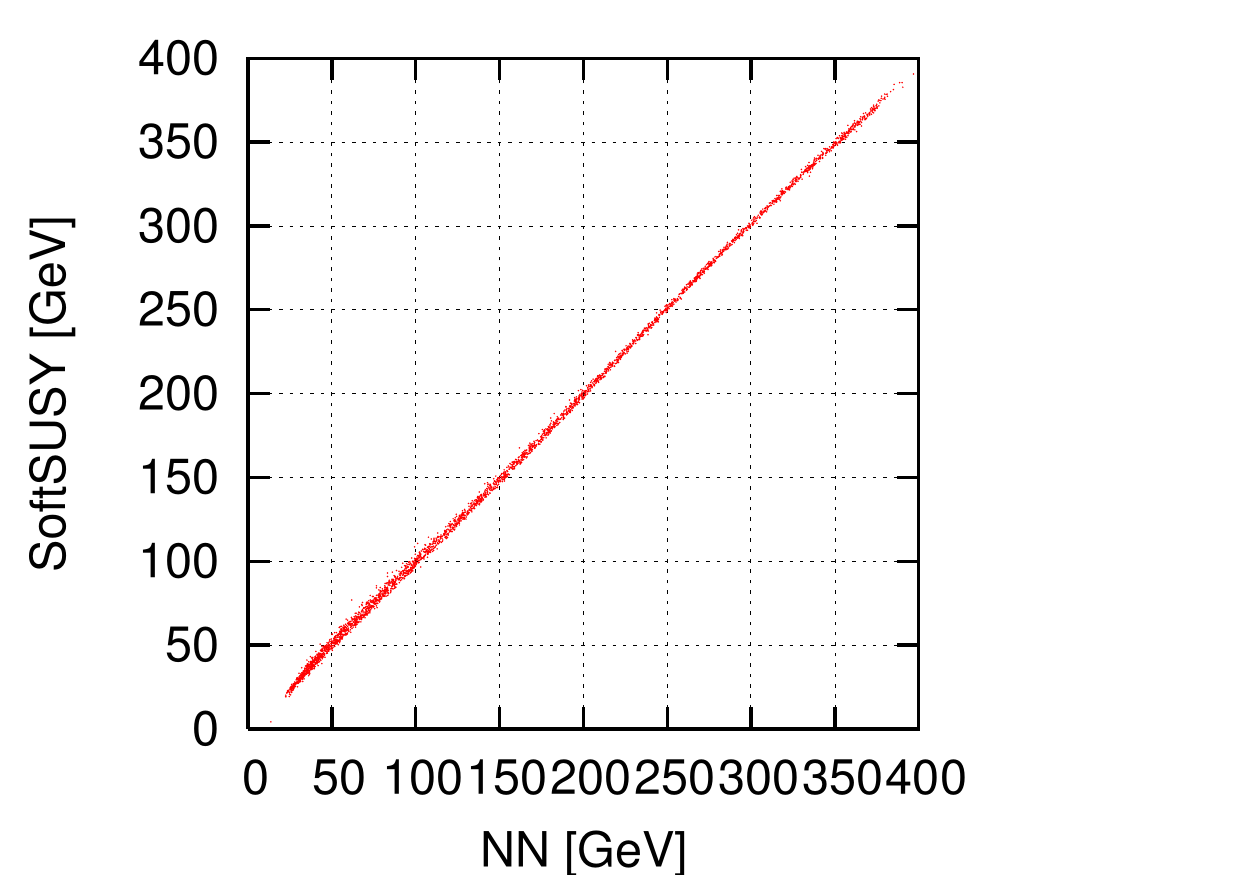}}
        \subfigure[$m_{\nu}^2$]{
          \includegraphics[width=.22\columnwidth,angle=-90]{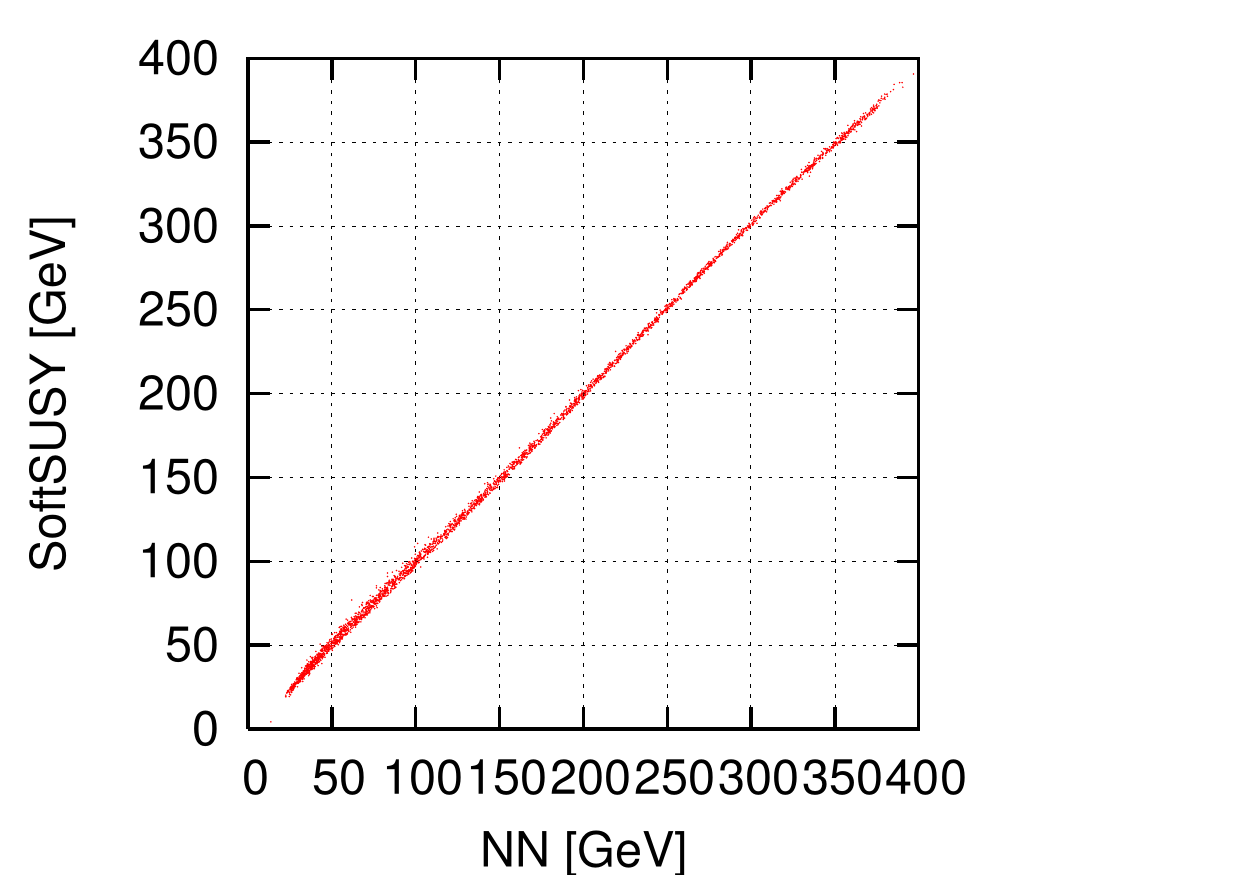}}
           \subfigure[$m_{se}^R$]{
          \includegraphics[width=.22\columnwidth,angle=-90]{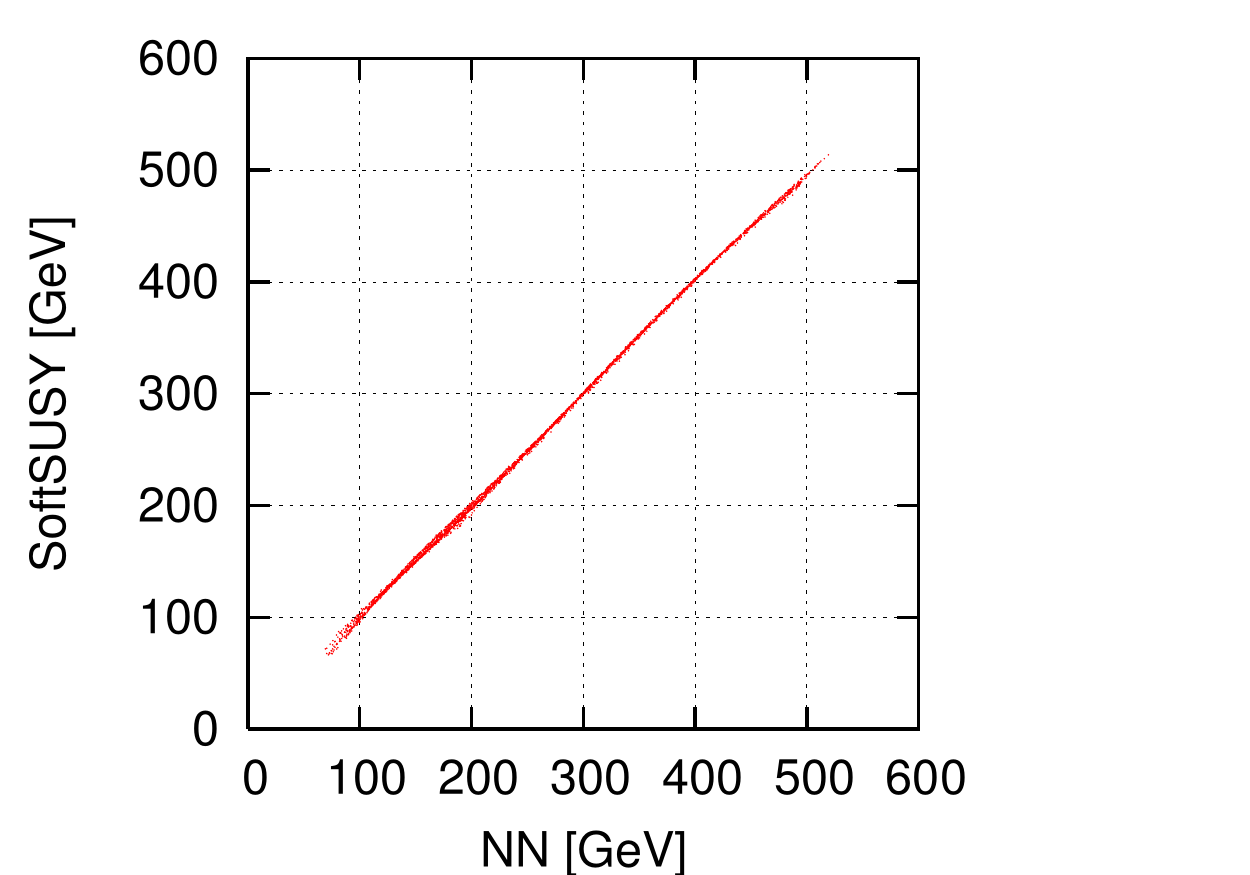}}\\
        \subfigure[$m_{se}^L$]{
          \includegraphics[width=.22\columnwidth, angle=-90]{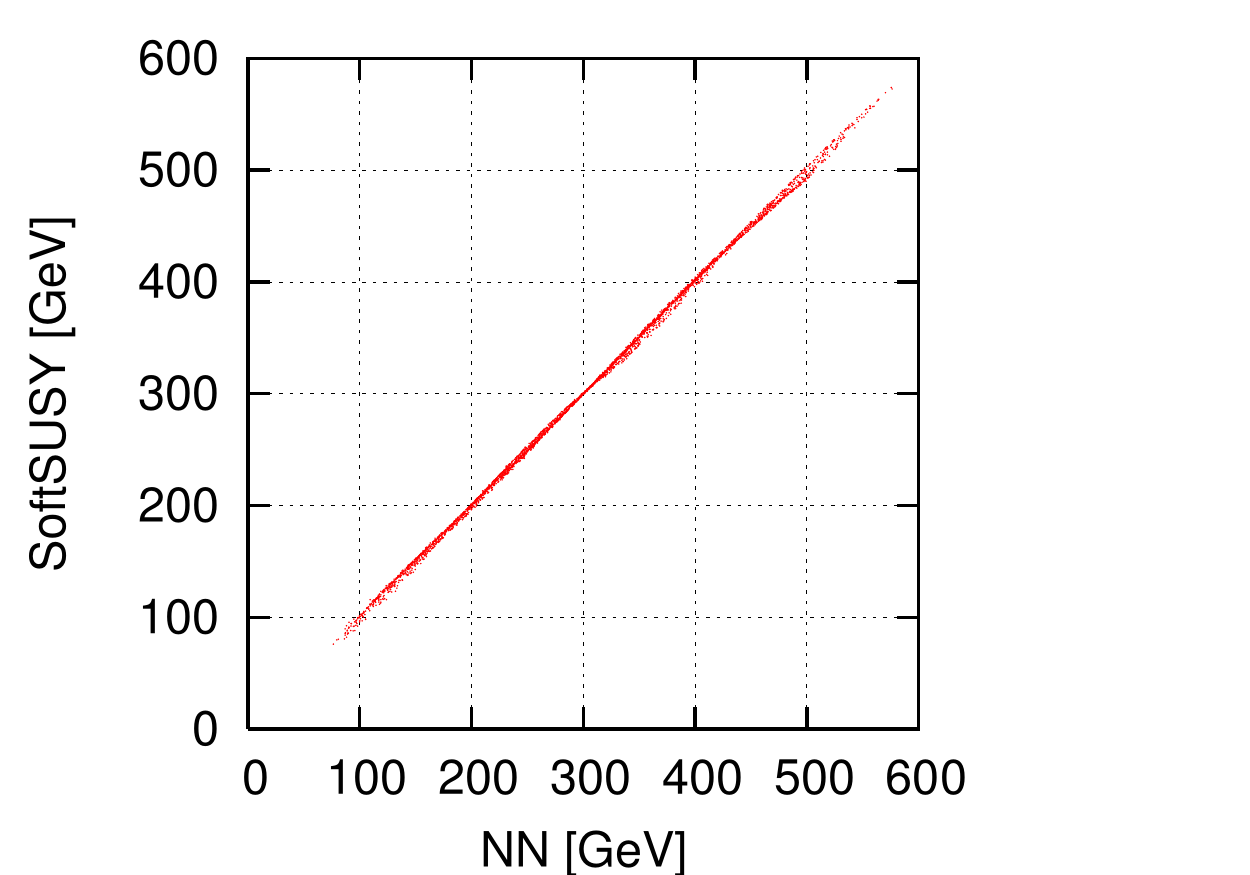}}
	         \subfigure[$m_{s\mu}^R$]{
          \includegraphics[width=.22\columnwidth, angle=-90]{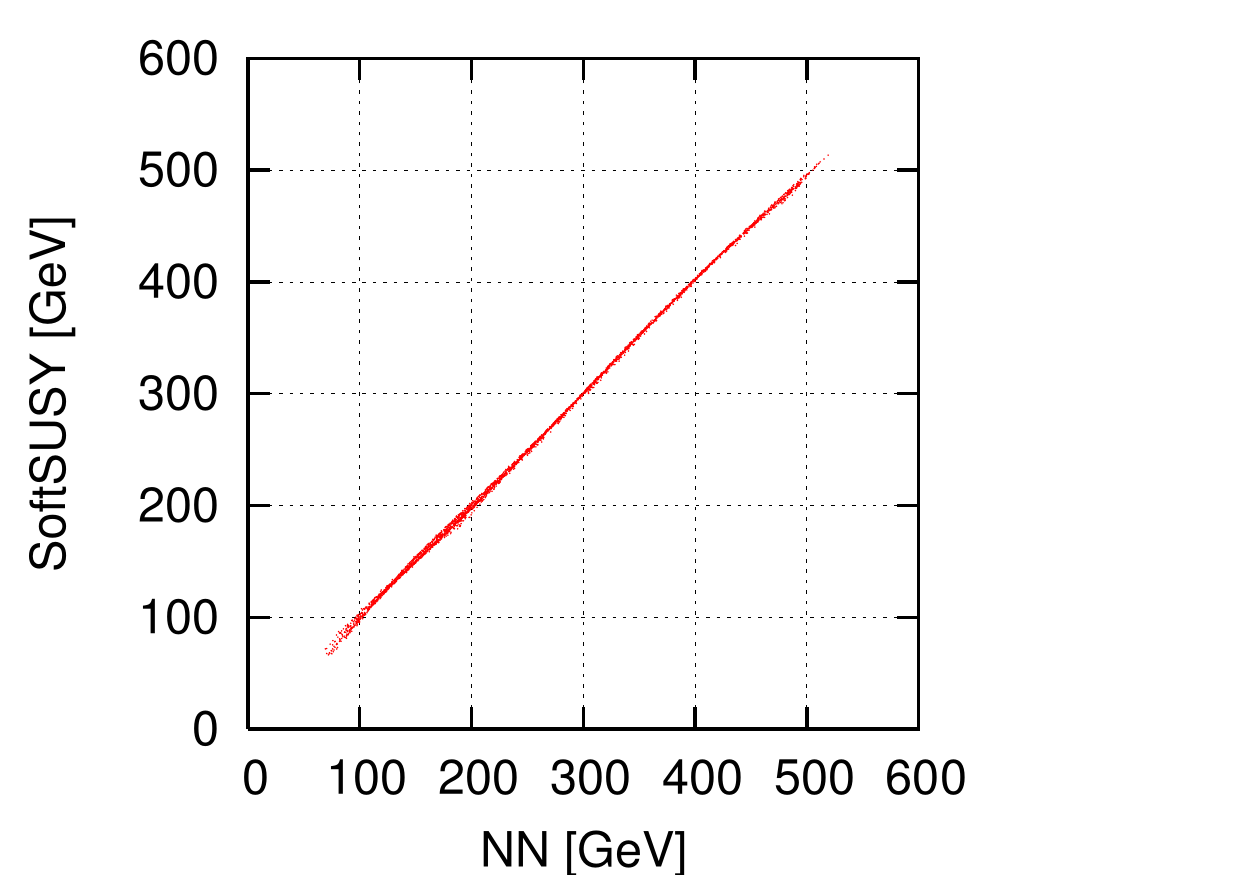}}
	          \subfigure[$m_{s\mu}^L$]{
          \includegraphics[width=.22\columnwidth, angle=-90]{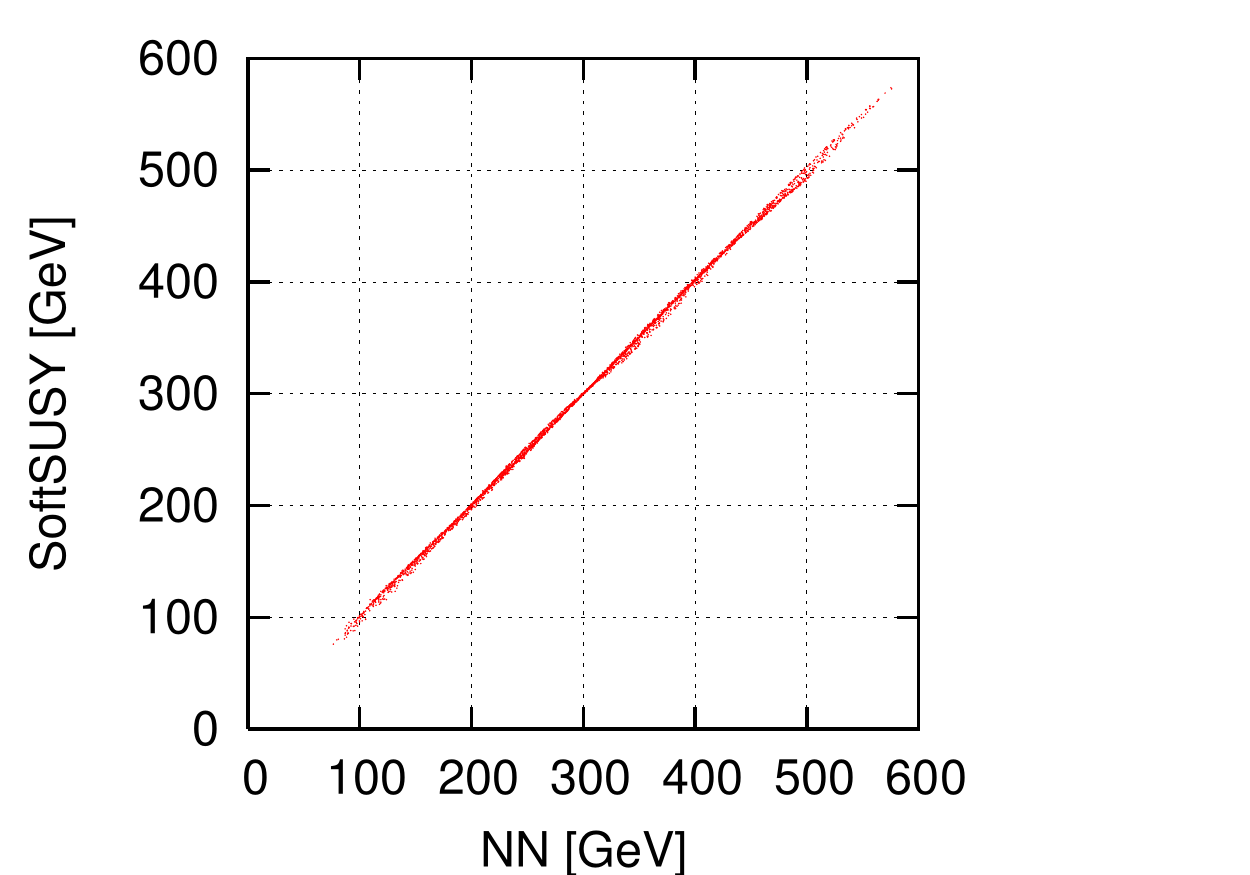}}\\
	         \subfigure[$m_{s\tau}^1$]{
          \includegraphics[width=.22\columnwidth, angle=-90]{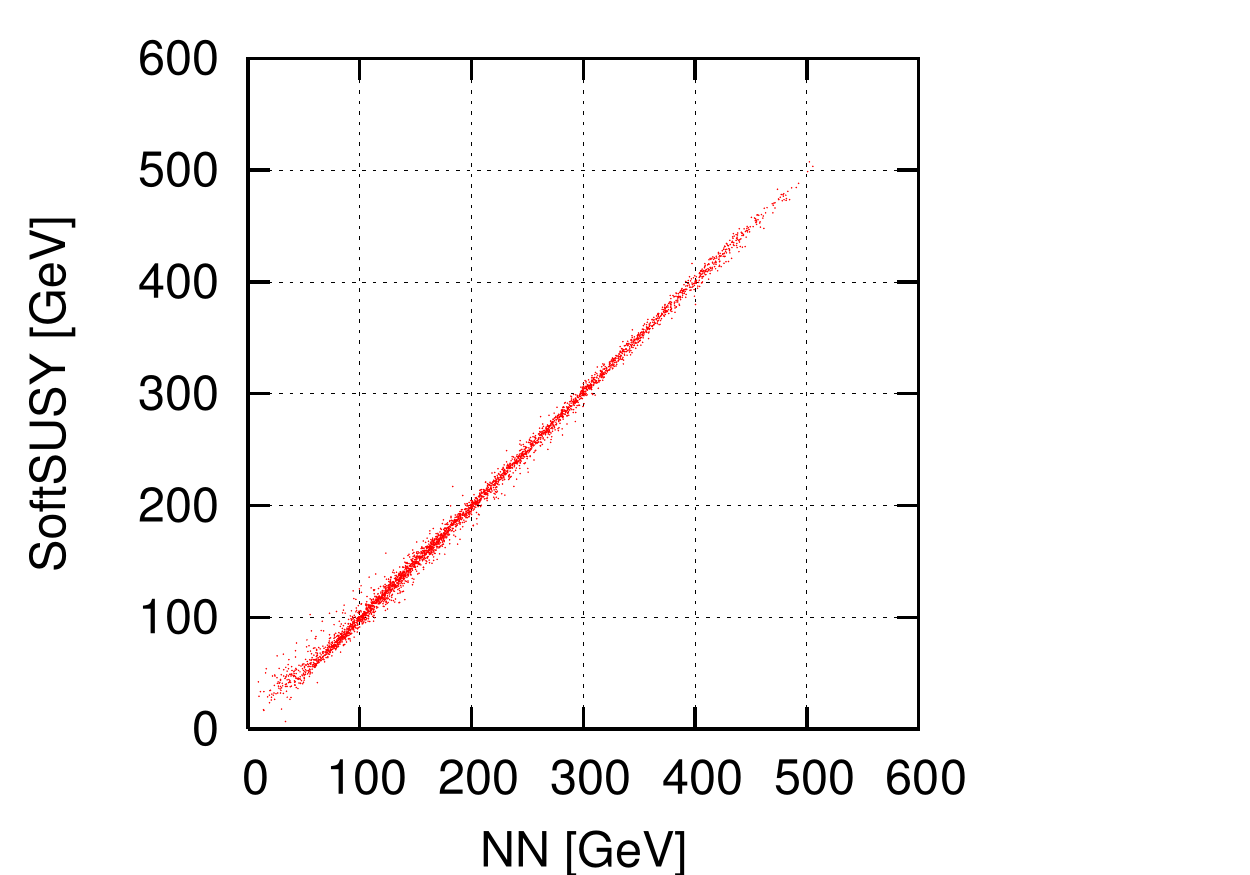}}
	         \subfigure[$m_{s\tau}^2$]{
          \includegraphics[width=.22\columnwidth, angle=-90]{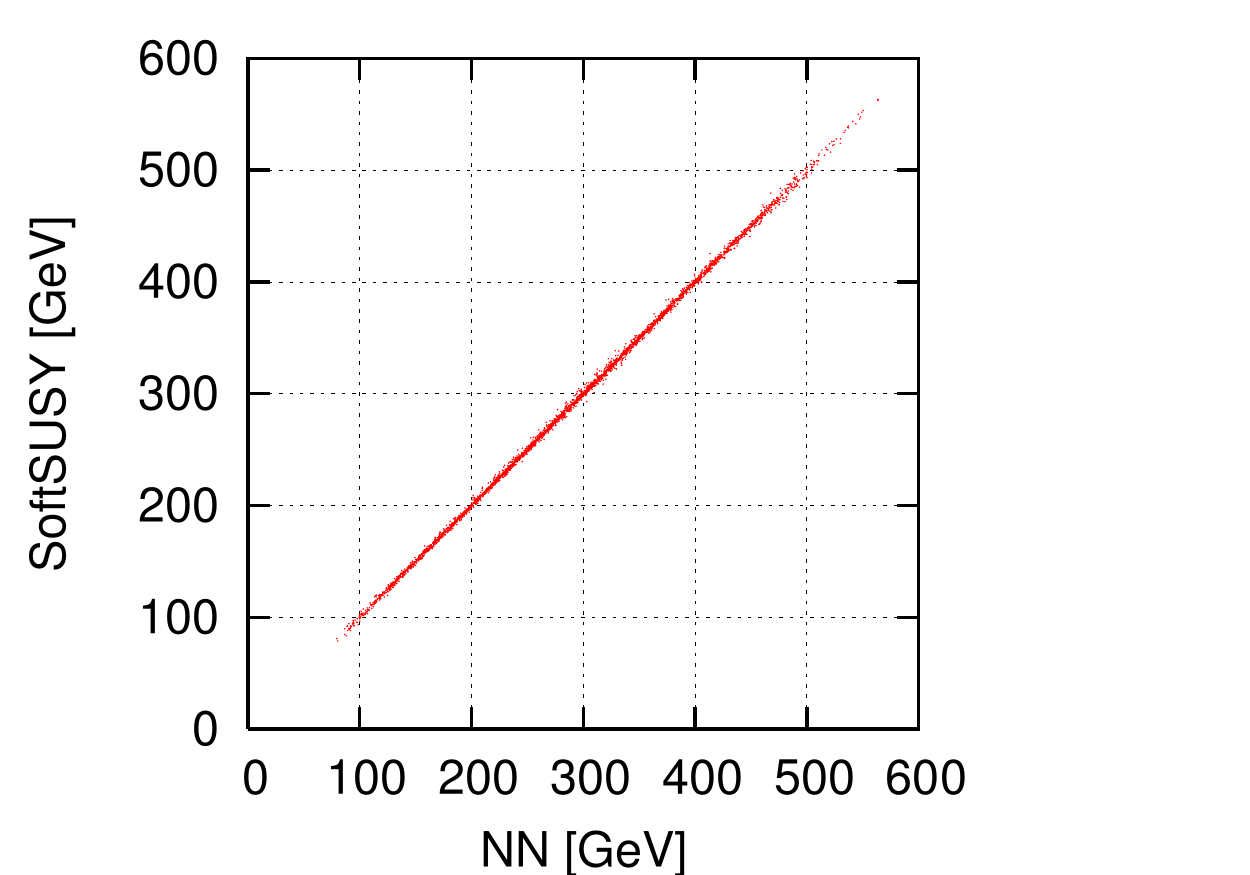}}
	          \subfigure[$m_{sd}^R$]{
          \includegraphics[width=.22\columnwidth, angle=-90]{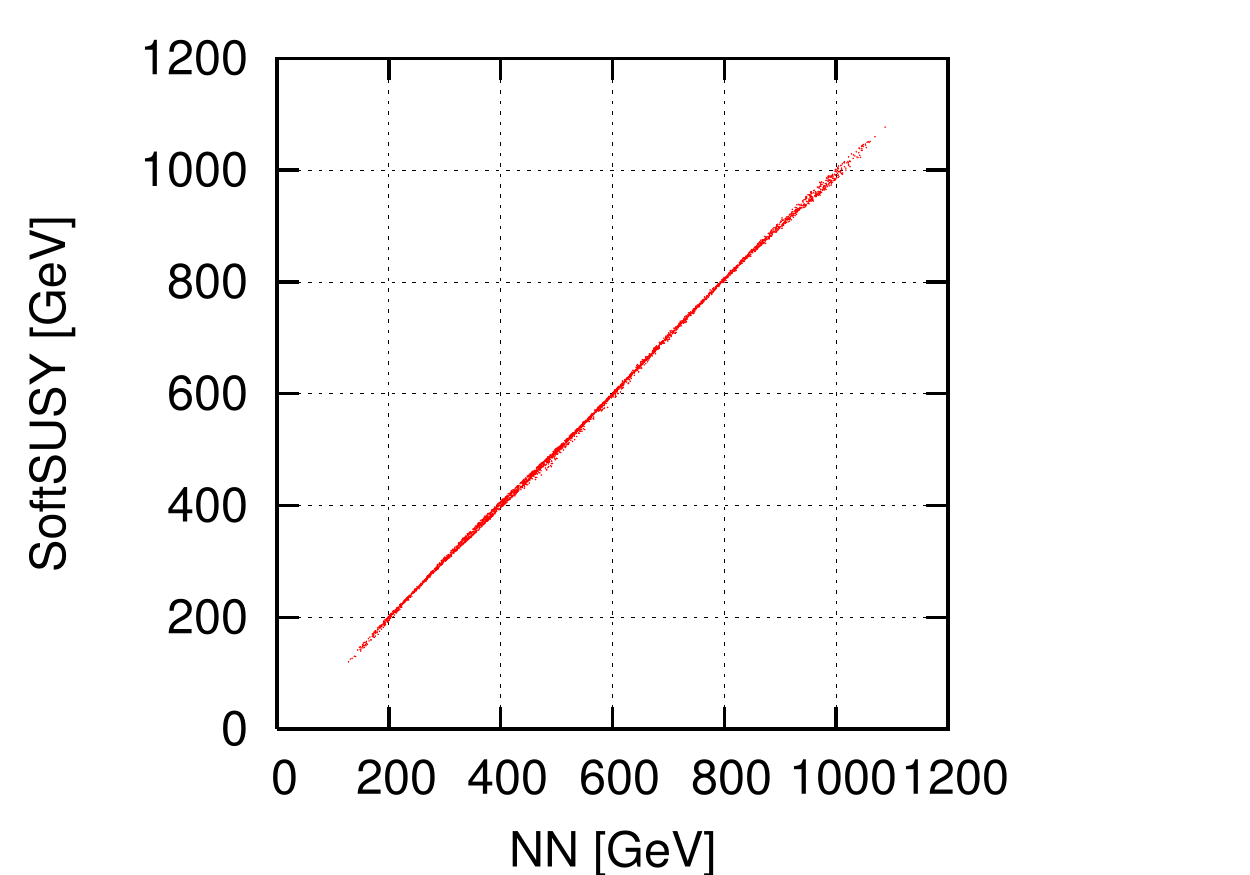}}\\
	          \subfigure[$m_{sd}^L$]{
          \includegraphics[width=.22\columnwidth, angle=-90]{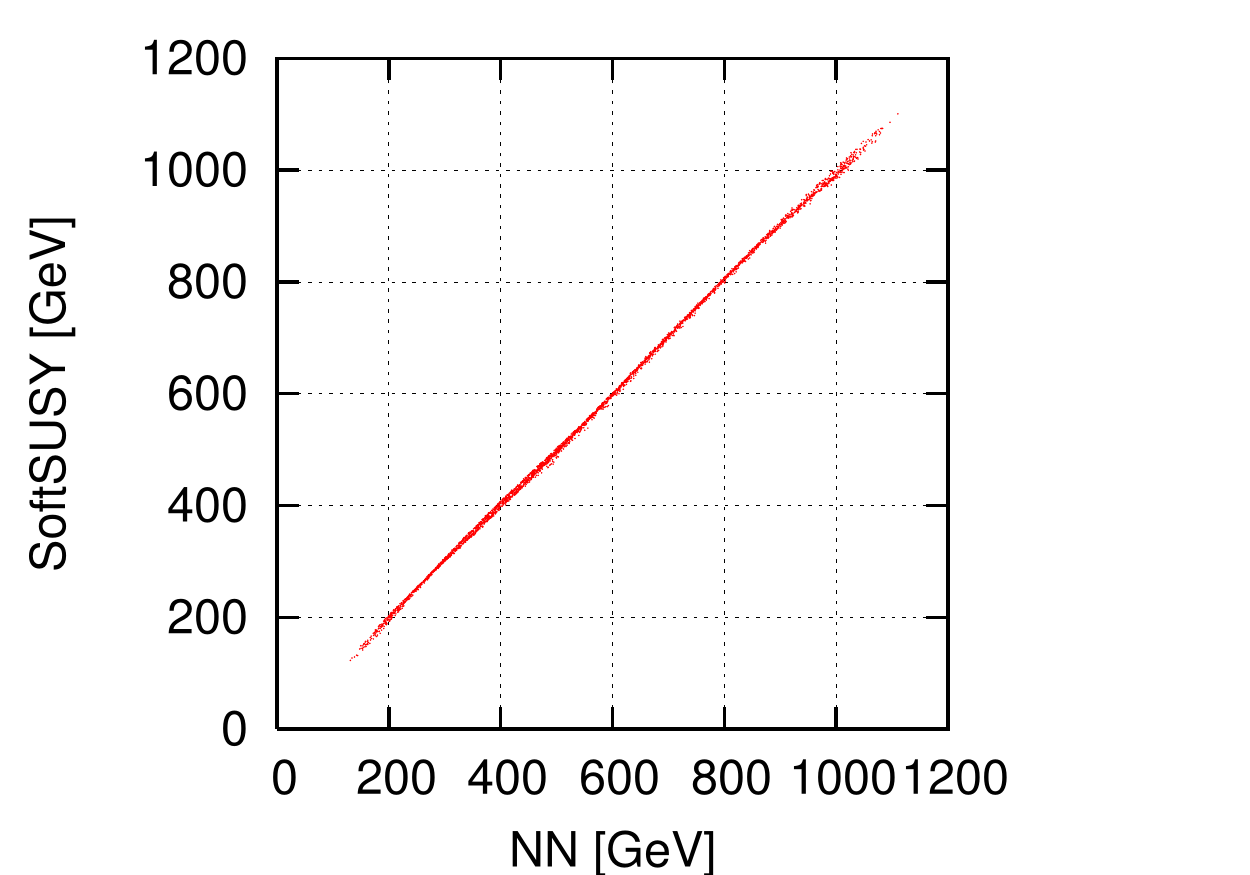}}
	         \subfigure[$m_{su}^R$]{
          \includegraphics[width=.22\columnwidth, angle=-90]{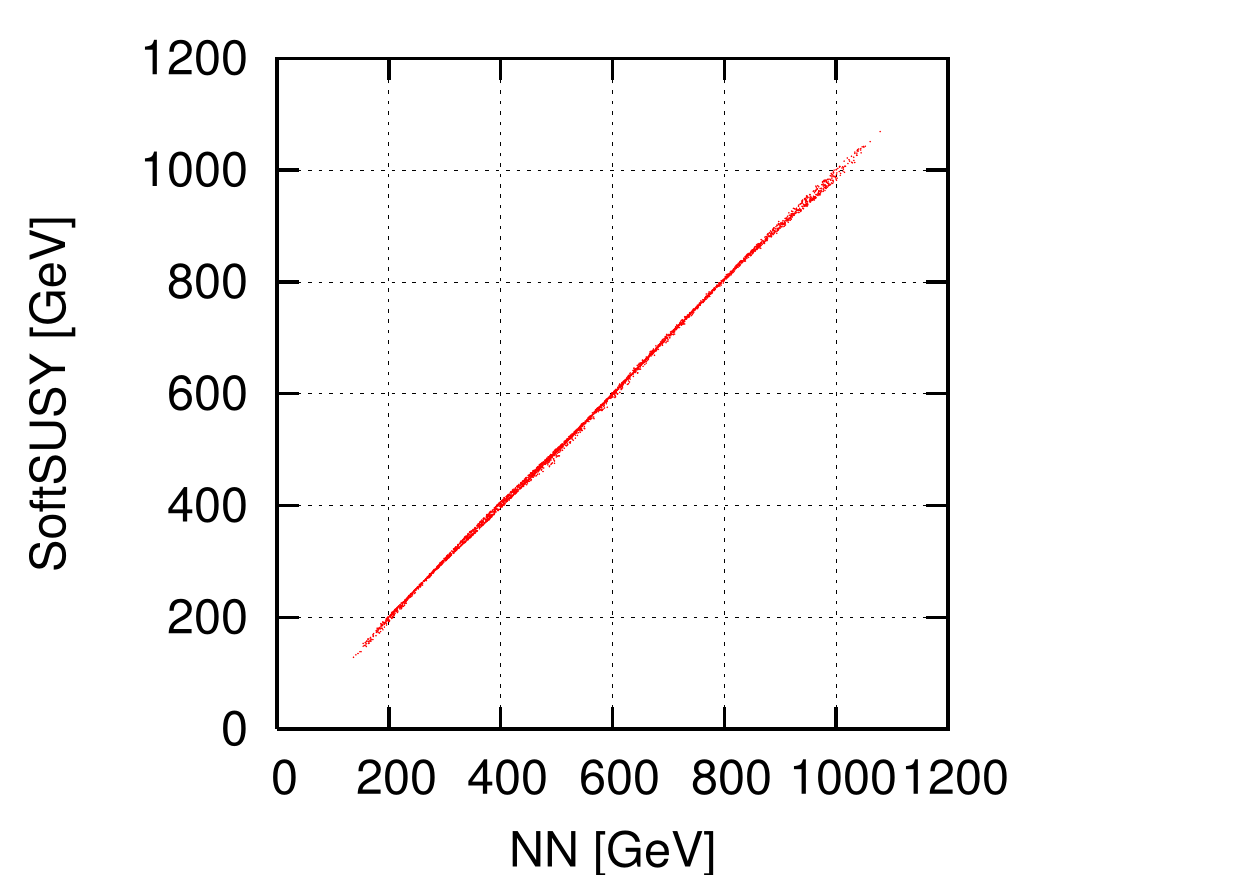}}
	          \subfigure[$m_{su}^R$]{
          \includegraphics[width=.22\columnwidth, angle=-90]{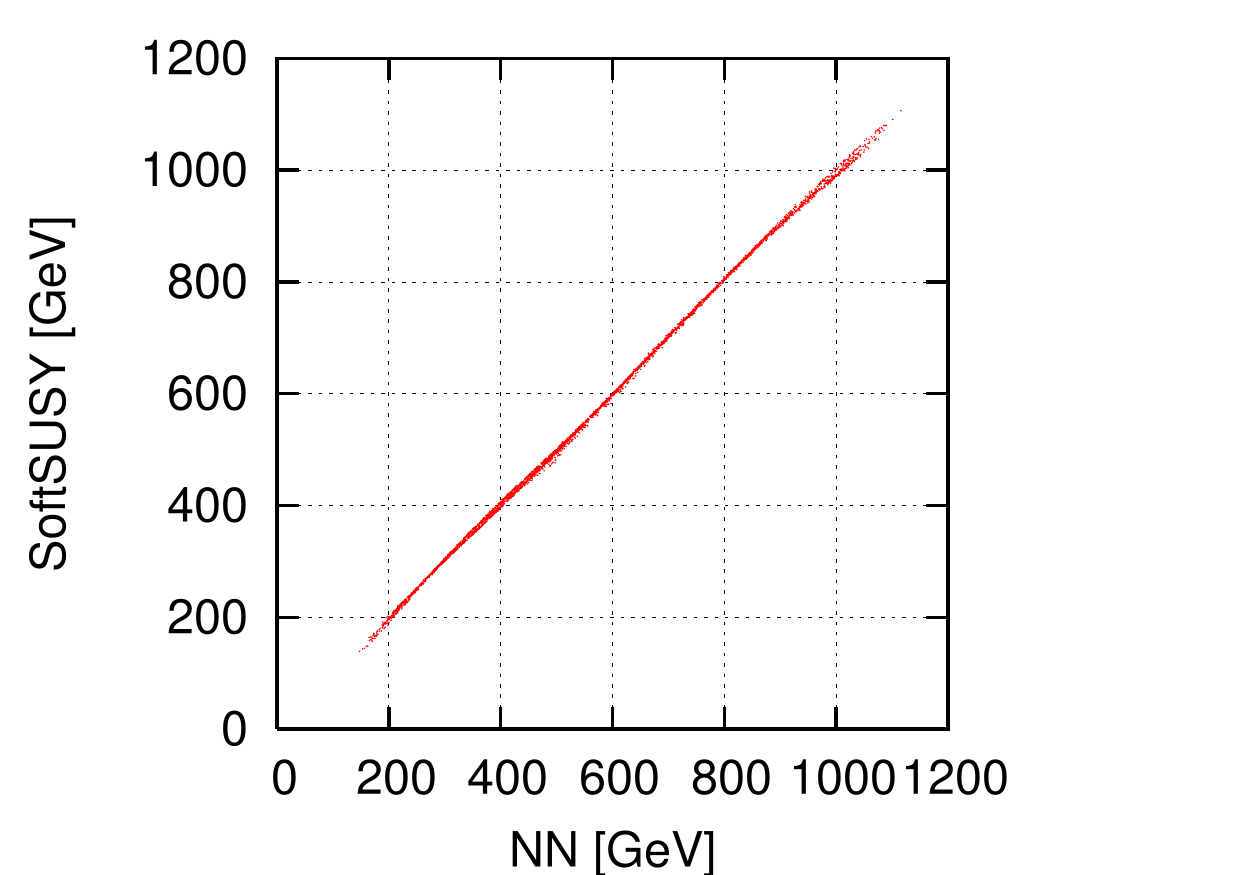}}\\
	          \subfigure[$m_{s\tau} (1,1)$]{
          \includegraphics[width=.22\columnwidth, angle=-90]{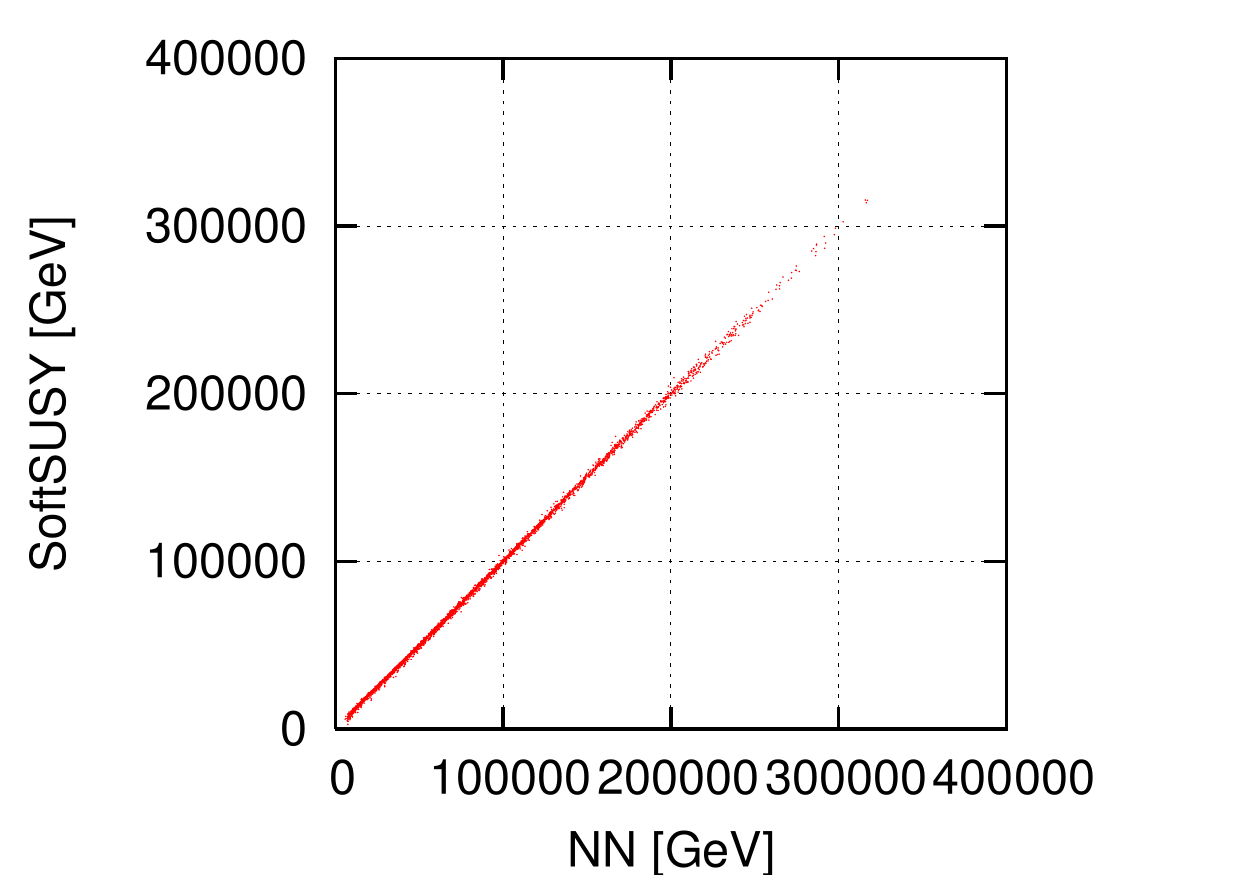}}
	         \subfigure[$m_{s\tau} (1,3)$]{
          \includegraphics[width=.22\columnwidth, angle=-90]{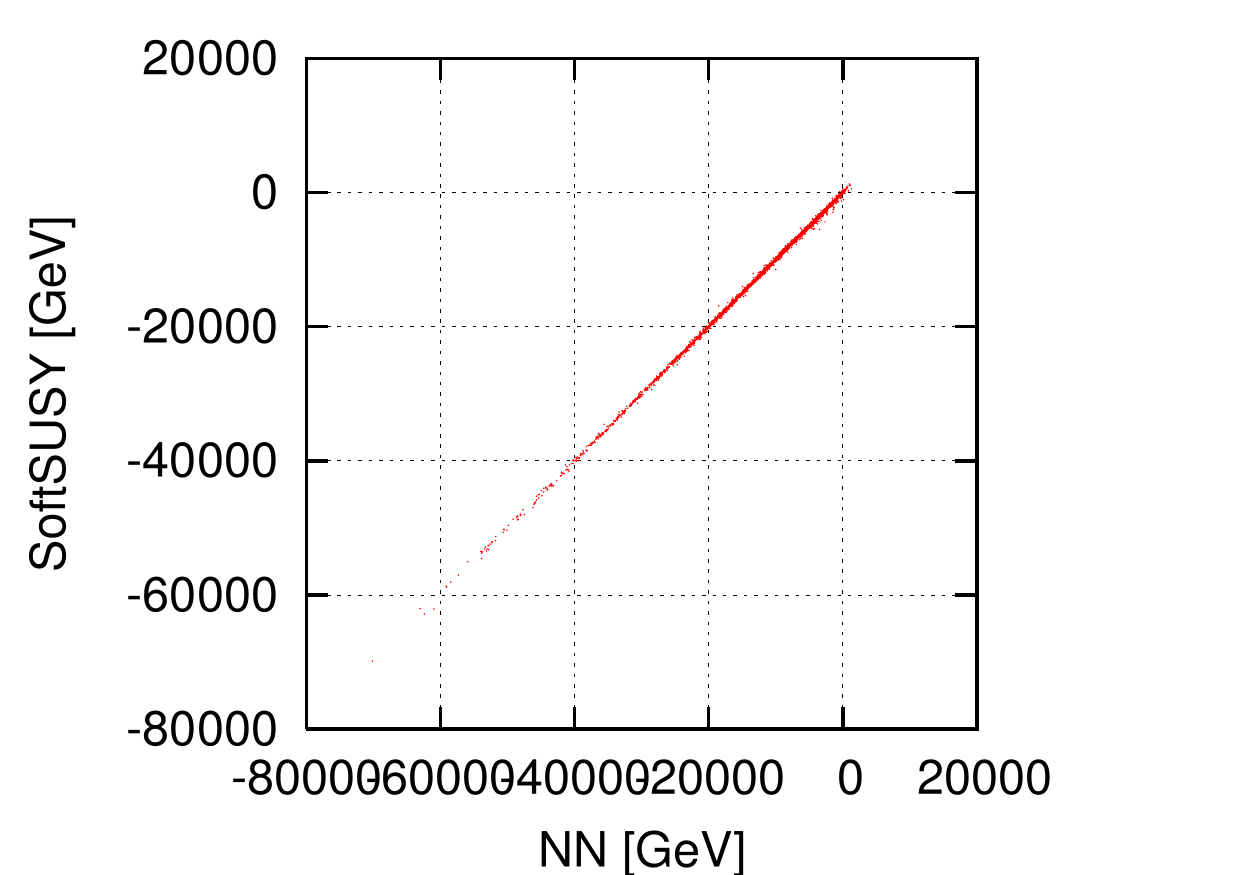}}
	         \subfigure[$m_{s\tau} (2,2)$]{
          \includegraphics[width=.22\columnwidth, angle=-90]{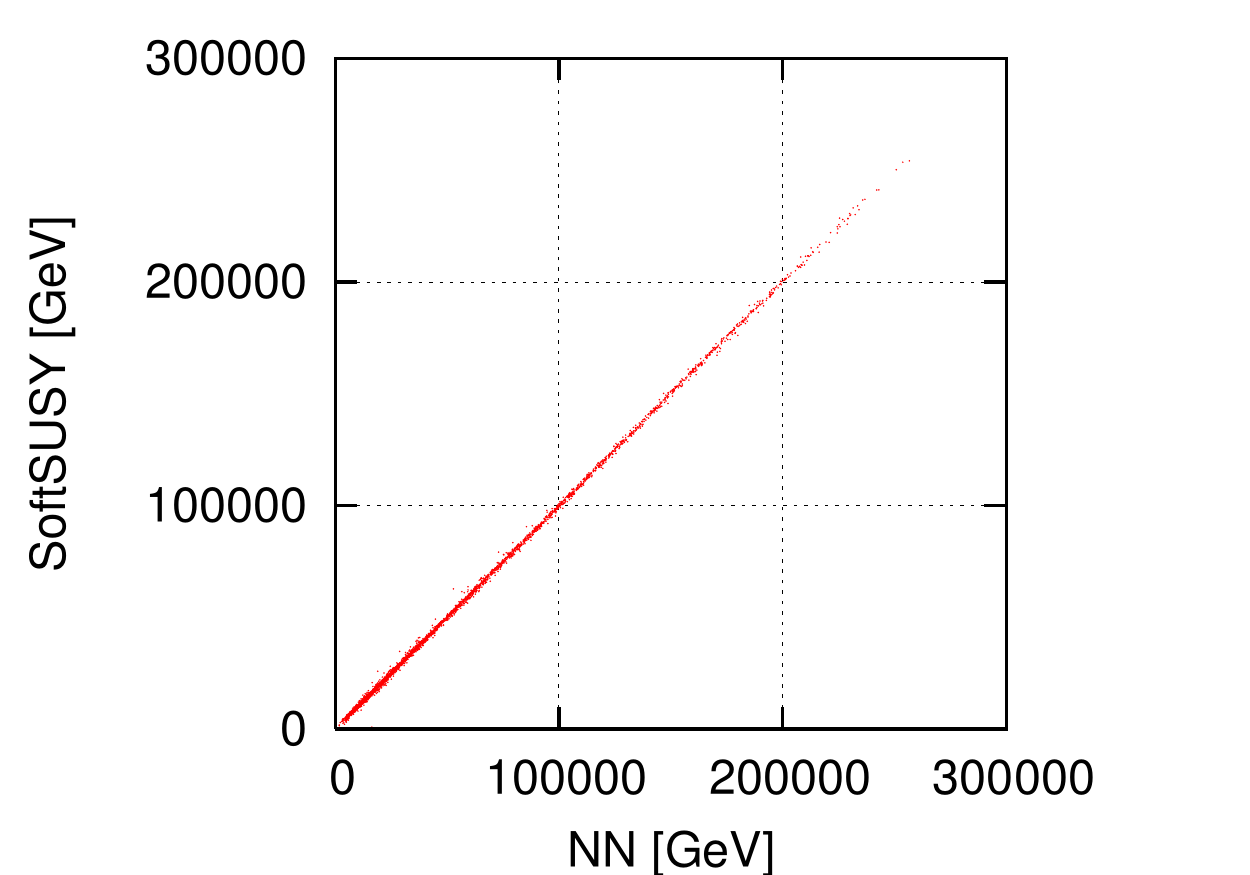}}
        \\	  
\caption{Correlation plots comparing the mass spectrum [GeV] output of 
{\sc SoftSUSY} with the neural network approximations {\sc NN}.  Subfigures (m)-(o) show the components of the mass matrix defining $m_{s\tau}^{1,2}$.}
\label{fig:correlations}
\end{figure}

The training data was produced by uniformly sampling roughly 4000 points within the
region of parameter space defined by Table \ref{table:priors}. Note that the region of
parameter space has been restricted to the vicinity of a SUSY benchmark point considered
in Section~\ref{sec:coverage}. A simple pre-processing step was used to speed up the
network optimization: this involves mapping all inputs and outputs linearly so they have
zero mean and a variance of one-half. While {\sc MemSys} provides algorithms for
optimising the network architecture, for our function approximation use-case we found
empirically that 10 hidden nodes provided a sufficiently accurate and computationally
efficient network architecture. 

Figure \ref{fig:correlations} show correlation plots of the {\sc SoftSusy} 
outputs with their neural network equivalents. Figures (a) - (l) illustrate 
the main mass spectrum components and all but $m_{s\tau}^1$ (g) demonstrated 
excellent correlation coefficients of $>$ 0.9999. The training problems for $m_{s\tau}^1$
(g), where the best correlation coefficient achieved was no greater than 0.998, 
was overcome by breaking the problem up into more basic components. 
In this case we noted that $m_{s\tau}^1$ and  $m_{s\tau}^2$ are eigenvalues 
of a (symmetric) $2 \times 2$ matrix. By asking the network
instead to interpolate over the 3 distinct matrix elements 
(shown in figures (m) -(o)) and then perform the eigenvalue decomposition 
independently,   it was possible to improve the final correlation on 
$m_{s \tau}^1$ to 0.9999. 

\begin{figure}
\begin{center}
\includegraphics[width=0.32\linewidth]{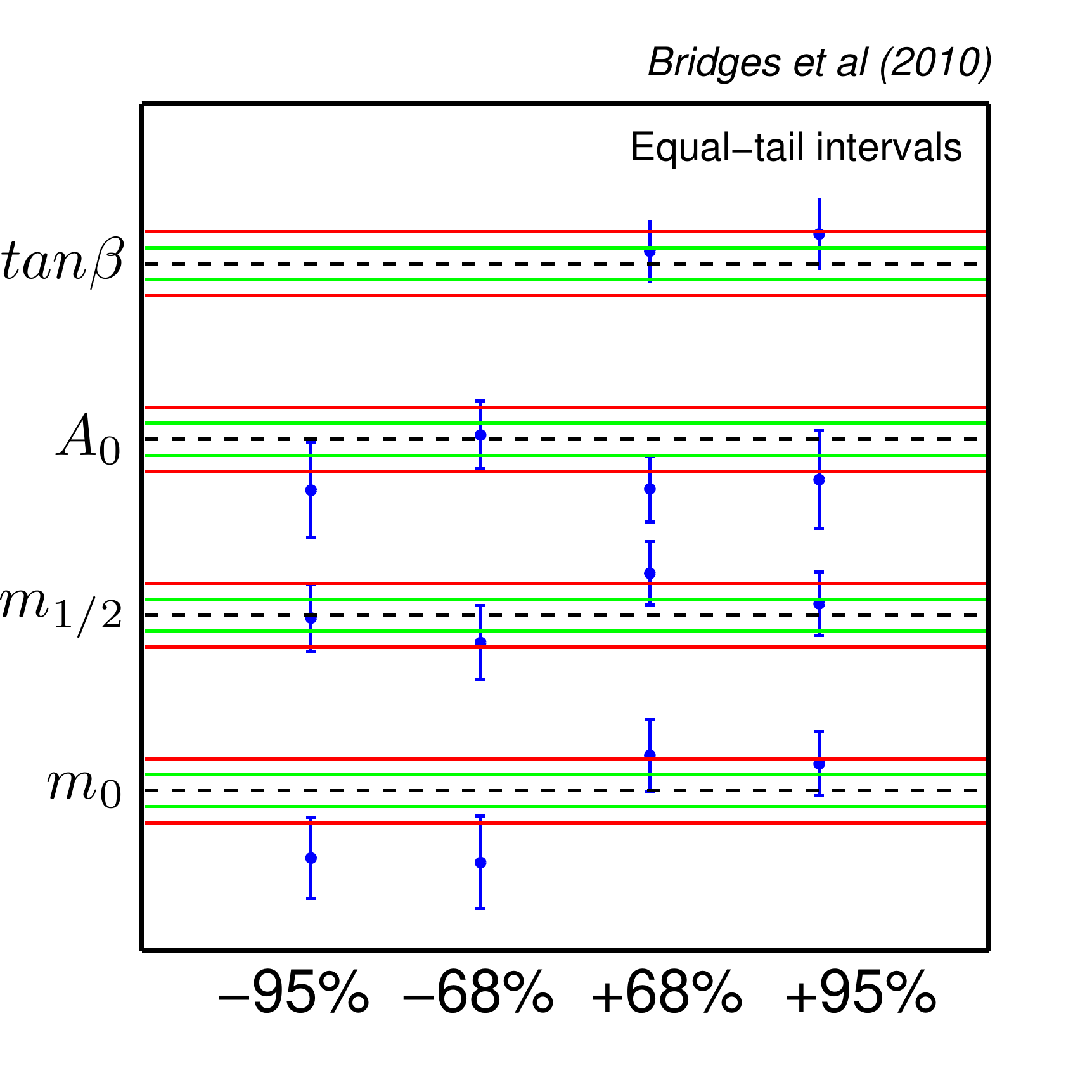}
\includegraphics[width=0.32\linewidth]{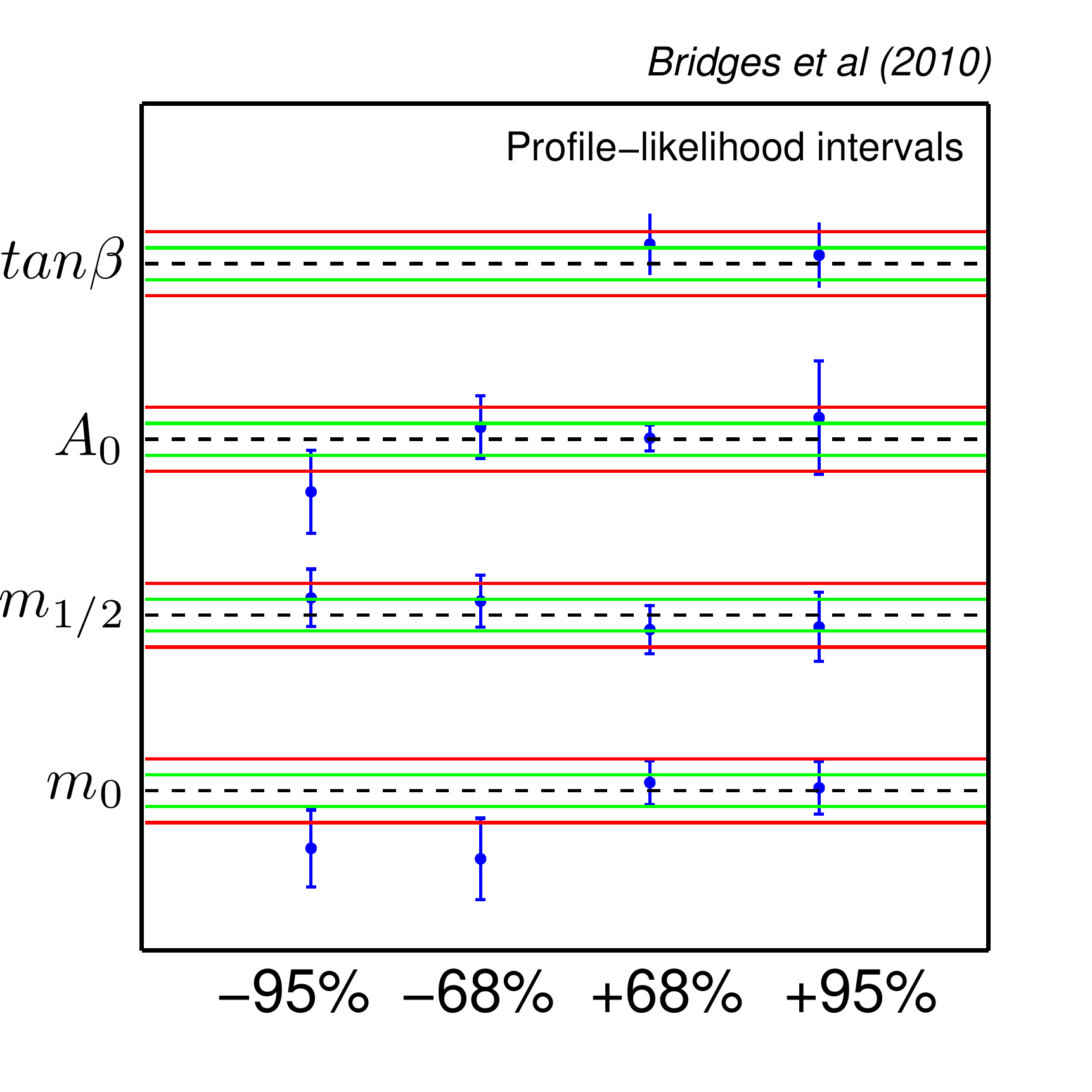}
\includegraphics[width=0.32\linewidth]{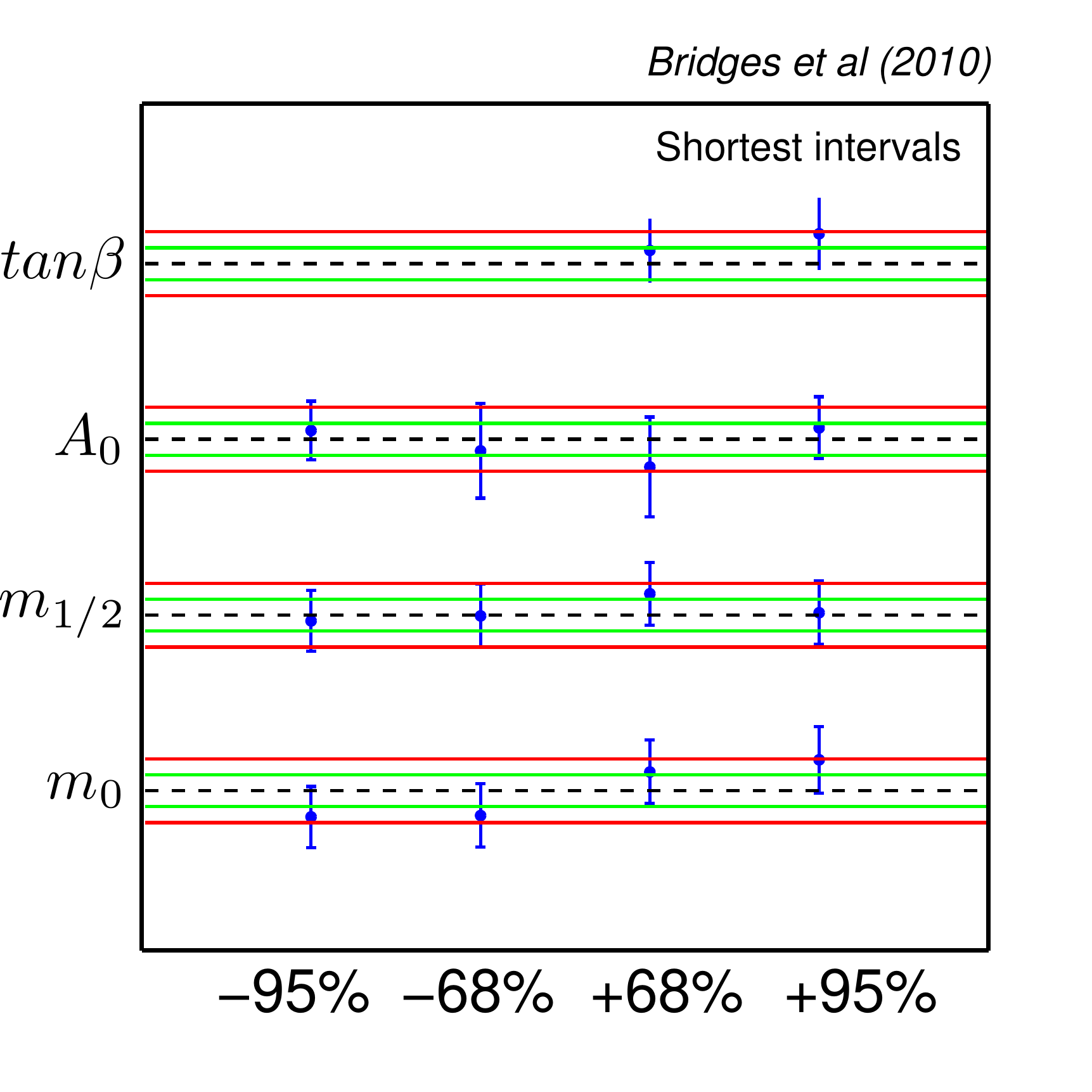}
\end{center}
\caption{\label{fig:noise} Evaluation of network noise. Horizontal red (green) lines denote error bands for the upper and lower boundaries of 68\% (95\%) intervals for the CMSSM parameters.  The red (green) lines denote $\pm 1\sigma$ ($\pm 2 \sigma$) from 100 runs of the {\sc SuperBayes} fitting package with {\sc SoftSusy} and quantify numerical/sampling noise. Units of the vertical axis have been rescaled appropriately for display purposes, so that all error bands have the same width. The blue error bars are the corresponding $2\sigma$ error bars based on $10^4$ runs using the neural network in place of {\sc SoftSusy} (with the corresponding rescaling). }
\end{figure}

\begin{figure}
\begin{center}
\includegraphics[width=0.32\linewidth]{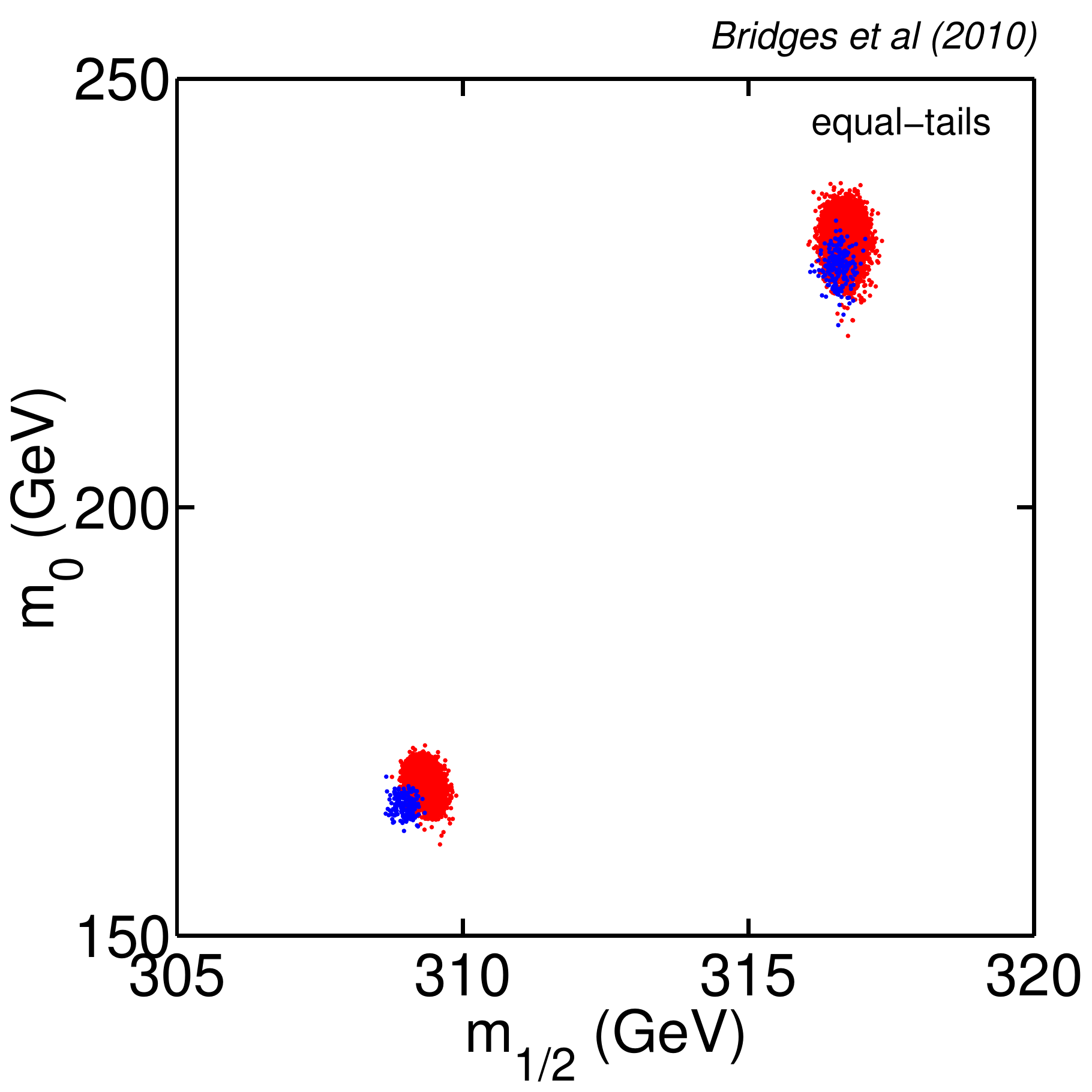}\includegraphics[width=0.32\linewidth]{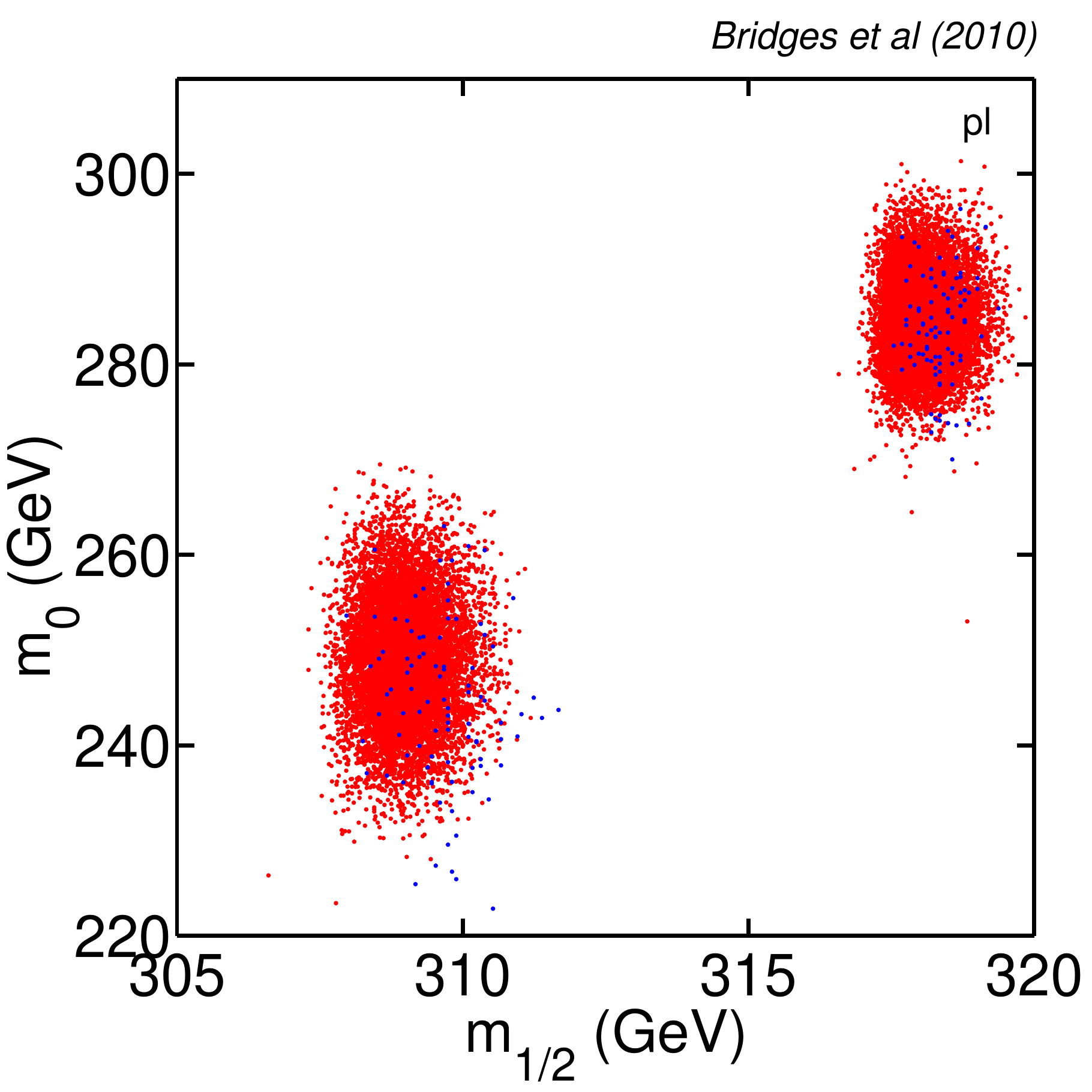}
\includegraphics[width=0.32\linewidth]{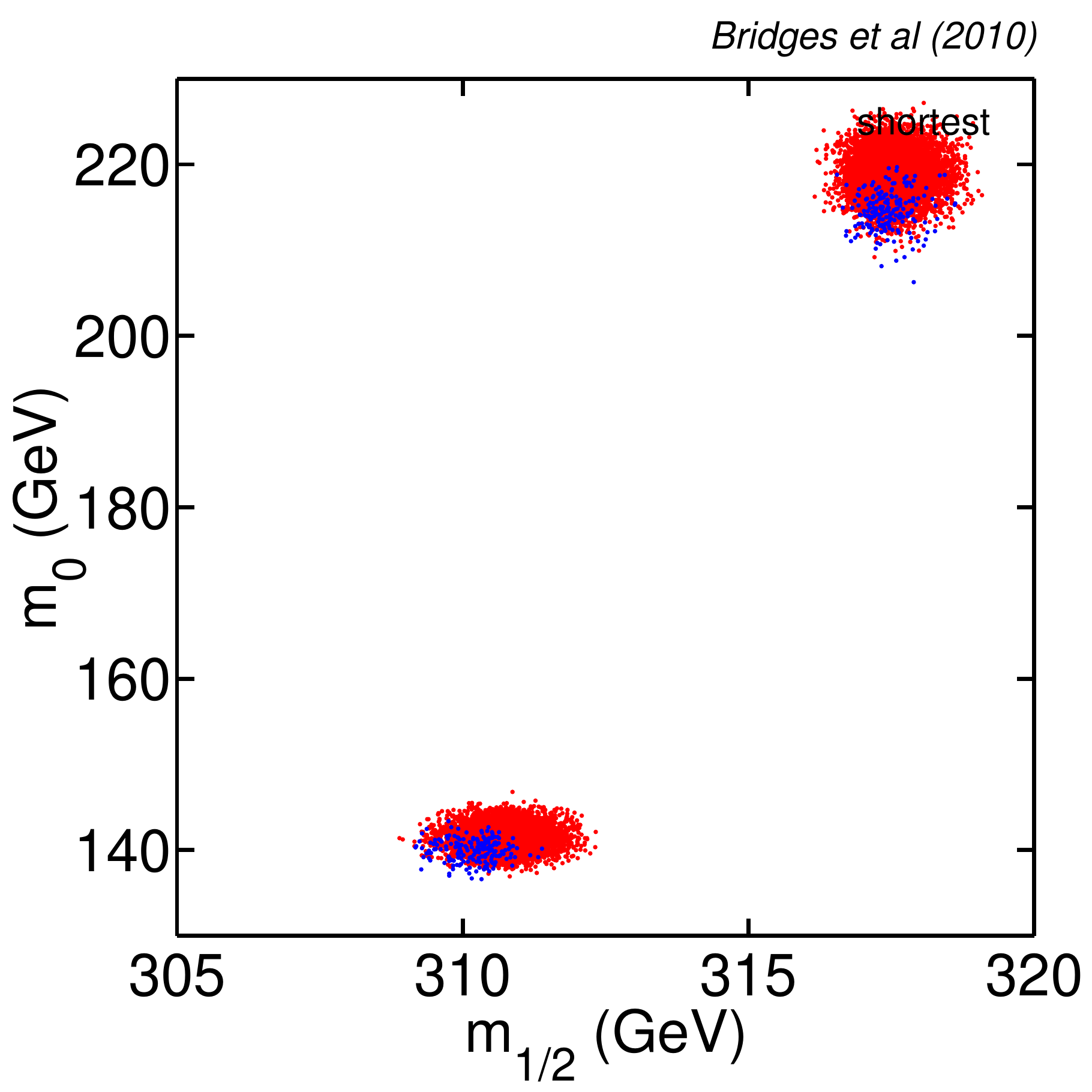}\\
\includegraphics[width=0.32\linewidth]{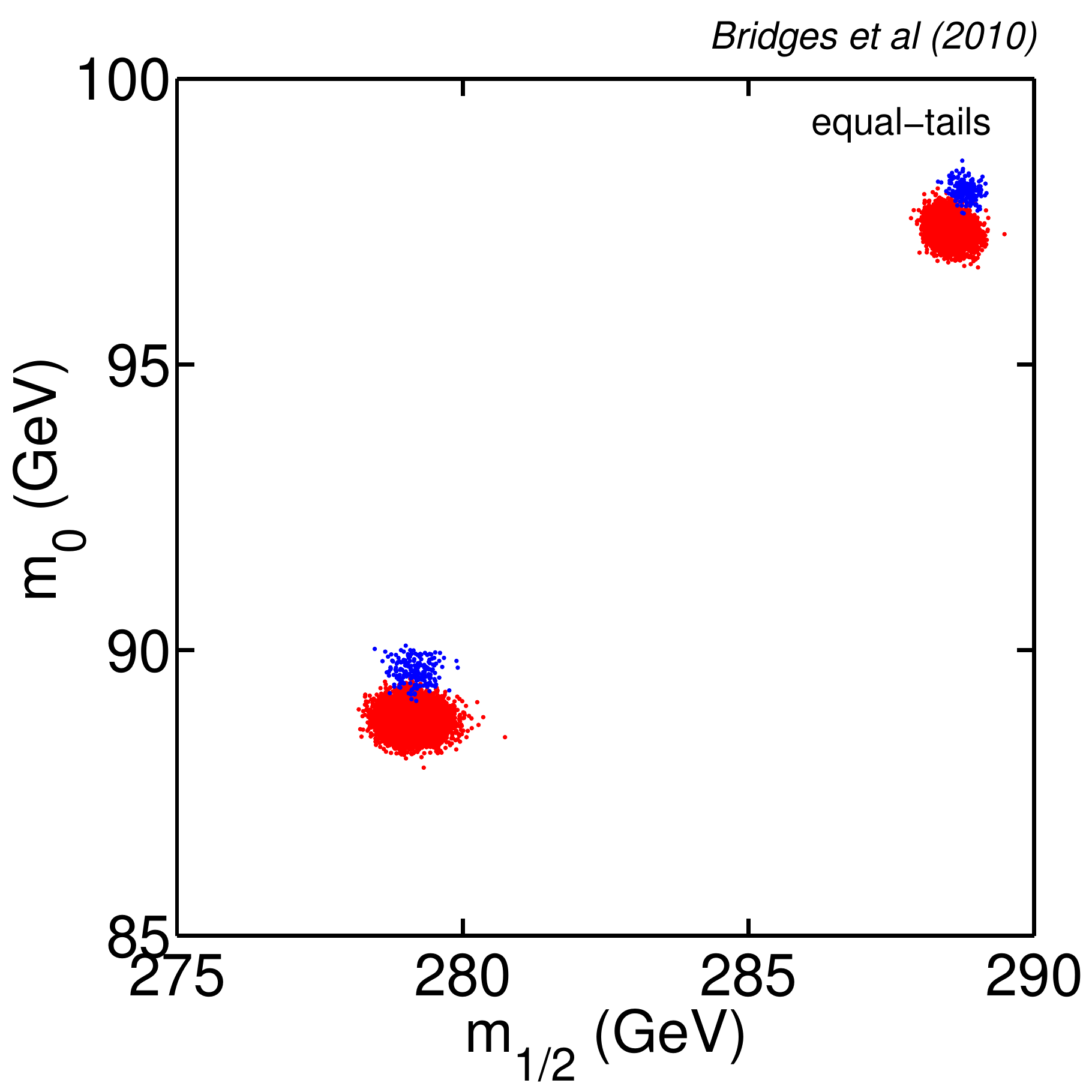} \includegraphics[width=0.32\linewidth]{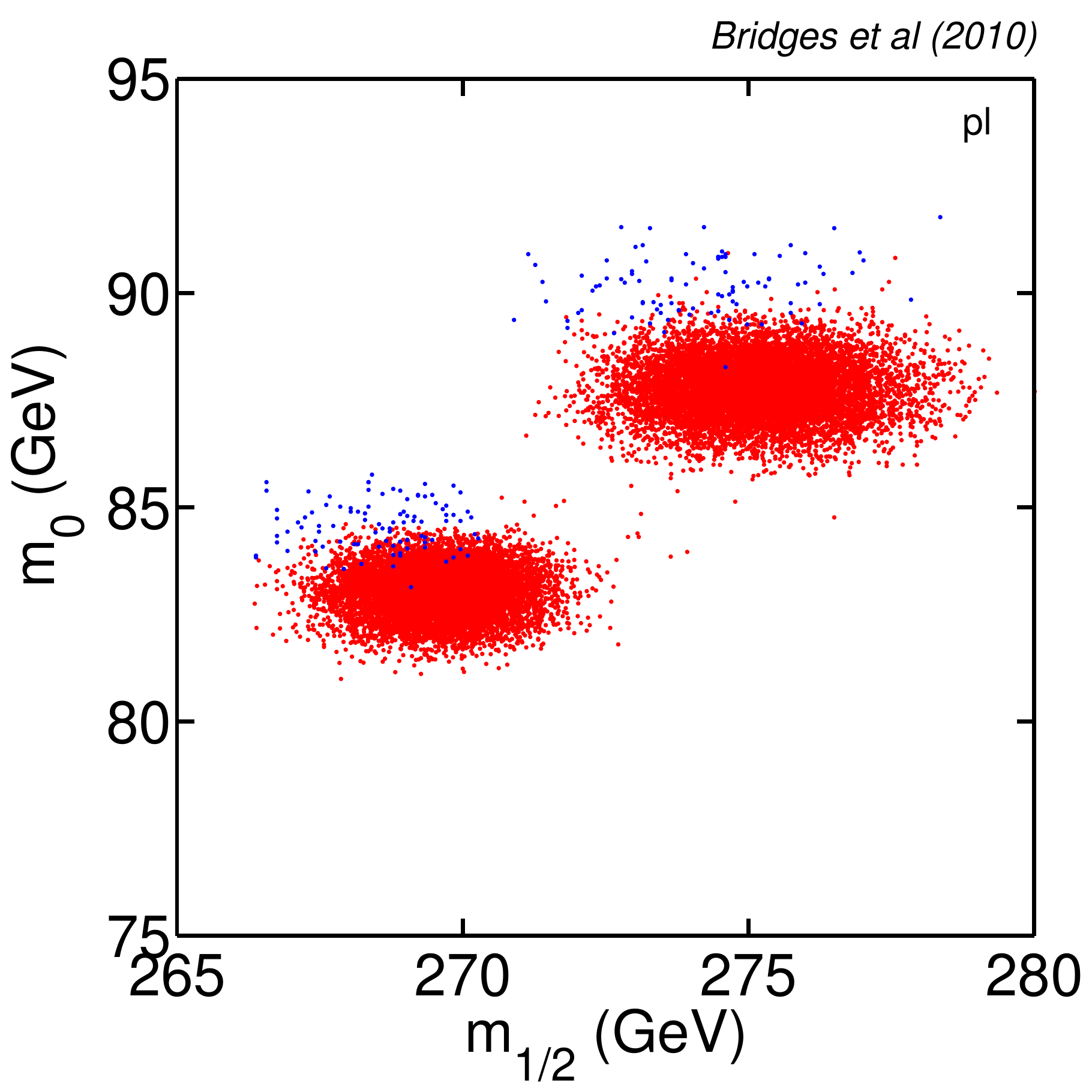}
\includegraphics[width=0.32\linewidth]{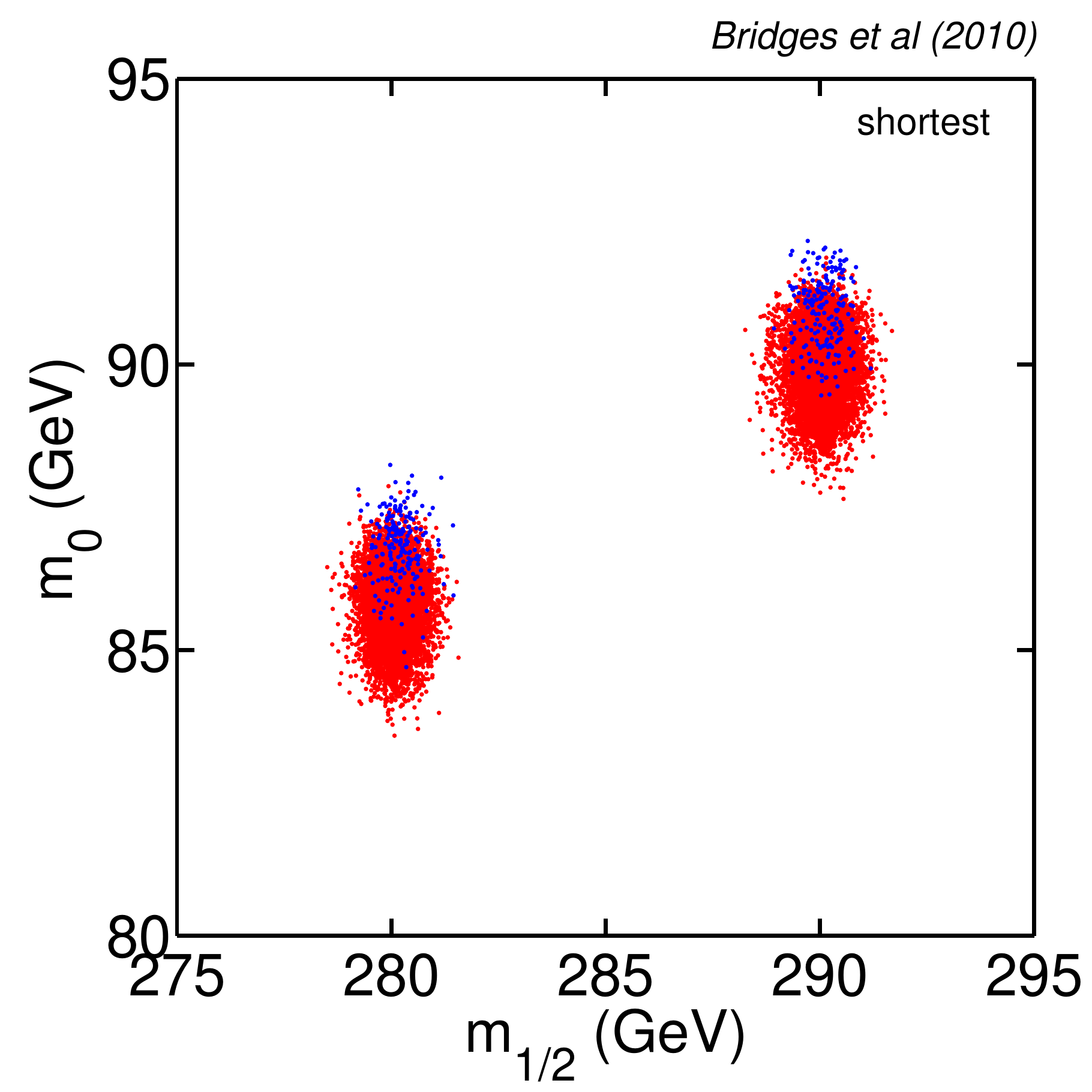} 
\end{center}
\caption{\label{fig:noise_2D} Evaluation of network noise. In the upper (lower) row, red dots give the location of the upper (lower) 68\% and 95\% intervals for $10^4$ runs using the neural network, while blue dots are the locations from 100 runs of the {\sc SuperBayes} fitting package with {\sc SoftSusy}.  }
\end{figure}

\begin{figure}
\begin{center}
\includegraphics[width=0.48\linewidth]{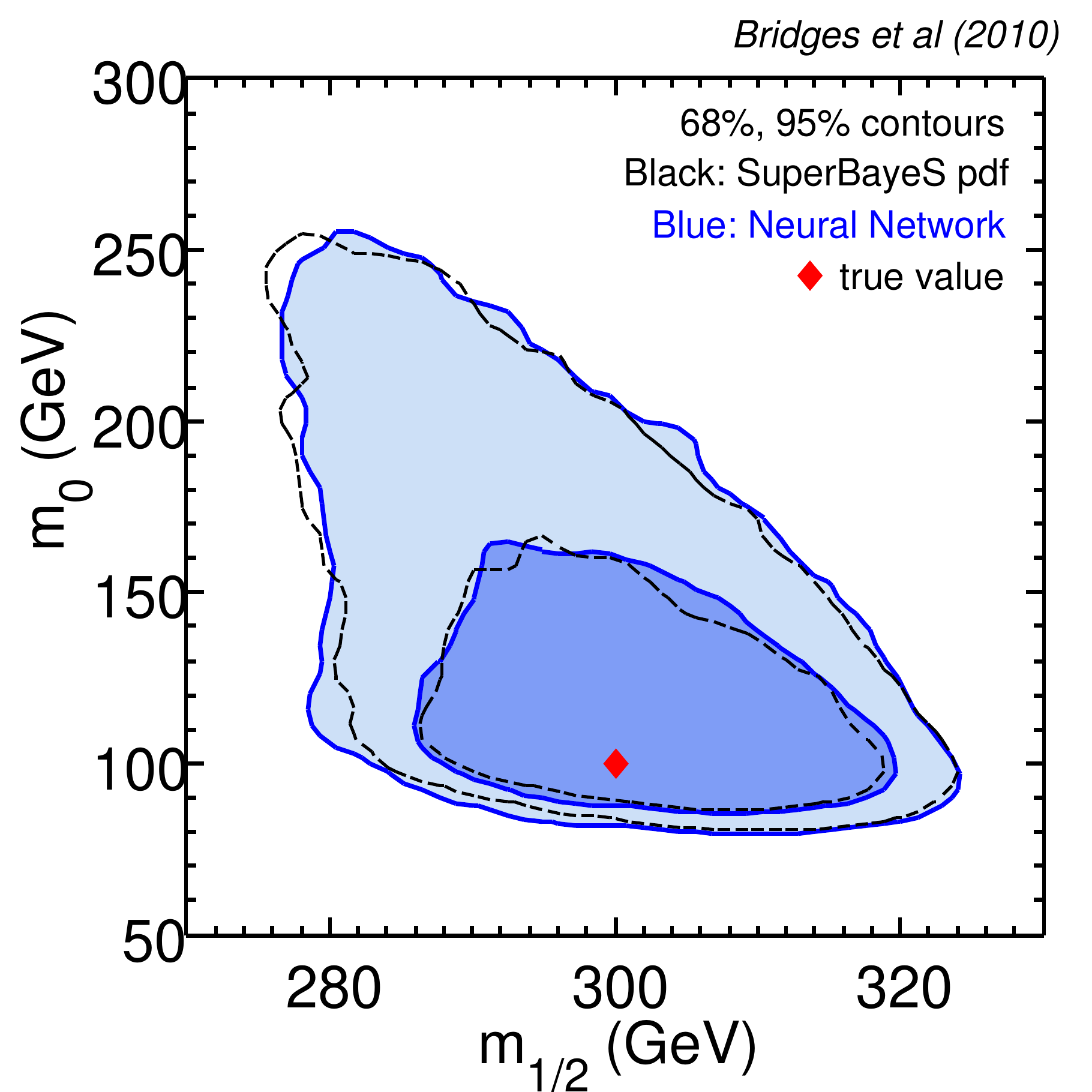}
\includegraphics[width=0.48\linewidth]{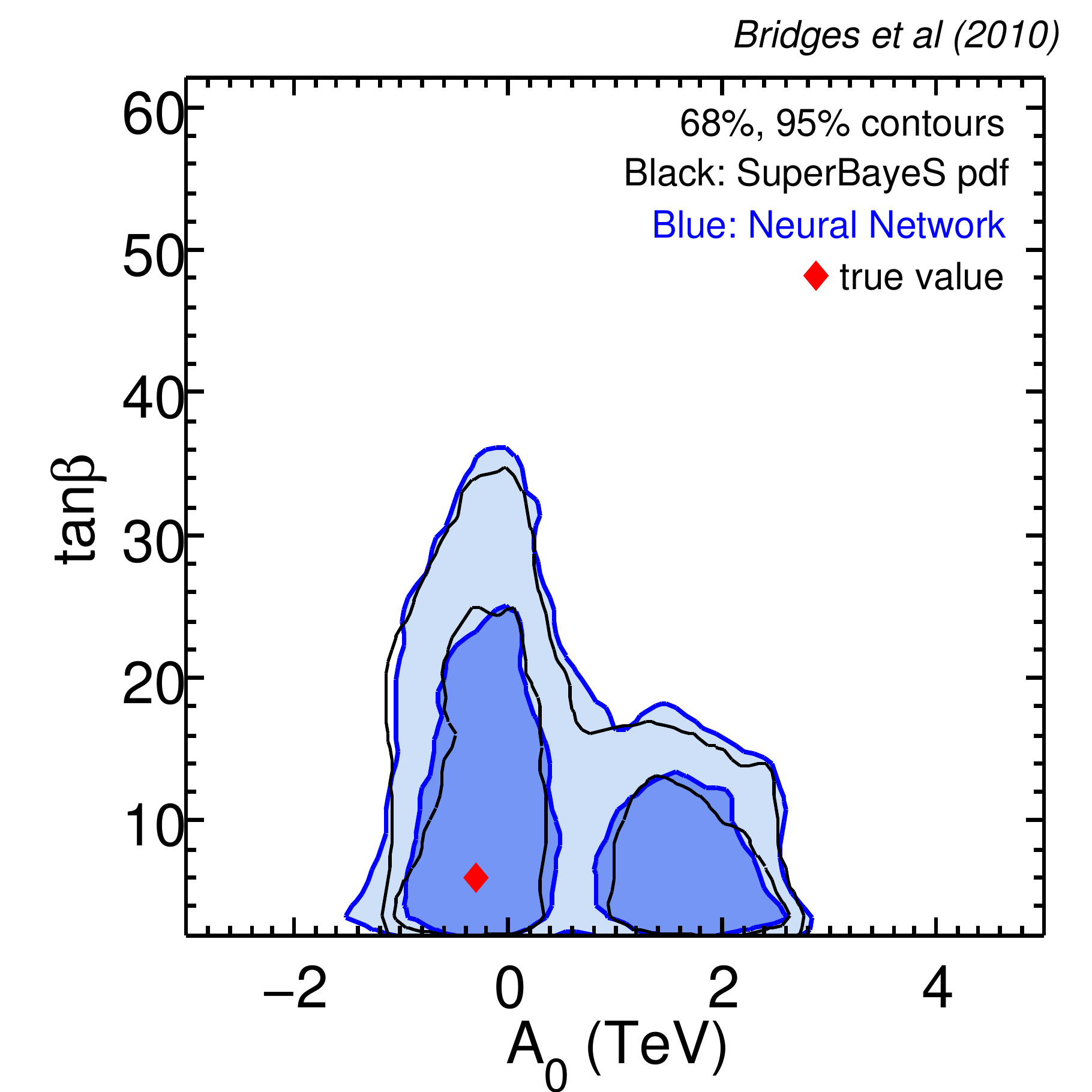}
\end{center}
\caption{\label{fig:nn_comparison} Comparison of Bayesian posteriors between runs of {\sc SuperBayes} with {\sc SoftSusy} (black lines, giving 68\% and 95\% regions) and neural networks (blue lines and corresponding filled regions) obtained using MultiNest, for a typical reconstruction. The agreement between the two methods is excellent, within numerical noise. The red diamond gives the true value for the benchmark point adopted.}
\end{figure}

The impact of replacing {\sc SoftSusy} with the network is summarized in Fig.~\ref{fig:noise} and \ref{fig:noise_2D}.  While the approximate mapping will slightly modify the obtained posterior distributions (network noise), this must be compared to the intrinsic variability (sampling noise) in the results due to the finite sampling of techniques such as MCMC and MultiNest. Figure~\ref{fig:noise} shows that the boundaries of univariate intervals from the two approaches tend to be compatible within about $2\sigma$ of the sampling noise. Figure~\ref{fig:noise_2D} shows two-dimensional scatter plots of  these boundaries.  Figure~\ref{fig:nn_comparison} compares the posterior obtained for a single run using the trained network with the posterior obtained with a run using {\sc SoftSusy}, showing the excellent agreement between the two. 

The neural network has been trained to perform to high accuracy only in the limited parameter range given in Table~\ref{table:priors}, since this was sufficient for the specific benchmark point being considered here. If one wanted to consider other regions of the CMSSM parameter space (e.g., the focus point), one would have to extend the network training to include such regions. While we have not experimented with this, based on our experience we believe that this should be straightforward to do.

\subsection{A classification network for unphysical parameter values}

A very high proportion of points sampled from within the full
parameter range are rejected by {\sc SoftSusy} as
\emph{unphysical}. Furthermore there is no trivial way to identify
such points when sampling without resorting to computationally
expensive calculations. This is demonstrated by the large degree of
overlap seen in Figure \ref{fig:holes} between physical (upper points, green) and unphysical (lower points, red) samples in the $m_0, m_{1/2}$ plane. Such
unphysical samples have no valid mass spectrum and so will lie outside of the
trained region of the function approximation networks described in the
previous section.  Thus, we require a means to exclude these points
while sampling. Neural networks can be trained to perform for
classification such as this, with modification of the training
objective function. In this section we will describe classification
networks and how they have been implemented for this application.

The aim of classification is to place members of a set into subsets
based on inherent properties or \emph{features} of the individuals,
given some pre-classified training data.  Formally, classification can
be summarised as finding a classifier $\mathcal{C} : \Theta
\rightarrow C$ which maps an object from some (typically
multi-dimensional) feature space $\Theta$ to its classification label
$C$, which is typically taken as one of $ \left\{1, ...,N \right\}$
where $N$ is the number of distinct classes.  Thus the problem of
classification is to partition feature space into regions, assigning
each region a label corresponding to the appropriate classification.
In our context, the aim is to classify points in the parameter space
$\Theta$ into one of two classes, physical or unphysical (hence
$N=2$), given a set of training data $\mathcal{D}=(\Theta^{(k)}, t^{(k)})$,
where $t^{(k)}$ is a vector of dimension $N$ that encodes the class label by placing unity in the $C^{\rm th}$ component and zero elsewhere.

In building a classifier using a neural network, it is convenient to
view the problem \emph{probabilistically}. To this end we again
consider a 3-layer MLP consisting of an input layer ($\Theta_l$), a
hidden layer ($h_j$), and an output layer ($u_i$). In classification networks,
however, the outputs are transformed according to the \emph{softmax}
procedure
\begin{equation}
u'_i = \frac{e^{u_i}}{\sum_i e^{u_i}},
\end{equation}
such that they are all non-negative and sum to unity. In this way
$u'_i$ can be interpreted as the probability that the input feature
vector $\Theta$ belongs to the $i$th class. A
suitable objective function for the classification problem is then
\begin{equation}
\mathcal{L}(\mathbf{b}) = \sum_{k} \sum_{i}
t^{(k)}_i \ln u'_i(\Theta^{(k)}, \mathbf{b}),
\end{equation}
where $k$ is an index over the training dataset and $i$ is an index over the $N$ classes.  One then wishes to choose
network parameters $\mathbf{b}$ so as to maximise this objective
function as the training progresses. Training of this network was
again performed using the {\sc MemSys} package as previously
described. In our context, the network has just two (transformed) 
outputs $u'_1$ and $u'_2$, which give the probabilities that the input
$\Theta$  is physical or unphysical. We partition $\Theta$ space, albeit
with some loss of information, by labeling regions according to
whether $u'_1$ or $u'_2$ exceeds some threshold value.

The training data was taken as a set of uniform samples of roughly
30,000 prior samples equally divided between physical and unphysical
status. Again a pre-processing step was carried out so that all inputs
were re-mapped to have zero mean and variance of one-half to improve
training efficiency. The number of hidden nodes was selected manually
and nine hidden nodes were found to perform well.

\begin{figure}  
\begin{center} 
{\includegraphics[width=0.33\linewidth, angle =-90]{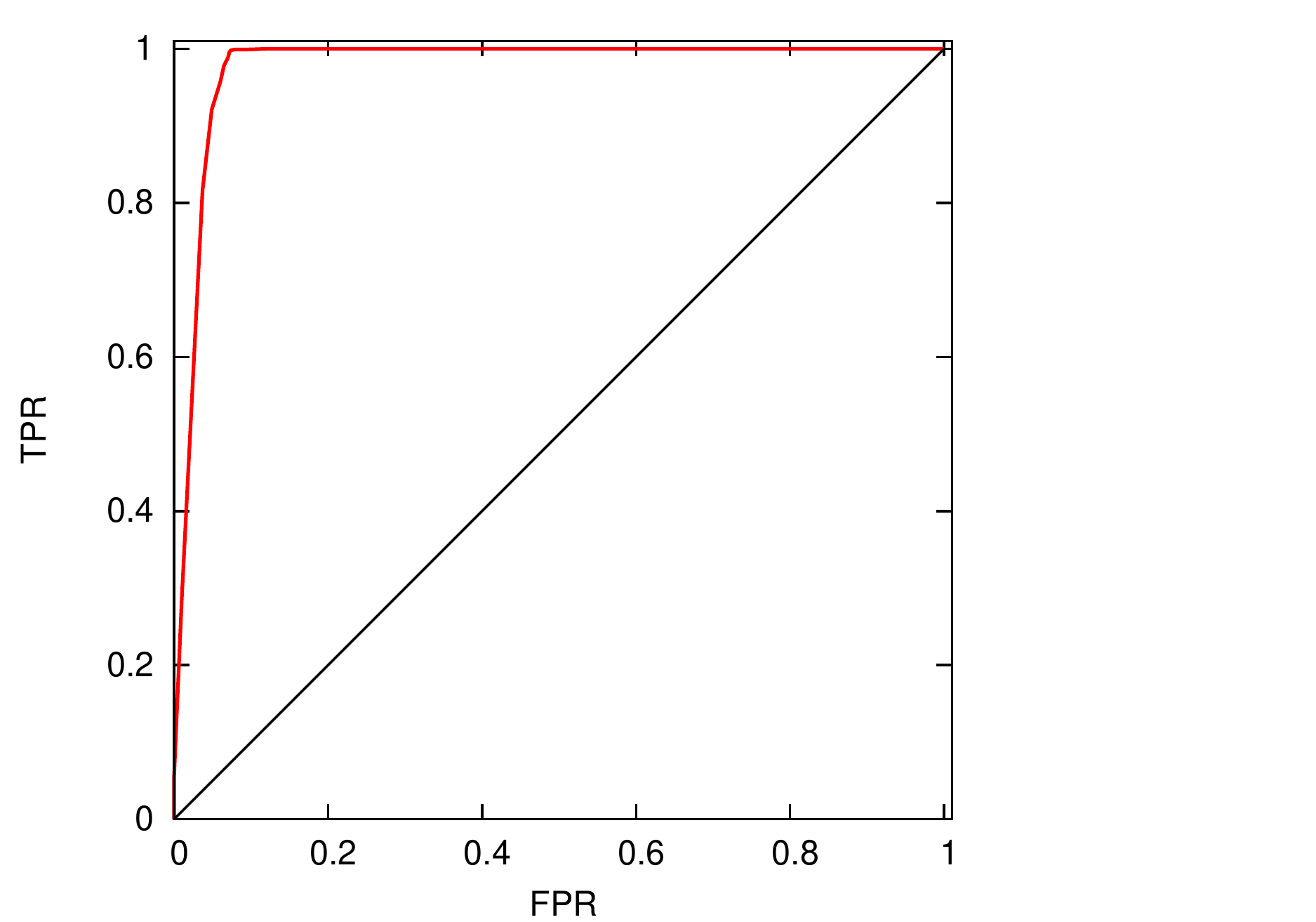} }
{\includegraphics[width=0.33\linewidth, angle = -90]{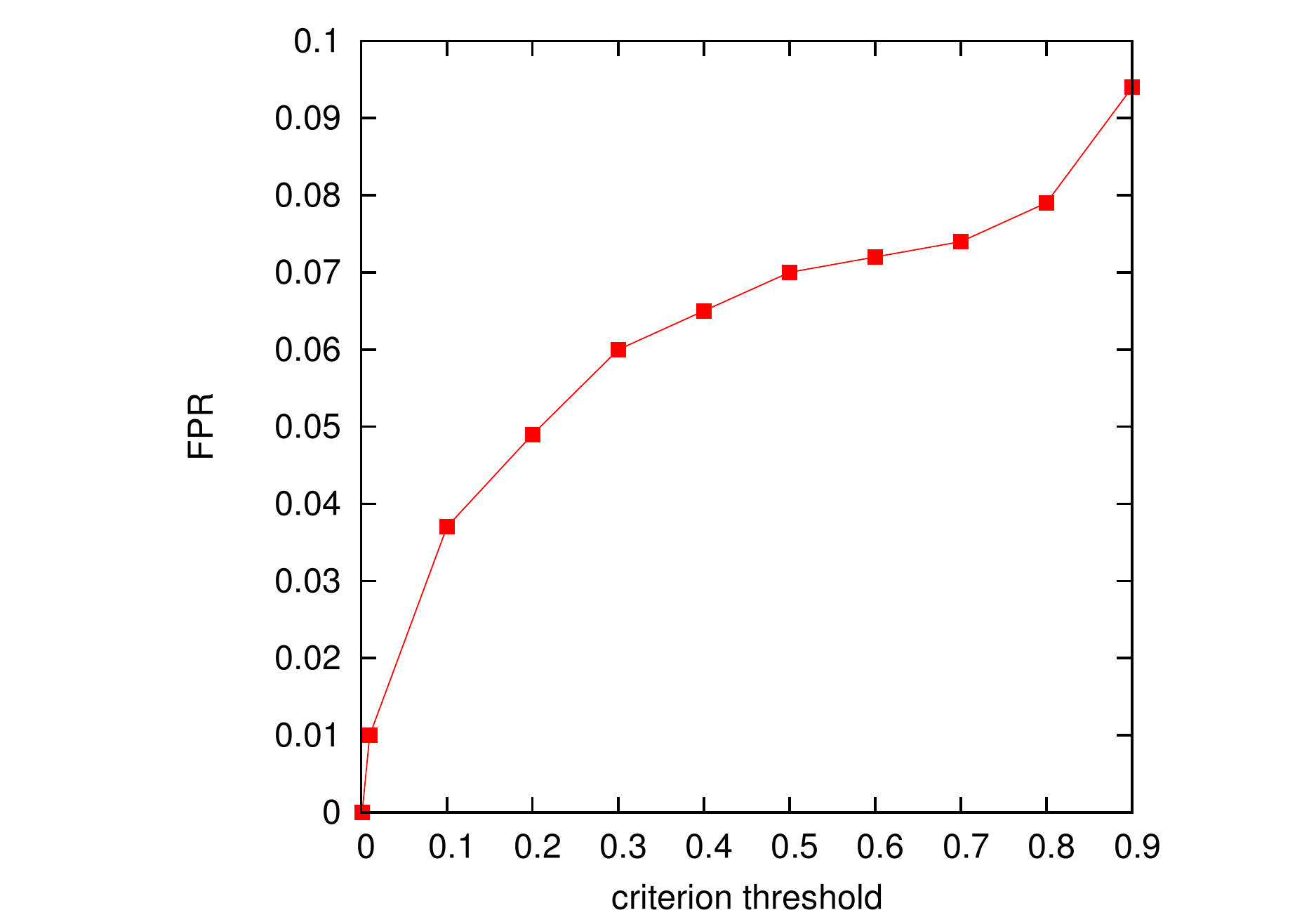} } 
\end{center} 
\caption{\label{fig:nn_classification_ROC} Left panel: Receiver Operating Characteristic (ROC) 
curve of the neural network classifier for physical and unphysical regions of parameter
space (red curve) and a random classifier (black curve). Right panel: False Positive Rate (FPR) as a function of criterion threshold. }  
\end{figure}

\begin{figure}  
\begin{center} 
\includegraphics[width=0.45\linewidth]{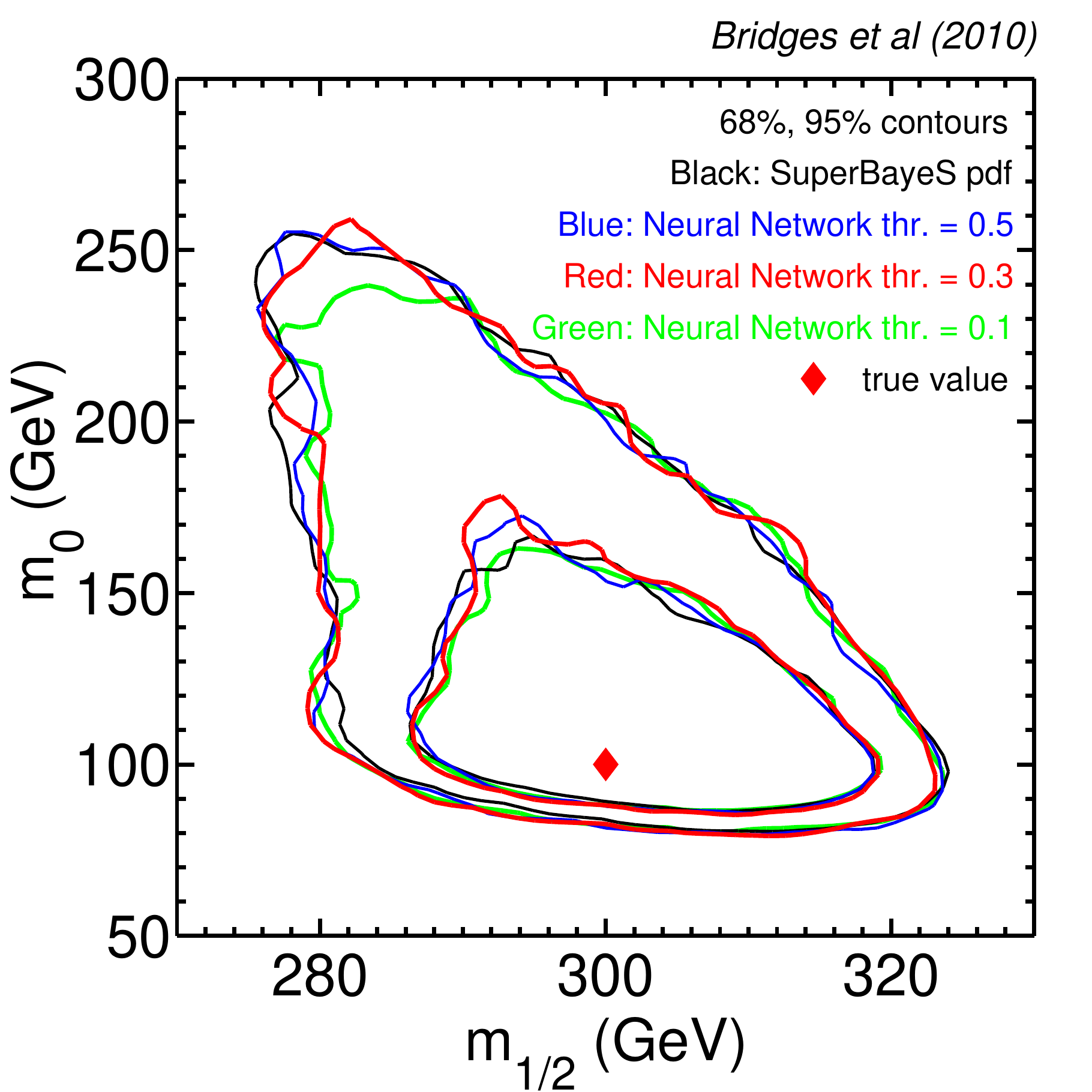}
\includegraphics[width=0.45\linewidth]{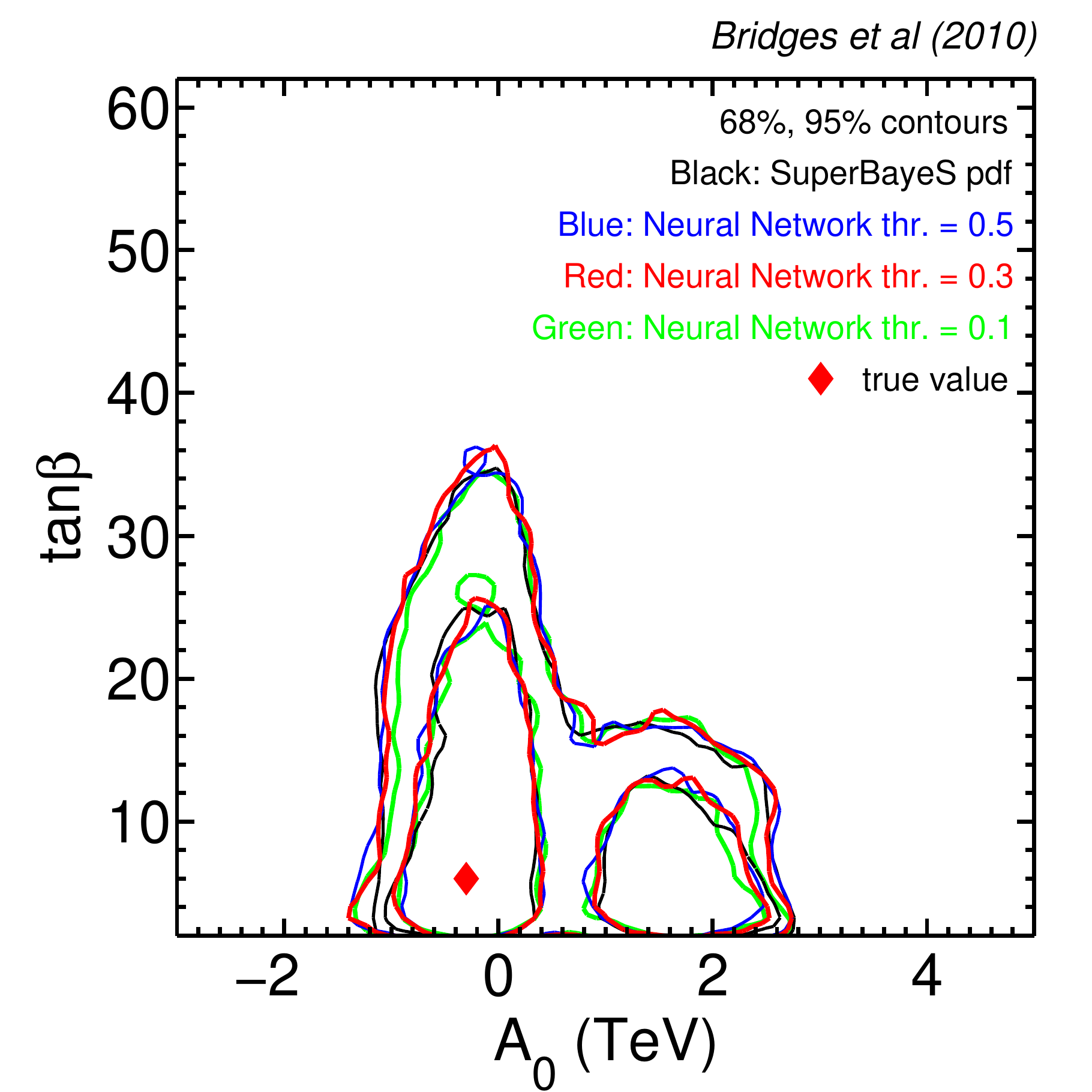}
\end{center} 
\caption{\label{fig:threshold_comparison} Comparison of Bayesian posteriors between runs of {\sc SuperBayes} with {\sc SoftSusy} (black contours, giving 68\% and 95\% regions) and neural networks for different values of the classification threshold adopted (coloured contours, threshold value according to the legend). The agreement between all posteriors (within numerical noise) demonstrates the classifier's resiliance to changes in the threshold value.}  
\end{figure}

It is common in signal processing to graphically represent the efficiency  
of a classifier using a receiver operating characteristic (ROC) curve 
which plots the true positive rate (TPR) versus the false positive rate 
(FPR) for increments of the classifiers discrimination threshold. In this 
case the network outputs $u'_i$ can be considered probabilities and so one 
can set a probability threshold above which the classifier should record a 
positive detection of membership. Variation of this criterion allows 
classifiers to make a tradeoff between the FPR and the TPR.  The left panel of figure 
\ref{fig:nn_classification_ROC} shows the ROC curve for the network 
classifier (solid line). An ideal classifier produces a step-function ROC 
curve with a TPR of 1 for all values of the threshold whereas a random 
classifier (black line) always provides equal numbers of true and 
false positives. For this analysis, adequate results were obtained using a 
threshold of 0.5, which produces a TPR of $99.8 \%$ and a FPR of less than $10\%$. 
It is possible to obtain arbitrarily small FPR by reducing the threshold (as shown in the right panel of figure~\ref{fig:nn_classification_ROC}), however this 
does result in smaller values of the TPR. Figure \ref{fig:threshold_comparison} demonstrates however that the precise choice of threshold has a
minimal effect on the final posteriors obtained.

\section{Coverage of intervals at the ATLAS benchmark}
\label{sec:coverage}

Recently the ATLAS collaboration has studied, 
the so-called ``SU3'' benchmark point, which is found in the bulk region, assuming an integrated 
luminosity of 1 $\fb^{-1}$~\cite{atlas09}. The analysis considered dilepton and lepton+jets final states from the decay chain 
$\squarkl \rightarrow \chi^0_2 (\rightarrow \slepton^{\pm} l^{\mp}) 
q \rightarrow\chi^0_1 l^+ l^- q$ and high-$p_T$ and large missing 
transverse energy from the decay chain $\squarkr \rightarrow \chi^0_1 q$. 
The aim was to reconstruct the CMSSM (mSUGRA in the ATLAS analysis) model parameters evaluating the 
accuracy in the reconstruction. For this they have done a MCMC 
exploration. In a recent work some of us have investigated the analogous constraints using the MultiNest algorithm and the profile likelihood ratio with and without constraints from dark matter relic abundance~\cite{Roszkowski:2009ye}.

\begin{table}
\centering    

\begin{tabular}{| l | l l l l |}
\hline
  & $\mneutone$ &  $\mneuttwo-\mneutone$ & $\msq-\mneutone$ &$\msl-\mneutone$ \\ \hline 
$\mneutone$ & $3.27$  & $-12.7$ &  $-4.89$ &   $-10.04$ \\
$\mneuttwo-\mneutone$ &   &    $476$ &    $4.23$ &  $128$ \\ 
$\msq-\mneutone$ &  & & $8.87$  &  $8.50$ \\
$\msl-\mneutone$ &  & & & $132$\\\hline
\end{tabular}
\caption{Fisher matrix for the ATLAS likelihood employed in the analysis. All entries have been multiplied by $10^3$ and are in units of GeV$^{-2}$.
\label{tab:covmat}}
\end{table}

Table~\ref{tab:covmat} shows the inverse of the covariance matrix (i.e., the Fisher matrix) for sparticle masses used in this study, which is based on the parabolic approximation of the log-likelihood function reported in Ref.~\cite{atlas09}.  This likelihood function is based on the measurement of edges and thresholds in the invariant mass distributions for various combinations of leptons and jets in final state of the selected candidate SUSY events. Note that the relationship between the sparticle masses and the directly observable mass edges is highly non-linear, so a Gaussian is likely to be a poor approximation to the actual likelihood function.  Furthermore, these edges share several sources of systematic uncertainties, such as jet and lepton energy scale uncertainties, which are only approximately communicated in Ref.~\cite{atlas09}.  While noting that these approximations to the likelihood function may omit essential information in the data, we proceed to assess the coverage of the intervals in this idealized setting.

In order to assess coverage, we not only need the likelihood function ($P(d|\Theta)$ with $d$ fixed), but we also need the ability to generate pseudo-experiments with $\Theta$ fixed.  Here we introduce an additional simplification that $P(d|\Theta)$ is also a multivariate Gaussian with the same covariance structure.  For a multivariate Gaussian, one would expect the profile likelihood to have nearly perfect coverage properties, unless there is a breakdown in one of the regularity requirements present in Wilks's theorem ~\cite{Wilks}. Two of the requirements that are most likely to be violated (or lead to a slow convergence to the asymptotic properties assured by Wilks's theorem) are related to the presence of boundaries on the parameter space or parameters that have no impact on the likelihood function in certain regions of the parameter space.  As discussed above, excluding unphysical regions from the CMSSM parameter space imposes complicated boundary conditions.  Similarly, in some regions of parameter space, certain interactions can be highly suppressed (either due to quantum effects or phase space considerations), in which case some parameters may have little effect on the likelihood function.  In the case of four CMSSM parameters and four non-degenerate measurements, the latter is not an issue, but boundary effects may be relevant.

In addition to the complications due to the structure of the model itself, there are other issues which could spoil the expected coverage properties. First, some algorithmic artifact or shortcoming associated with the procedure that creates the intervals. These are complex, high dimensional problems, so it is possible that the algorithms may not perform in practice as well as they should in principle.  Secondly, if the likelihood function is not strong enough to overcome the prior and they prefer different regions of parameter space, one may expect poor coverage properties.  

The sensitivity to priors and comparisons with the profile likelihood ratio were considered for the same ATLAS benchmark point in Ref.~\cite{Roszkowski:2009ye};  reasonable agreement between the different approaches was found, indicating the likelihood function is dominating the inference.  Thus, we focus our attention on effects from algorithmic artifacts and the complicated boundaries of the physical CMSSM.

\begin{figure}
\begin{center}
\hspace{-.in}\includegraphics[width=0.3\linewidth]{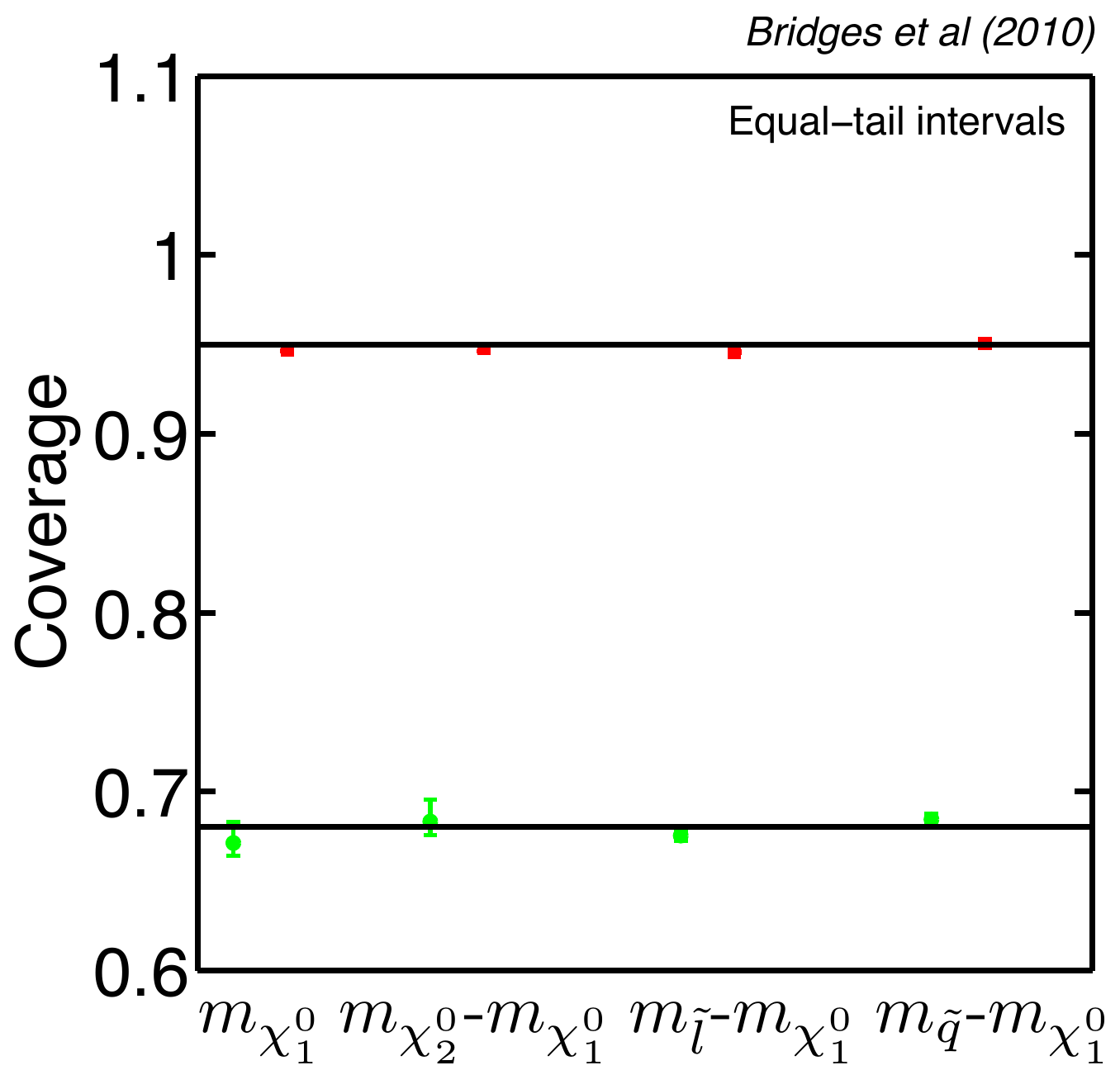}\hspace{.2in}
\includegraphics[width=0.3\linewidth]{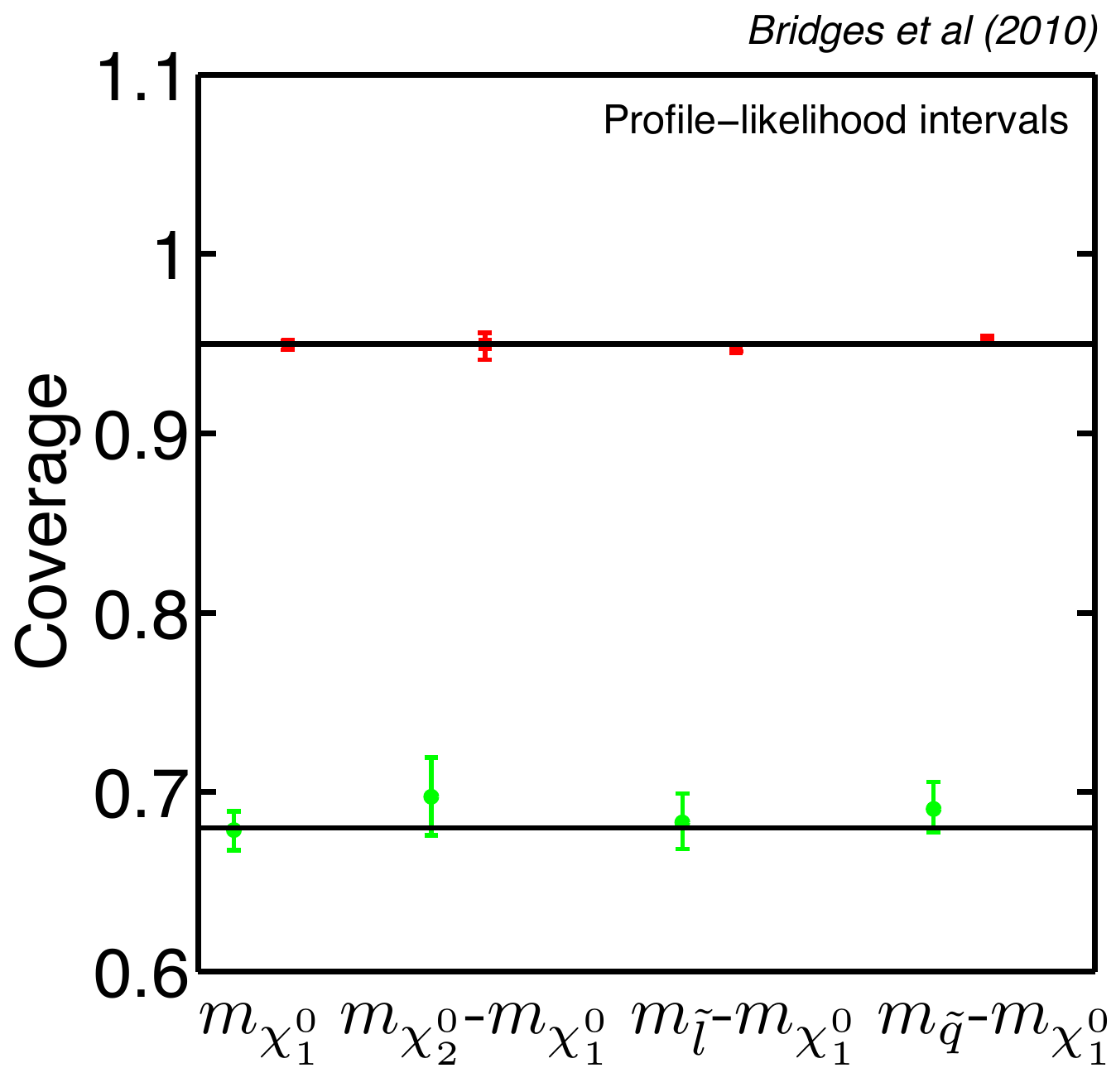}\hspace{.2in}
\includegraphics[width=0.3\linewidth]{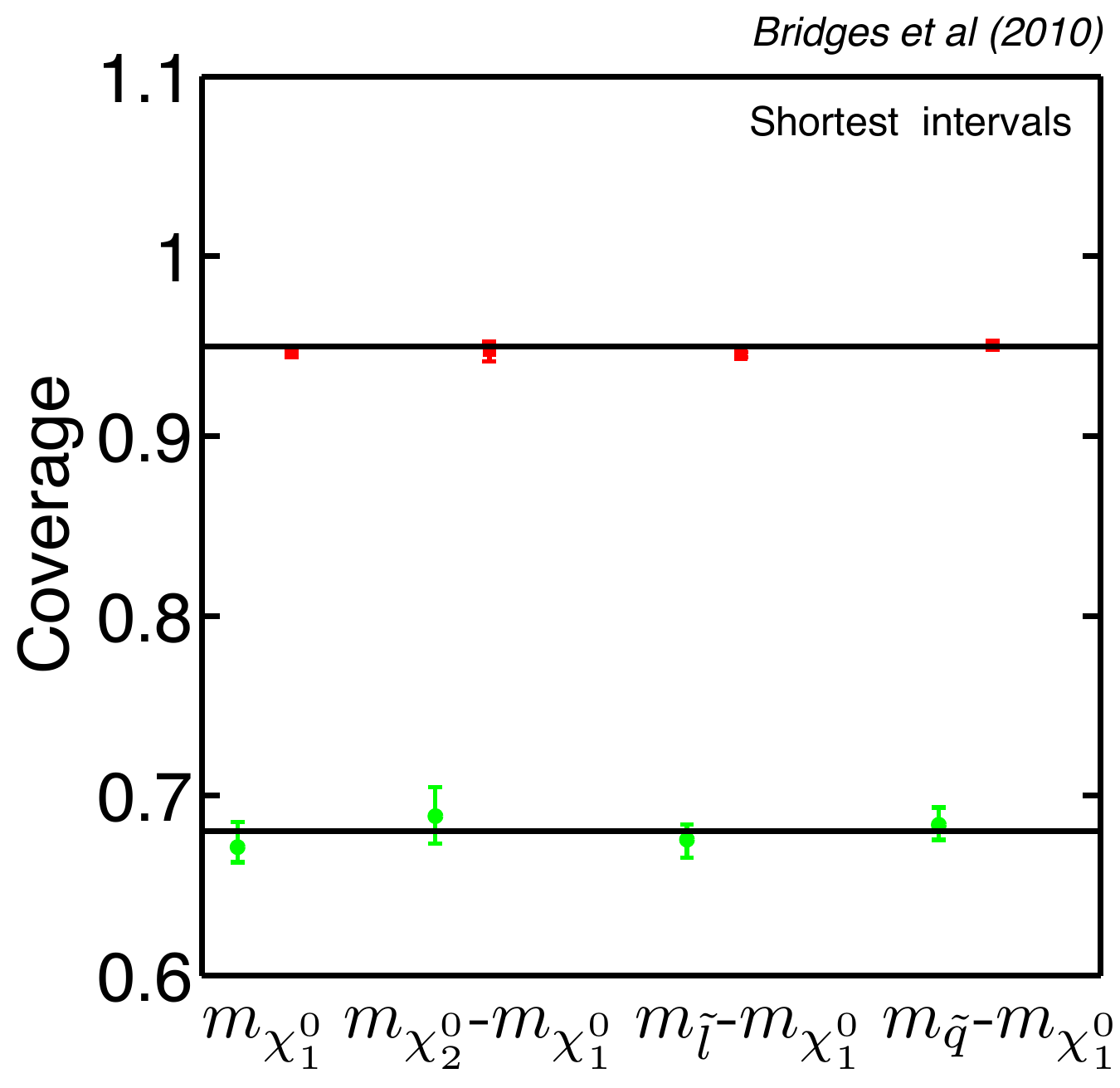}\hspace{.2in}
\end{center}
\caption{\label{fig:test} Test of coverage for a toy 8D MCMC model. Green (red) is for the nominal 68\% (95\%) error. The recovered coverage is compatible with the exact value within the errors due to numerical/sampling noise. Error bars denote the standard deviation from $10^4$ reconstructions with MCMC.}
\end{figure}

In order to isolate algorithmic effects, we begin by constructing intervals in the weak scale masses $\bfm{m}$ (as opposed to the more fundamental CMSSM parameters $\theta$).  The likelihood function is invariant under these reparametrizations, and it is trivial to extend the domain of the weak scale masses so that boundary effects are absent.  In total $10^4$ pseudo-experiments were analyzed using MCMC using $10^6$ samples each. A flat prior was used on the space of $\bfm{m}$.   From the MCMC samples, equal-tail and shortest credible intervals were constructed by marginalizing the posterior distribution to each of the four parameters.  Additionally, profile likelihood intervals were constructed by maximizing the likelihood across the samples.  Figure~\ref{fig:test} shows that each of the restricted 1-d intervals covers with the stated rate for each of the methods.  Thus, we conclude that our MCMC algorithm is performing as expected for this likelihood function.

To assess the coverage in the restricted space of the physically realizable CMSSM requires several additional ingredients\footnote{As this work was being finalized, we learnt of a similar study being carried out in the context of a coverage study of the CMSSM from future direct detection experiments~\cite{Yashar}. While the spirit of the present paper and of Ref~\cite{Yashar} is broadly similar, our approach takes advantage of the large computational speed-up afforded by neural networks. Both works clearly show the timeliness and importance of evaluating the coverage properties of the reconstructed intervals for future data sets.}. %
First, we must introduce the mapping $\Theta \to \bfm{m}$ and the boundaries on $\Theta$ corresponding to physically realizable parameter points.  It is not computationally feasible to use {\sc SoftSusy} for so many repeated likelihood evaluations, thus we rely on the neural network approximation to the mapping and a second network for classification of the physical points.  Additionally, we must specify a prior on the space of $\Theta$; for this we used log priors in $m_0, m_{1/2}$ and note there was little sensitivity to this choice in previous studies based on the same ATLAS likelihood function~\cite{Roszkowski:2009ye}.  We construct $10^4$ pseudo-experiments and analyze them with both MCMC (using a Metropolis-Hastings algorithm) and MultiNest.  Each realization is run as an independent chain until we gather a total of $10^6$ posterior samples in the MCMC chain. For MultiNest, we run until the tolerance criterium is satisfied (with a tolerance parameter 0.5), which results in about 50k posterior samples. Altogether, our neural network MCMC runs have performed a total of $4 \times 10^{10}$ likelihood evaluations, in a total computational effort of approximately  $2\times 10^4$ CPU-minutes. We estimate that the same computation using {\sc SoftSusy} would have taken about 1100-CPU years, which is at the boundary of what is feasible today, even with a massive parallel computing effort.
By using the neural network, we observe  a speed-up factor of about $3 \times 10^4$ compared with scans using the explicit spectrum calculator.  

As above, the posterior samples were used to form equal-tailed, central, and profile-likelihood based intervals for each of the four fundamental CMSSM parameters. Coverage was then computed for each type of interval. In order to quantify the uncertainty in the coverage introduced by sampling noise (i.e., the finite length of our MCMC chains) and by the neural network approximation to the full RGE calculation, we proceeded as follows. First, for each lower/upper limit (at a given confidence level) we built the quantity $\sigma_\text{tot} = (\sigma^2_\text{SN} + \sigma^2_\text{NN})^{1/2}$, where  $\sigma_\text{SN}$ is the sampling noise standard deviation while $\sigma_\text{NN}$ is the standard deviation from the neural network noise for that limit, both of which have been estimated by the pseudo-experiments of section~\ref{sec:nn_performance}. We then shifted the intervals uniformly downwards or upwards (keeping the intervals' length constant) by an amount given by the appropriate $\sigma_\text{tot}$. This leads to a higher/lower coverage (depending on the direction of the shift and the limit being considered), which we used as an estimation of the standard deviation of the coverage value itself. This quantity is shown as an errorbar on the coverage values. We note that because of the large number of pseudo-experiments employed, the statistical noise from the binomial process itself is negligible compared with the above uncertainties. For the 2-d intervals we proceeded in a similar fashion, with the direction of the shift (in the 2-d plane) chosen to be parallel to the line connecting the mean value of the 2-d upper/lower limit (i.e., this is the most conservative choice for the error).

\begin{figure}
\begin{center}
\includegraphics[width=0.32\linewidth]{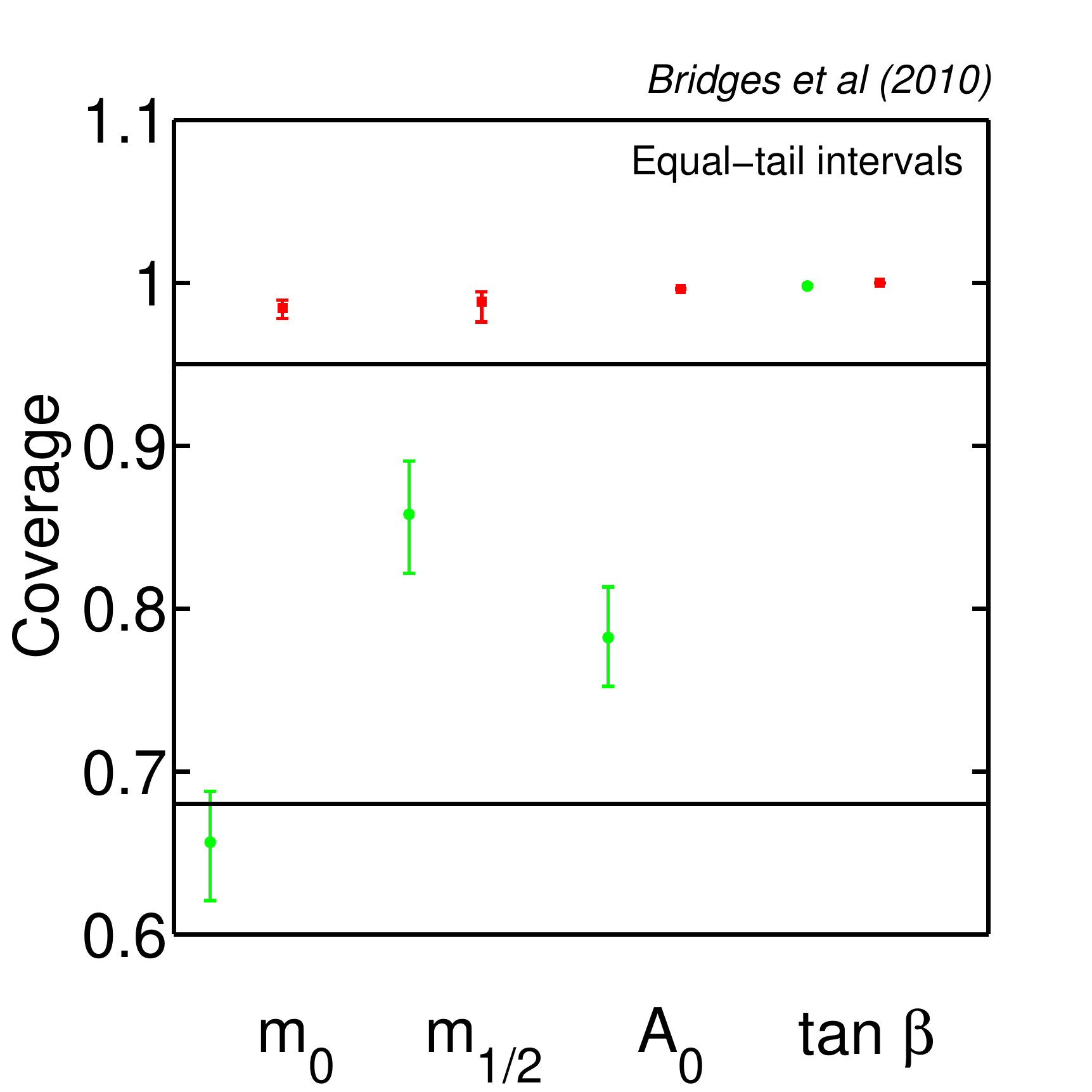}
\includegraphics[width=0.32\linewidth]{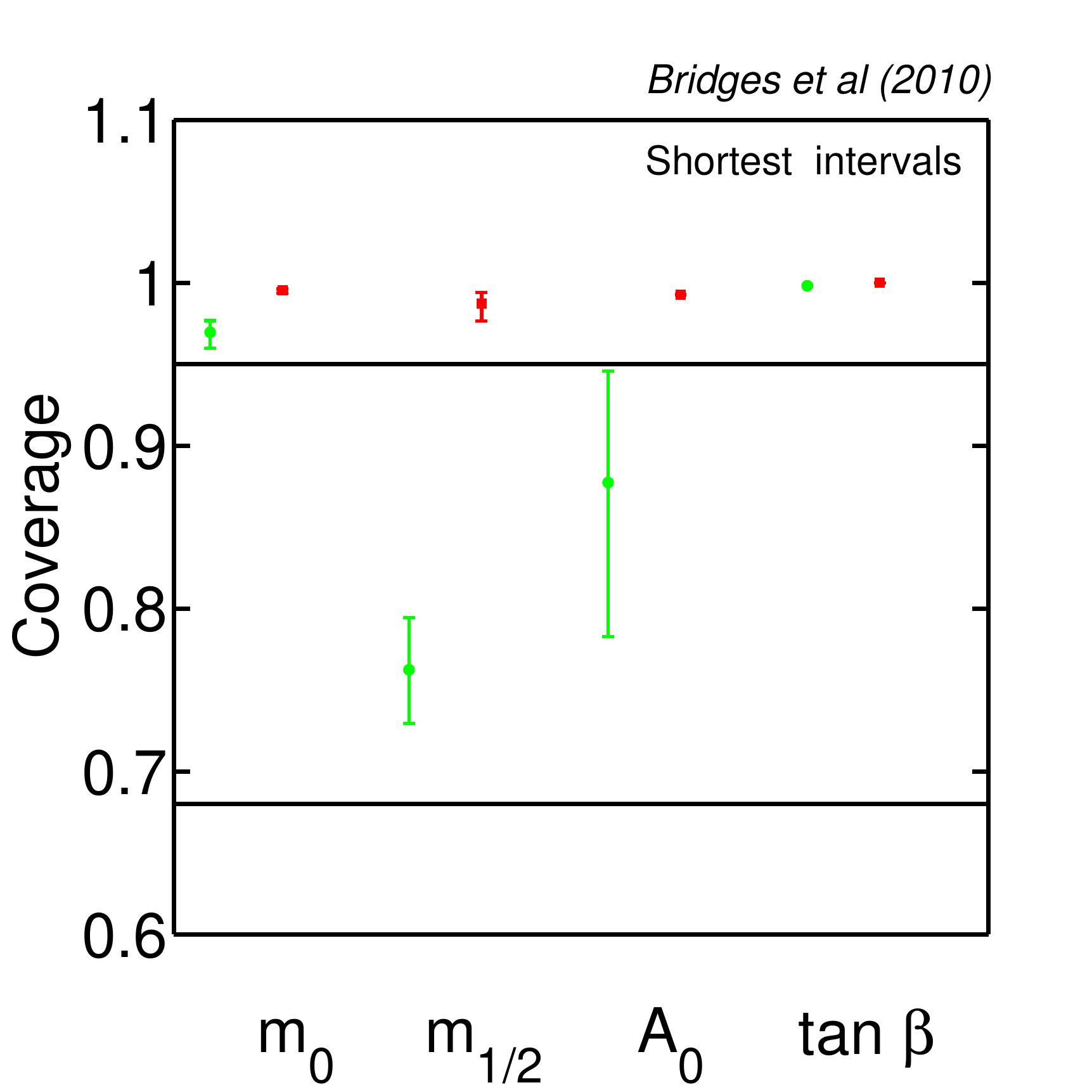} 
\includegraphics[width=0.32\linewidth]{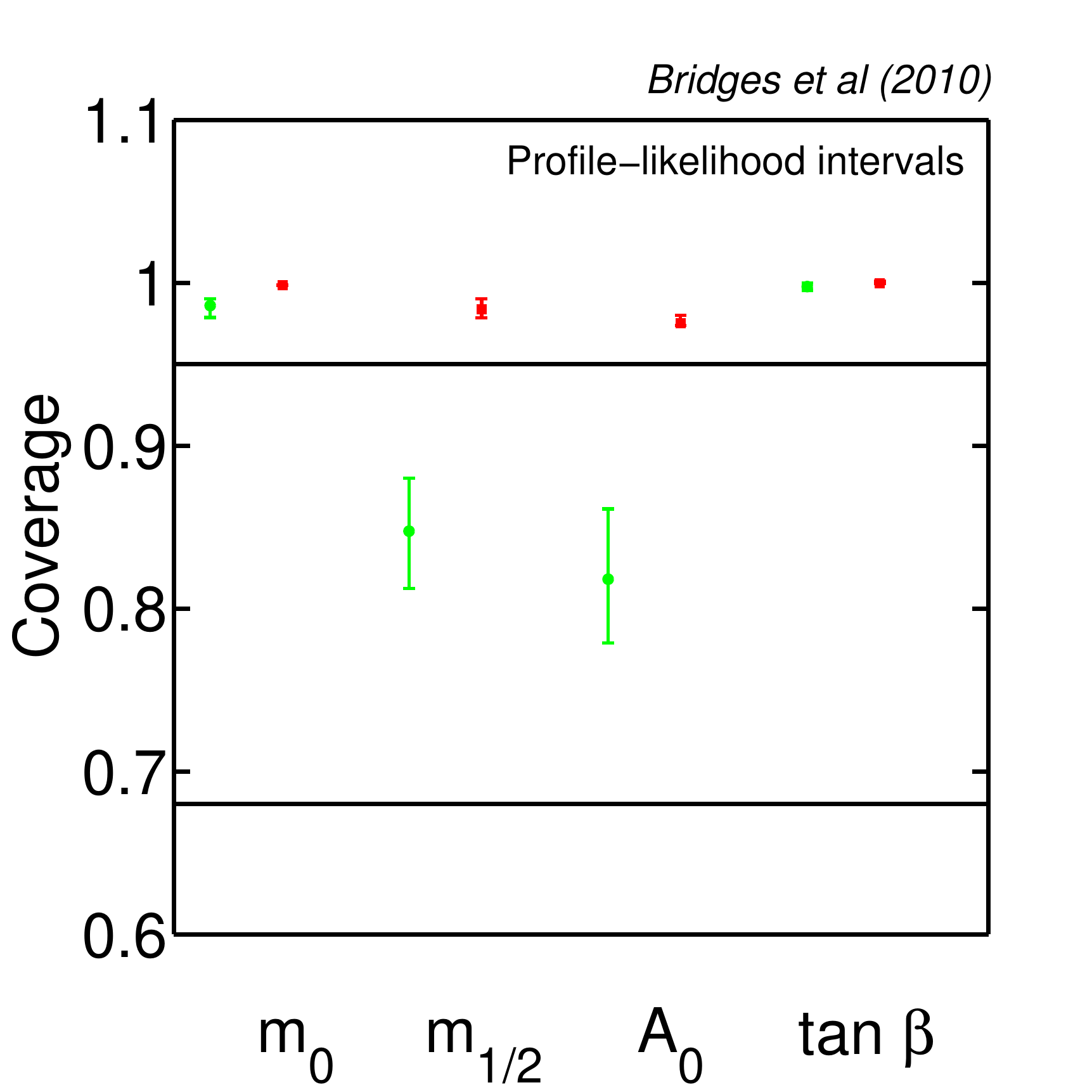} \\ 
\includegraphics[width=0.32\linewidth]{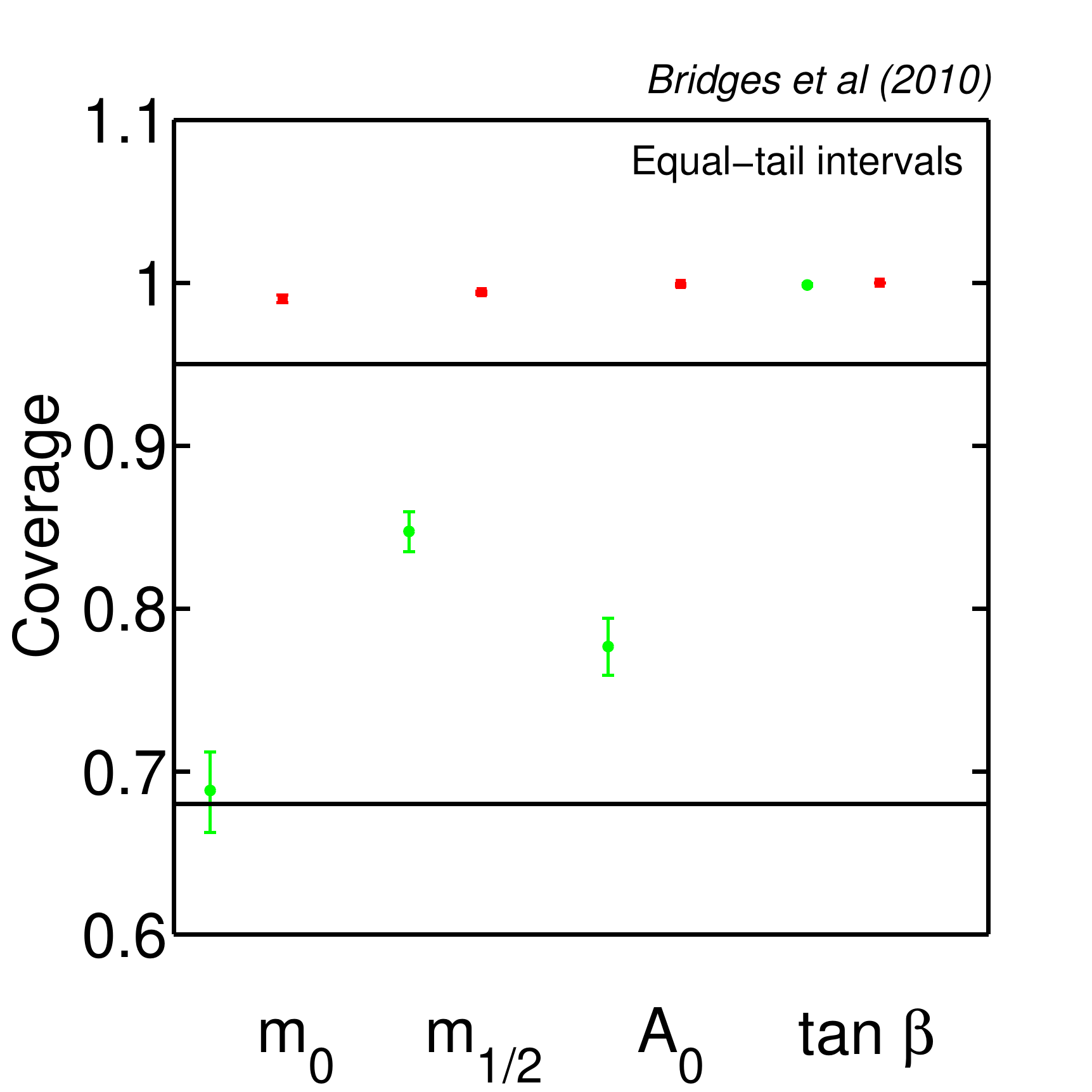}
\includegraphics[width=0.32\linewidth]{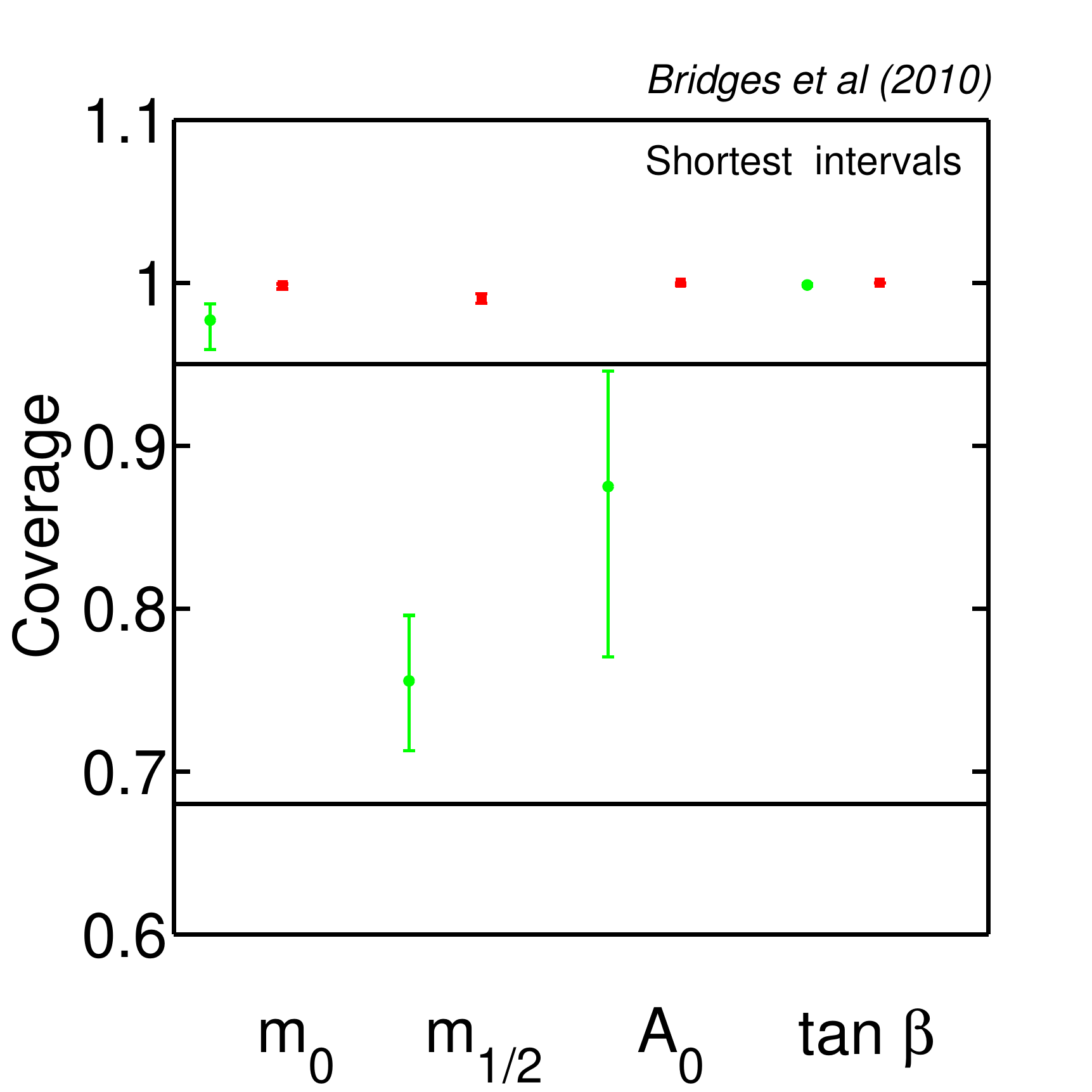} 
\includegraphics[width=0.32\linewidth]{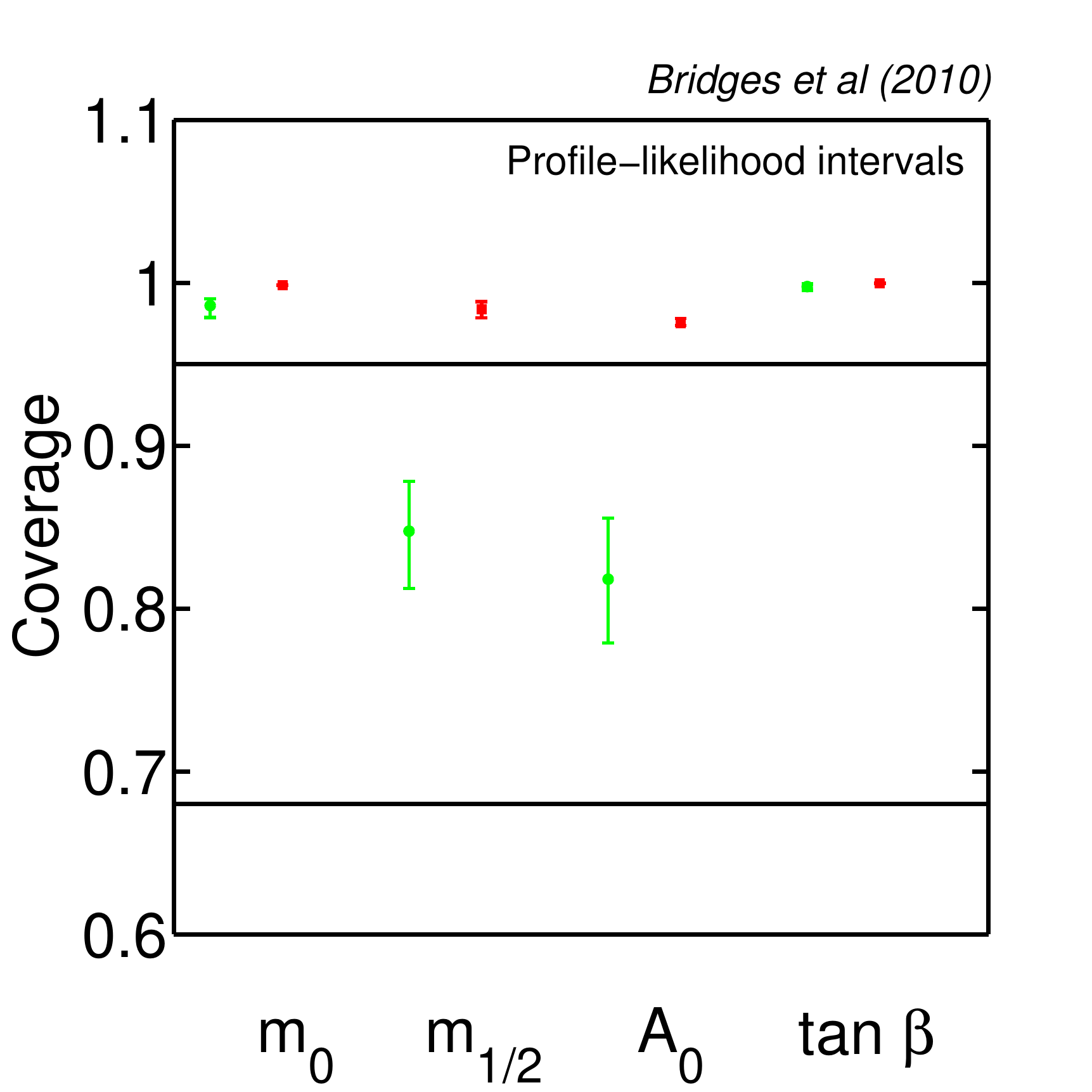} 

\end{center}
\caption{\label{fig:mcmc_coverage} Coverage for various types of intervals for the CMSSM parameters, from $10^4$ realizations. The top row employs MCMC for the reconstruction (each pseudo-experiment is reconstructed with $10^6$ samples), while to bottom row uses MultiNest (with $5\times 10^4$ samples per pseudo-experiment).  Green (red) is for the 68\% (95\%) error. }
\end{figure}

\begin{figure}
\begin{center}
\includegraphics[width=0.32\linewidth]{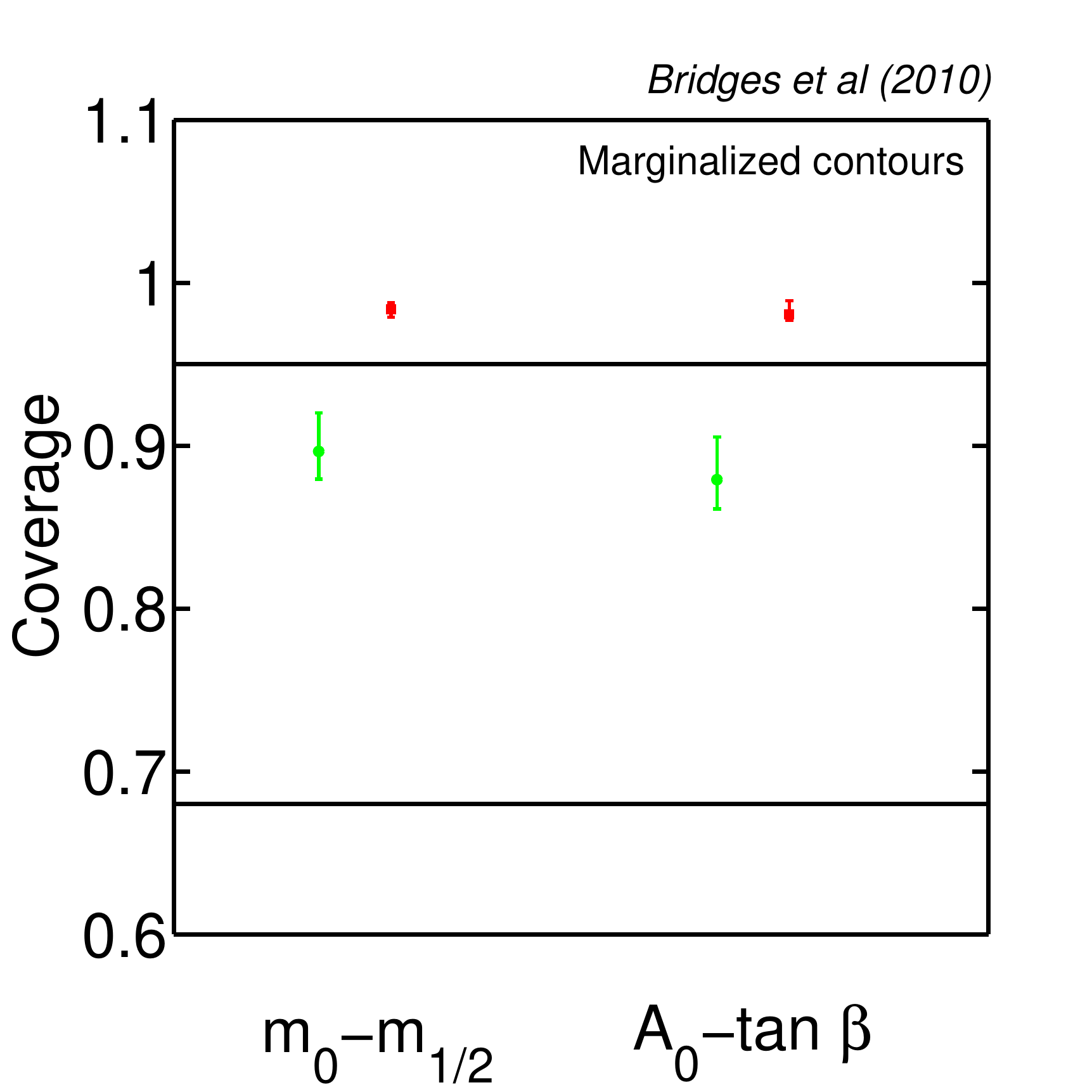}
\includegraphics[width=0.32\linewidth]{coverage_marg_2d} 
\includegraphics[width=0.32\linewidth]{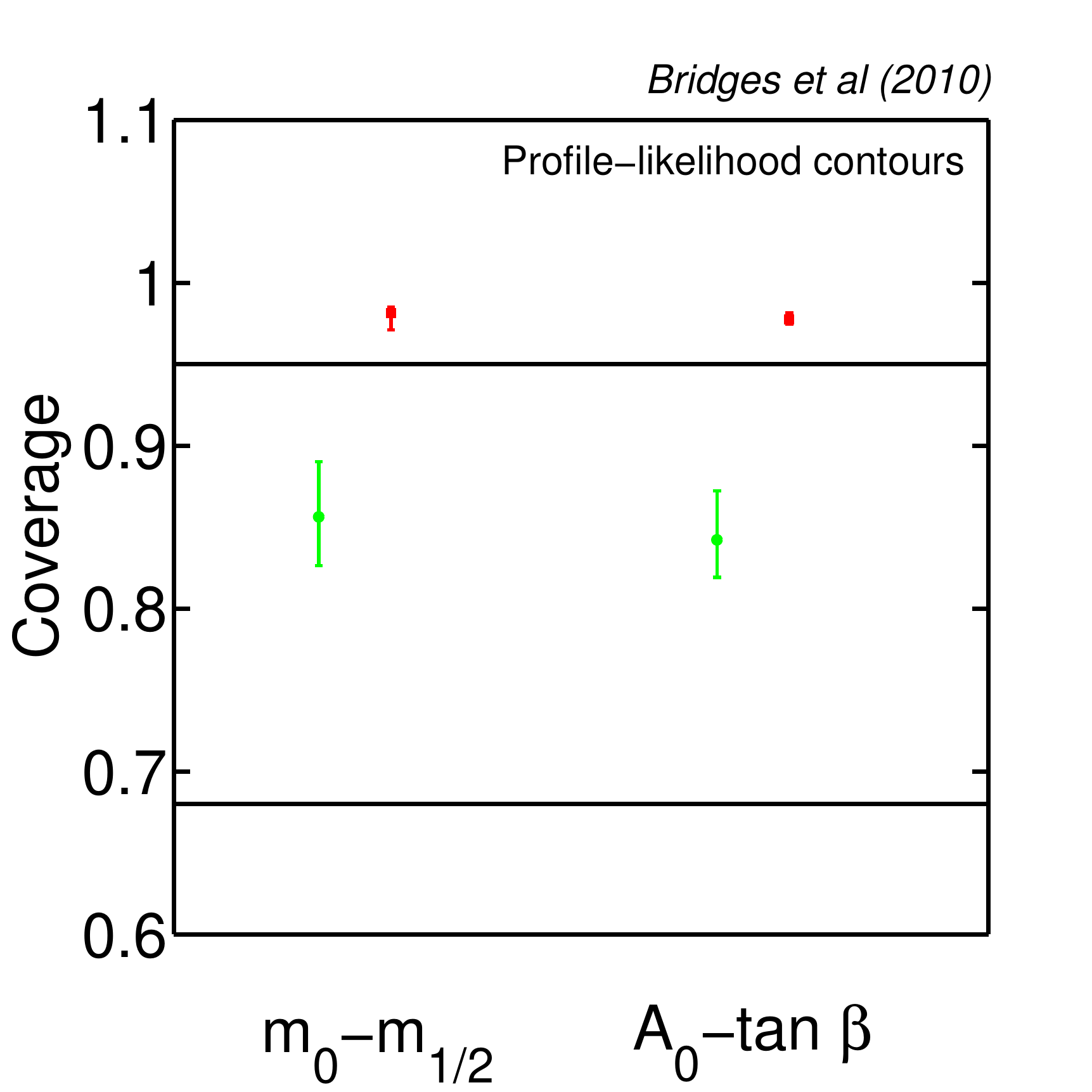} 

\end{center}
\caption{\label{fig:mcmc_coverage_2d} Coverage for various types of two-dimensional intervals for the CMSSM parameters, from $10^4$ realizations based on MCMC (with $10^6$ samples per pseudo-experiment). Green (red) is for the 68\% (95\%) error.  }
\end{figure}

The results are shown in Fig.~\ref{fig:mcmc_coverage}, where it can be seen that the methods have substantial over-coverage (are conservative).  The coverage of 2-d intervals is also shown in  Fig.~\ref{fig:mcmc_coverage_2d}.

\subsection{Discussion of coverage results}

\begin{figure}[tbh!]
\begin{center}
\includegraphics[width=.48\linewidth]{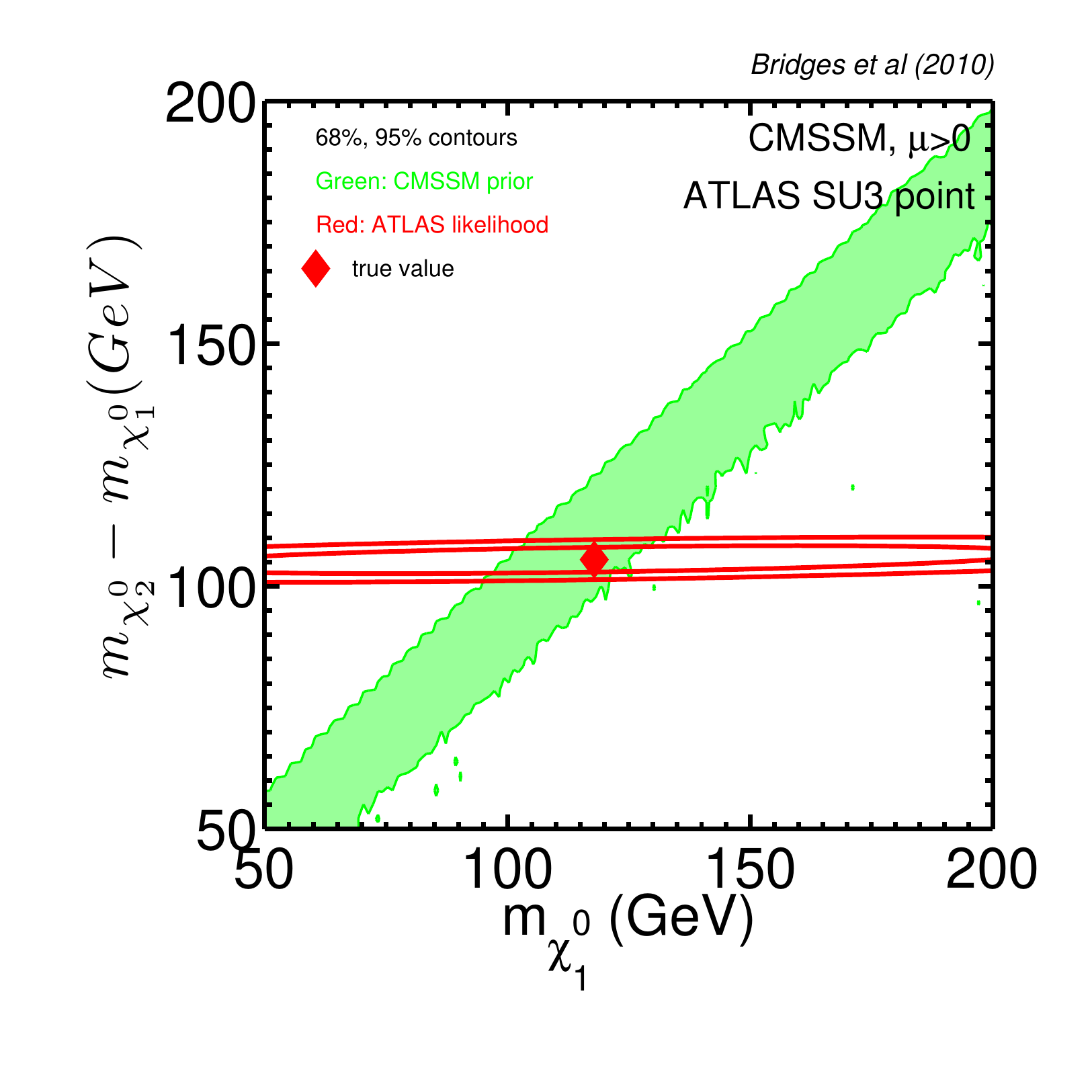}
\includegraphics[width=.48\linewidth]{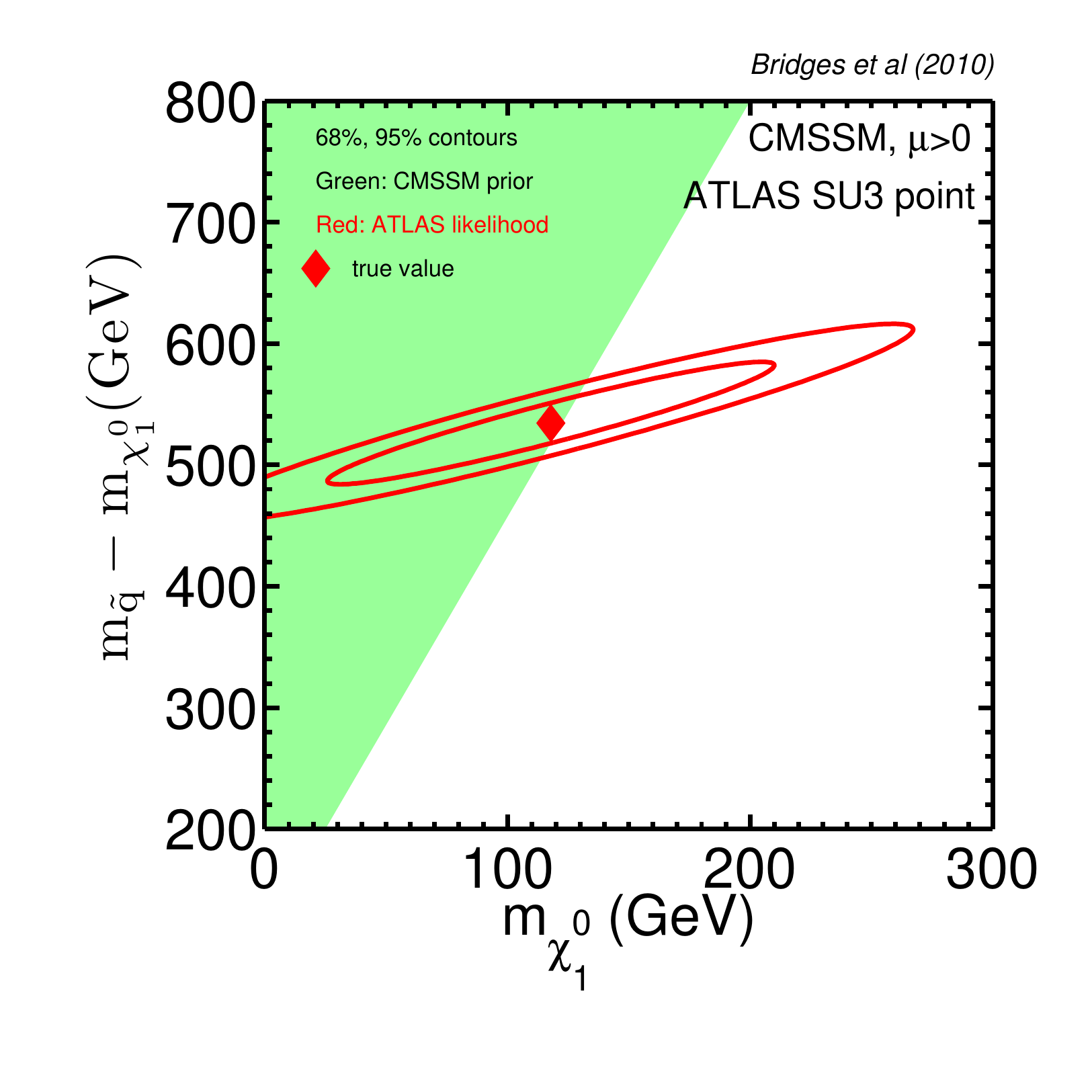}
\caption[test]{Illustration of the boundary effects coming from the physically realizable CMSSM (green region) with respect to the unconstrained likelihood expressed in the weak-scale masses (red ellipses).  The likelihood extends beyond the boundaries imposed by the CMSSM, which leads to a slow convergence of the profile likelihood ratio to the asymptotic chi-square distribution. }
\label{fig:CMSSM_correlation} 
\end{center}
\end{figure}

While it is difficult to unambiguously attribute the over-coverage to a specific cause, the most likely cause is the effect of boundary conditions imposed by the CMSSM.  An example of such boundaries and how they compare with the extent of the likelihood function is shown in Fig.~\ref{fig:CMSSM_correlation}. This figure shows regions in some of the weak-scale mass combinations entering the likelihood which can be realized within the CMSSM (green shades). The red ellipses are $1\sigma$ contours from the likelihood, and are centered around the benchmark value. It is clear that likelihood extends beyond the boundaries imposed by the CMSSM, which leads to a slow convergence of the profile likelihood ratio to the asymptotic chi-square distribution.

\begin{figure}[tbh!]
\begin{center}
\includegraphics[width=.24\linewidth]{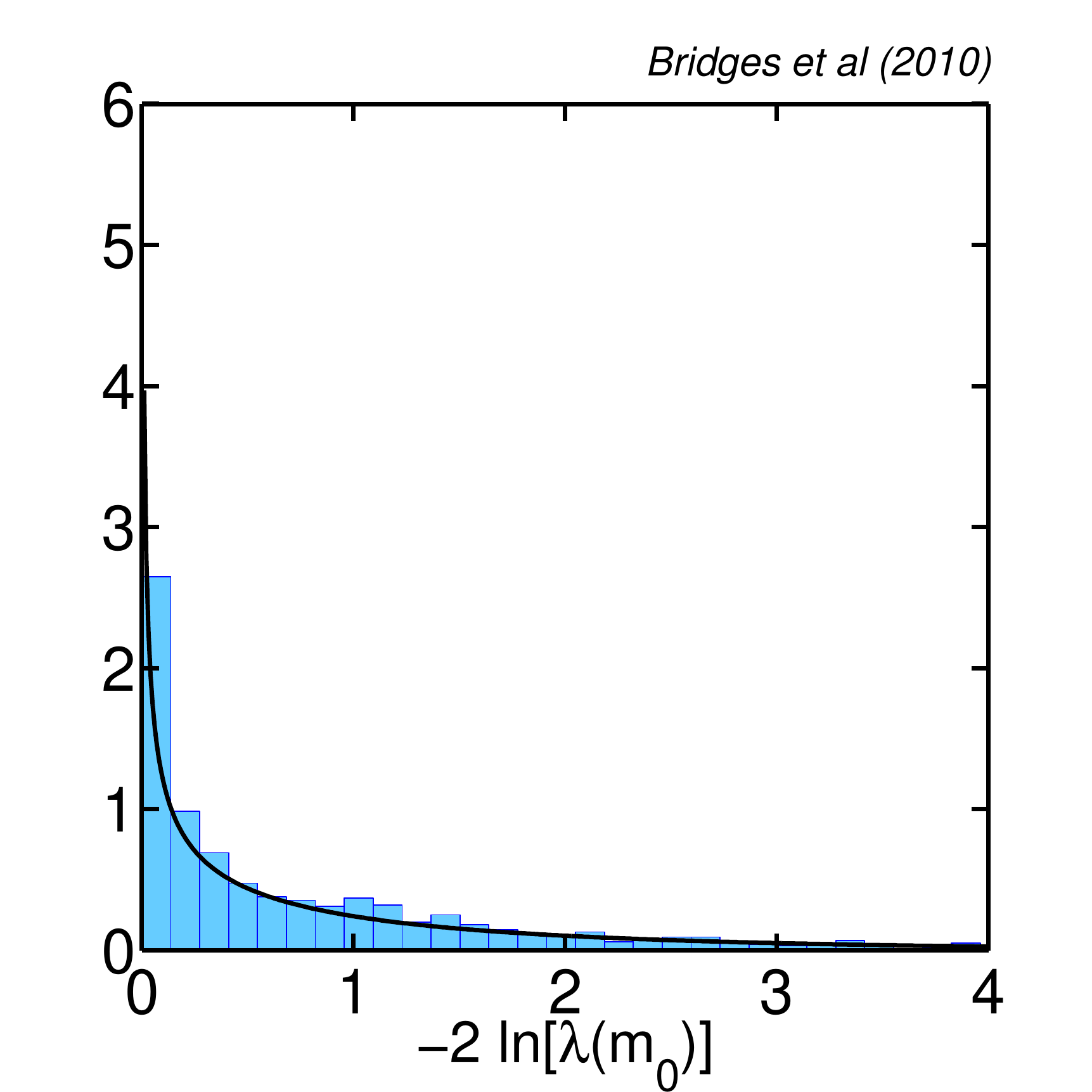}
\includegraphics[width=.24\linewidth]{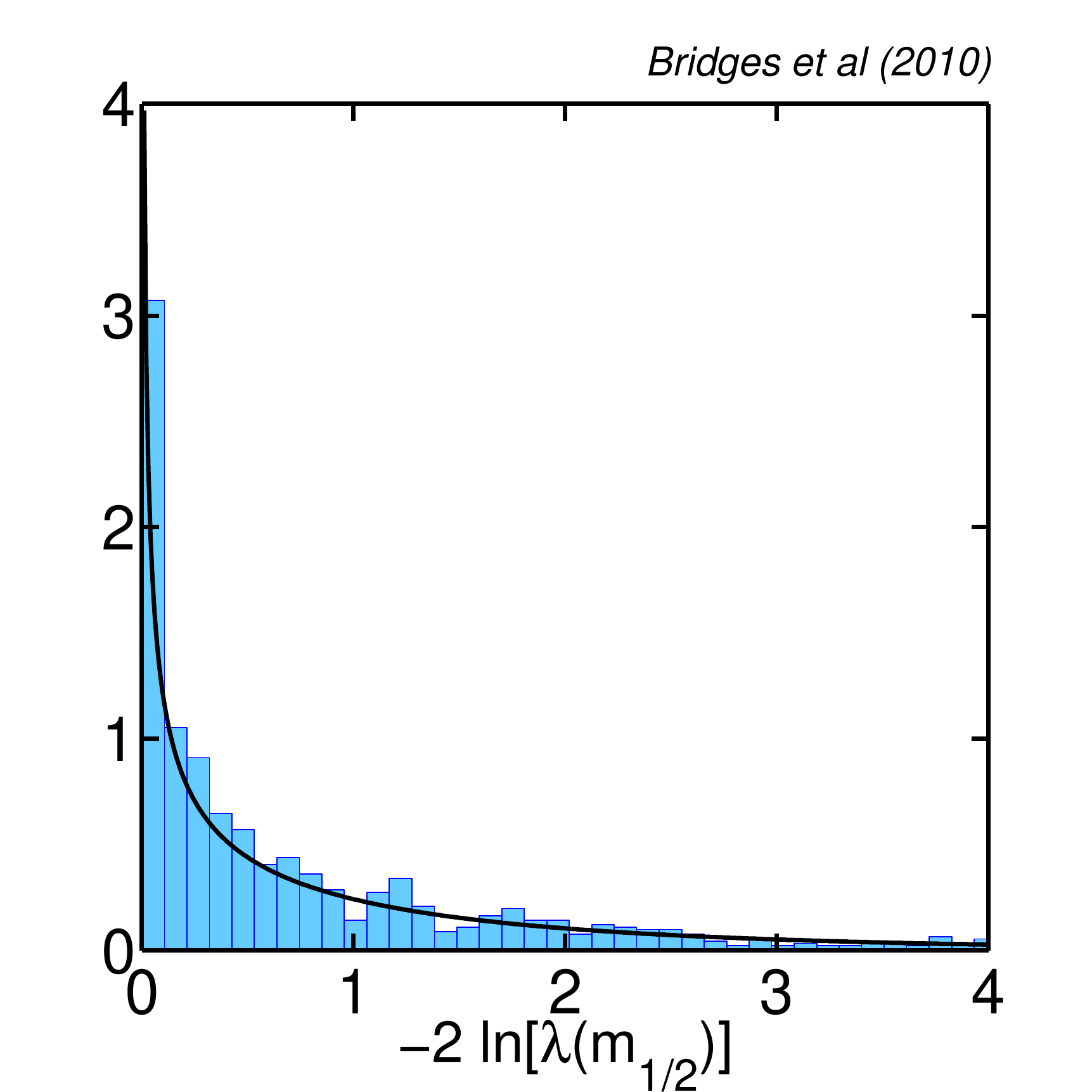}
\includegraphics[width=.24\linewidth]{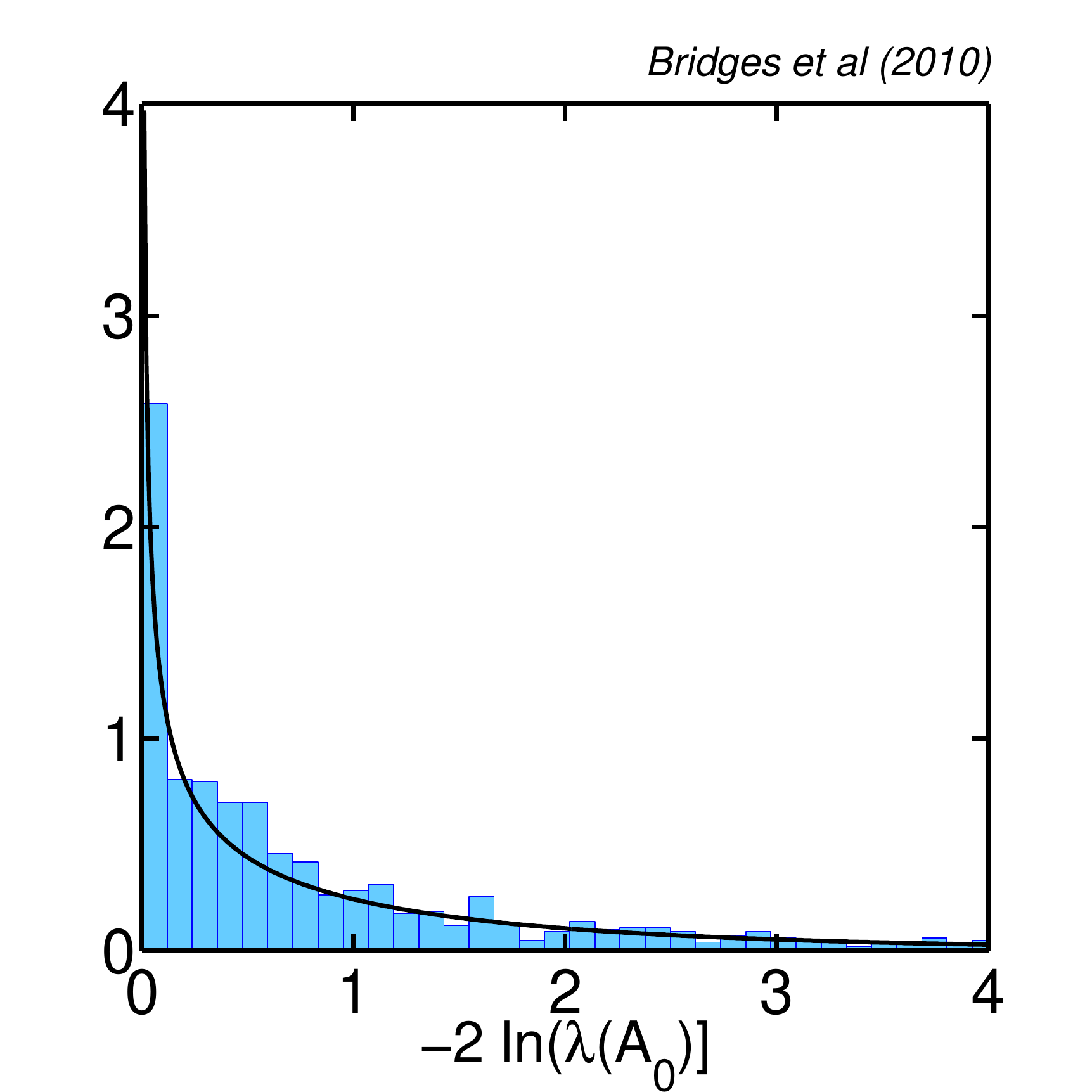}
\includegraphics[width=.24\linewidth]{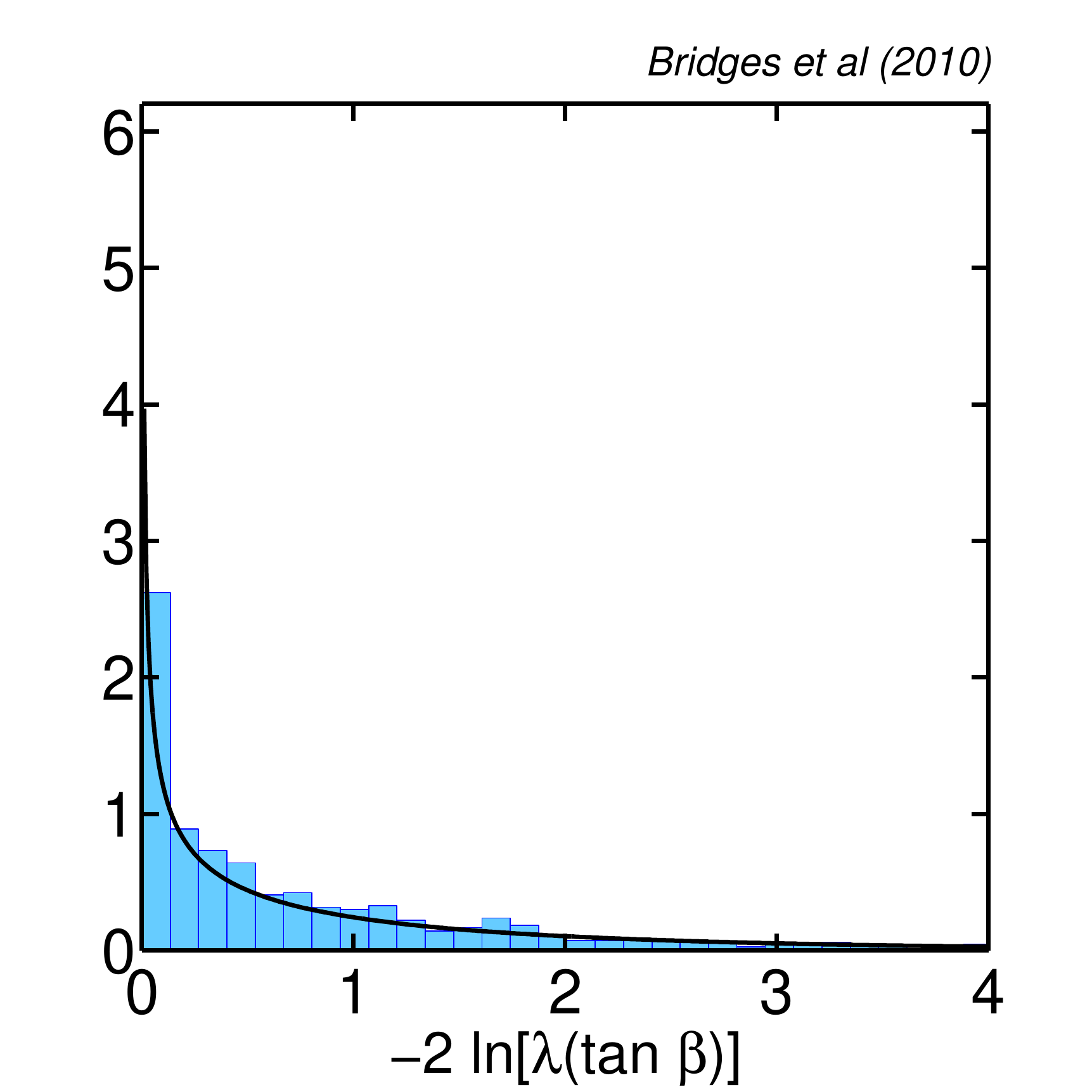}\\
\includegraphics[width=.24\linewidth]{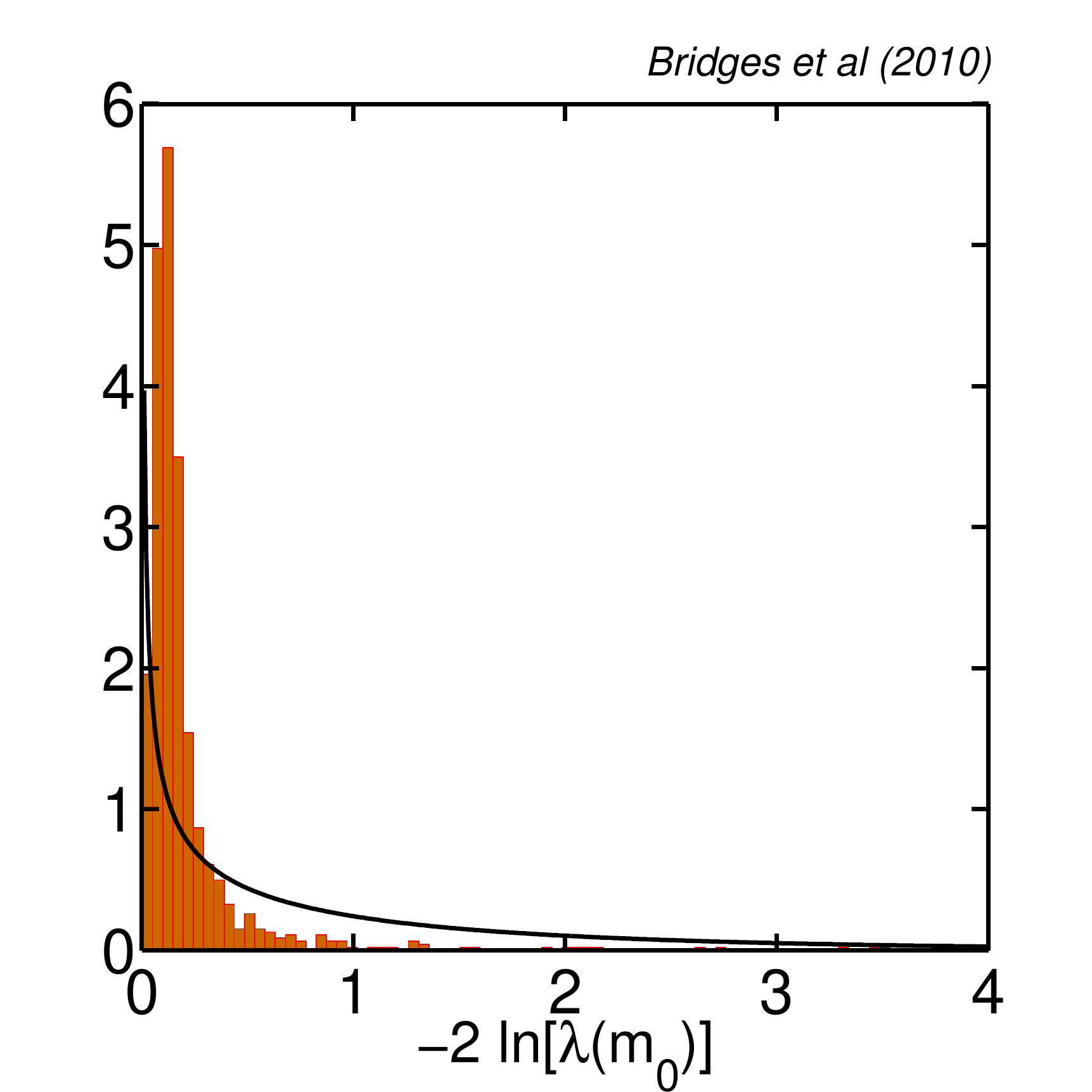}
\includegraphics[width=.24\linewidth]{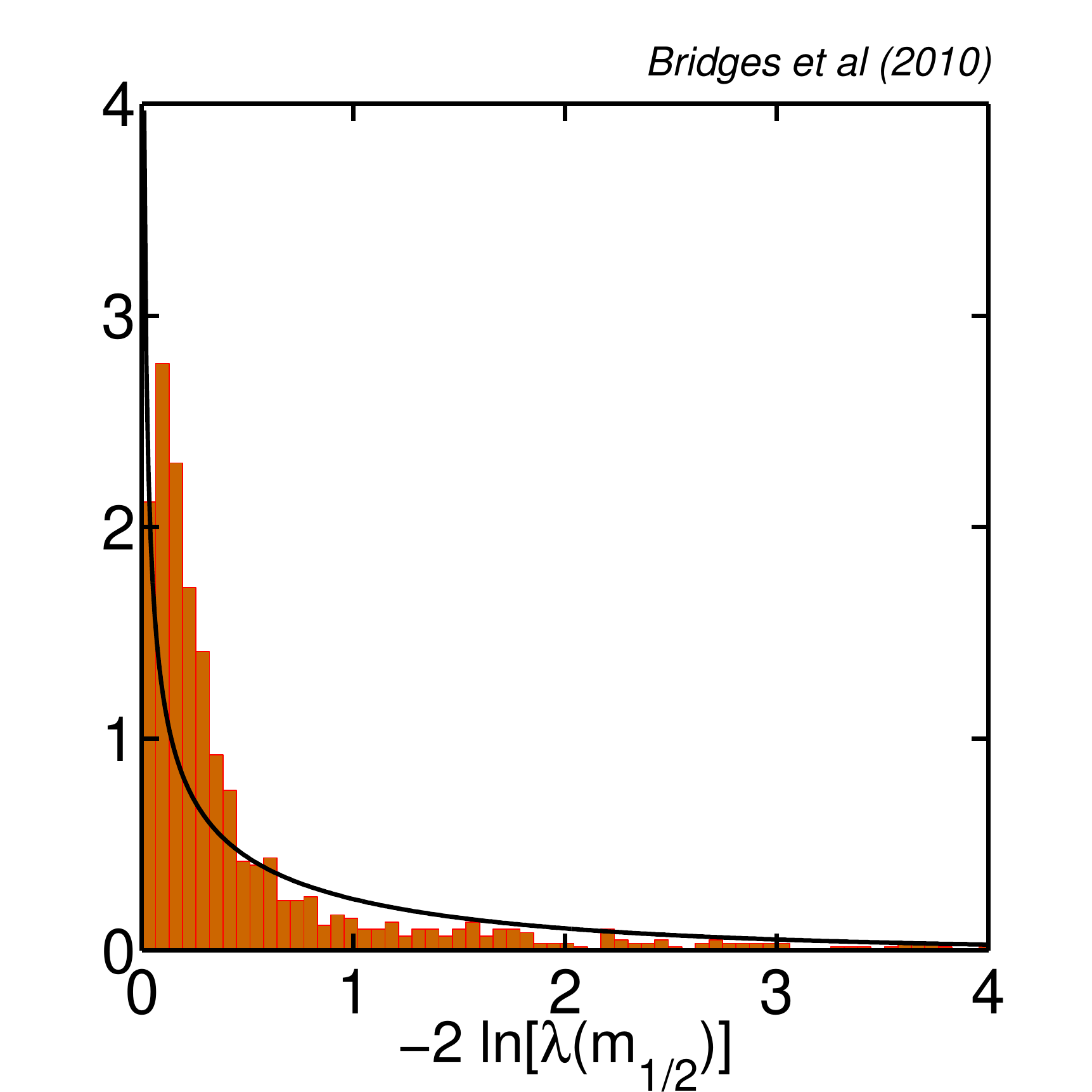}
\includegraphics[width=.24\linewidth]{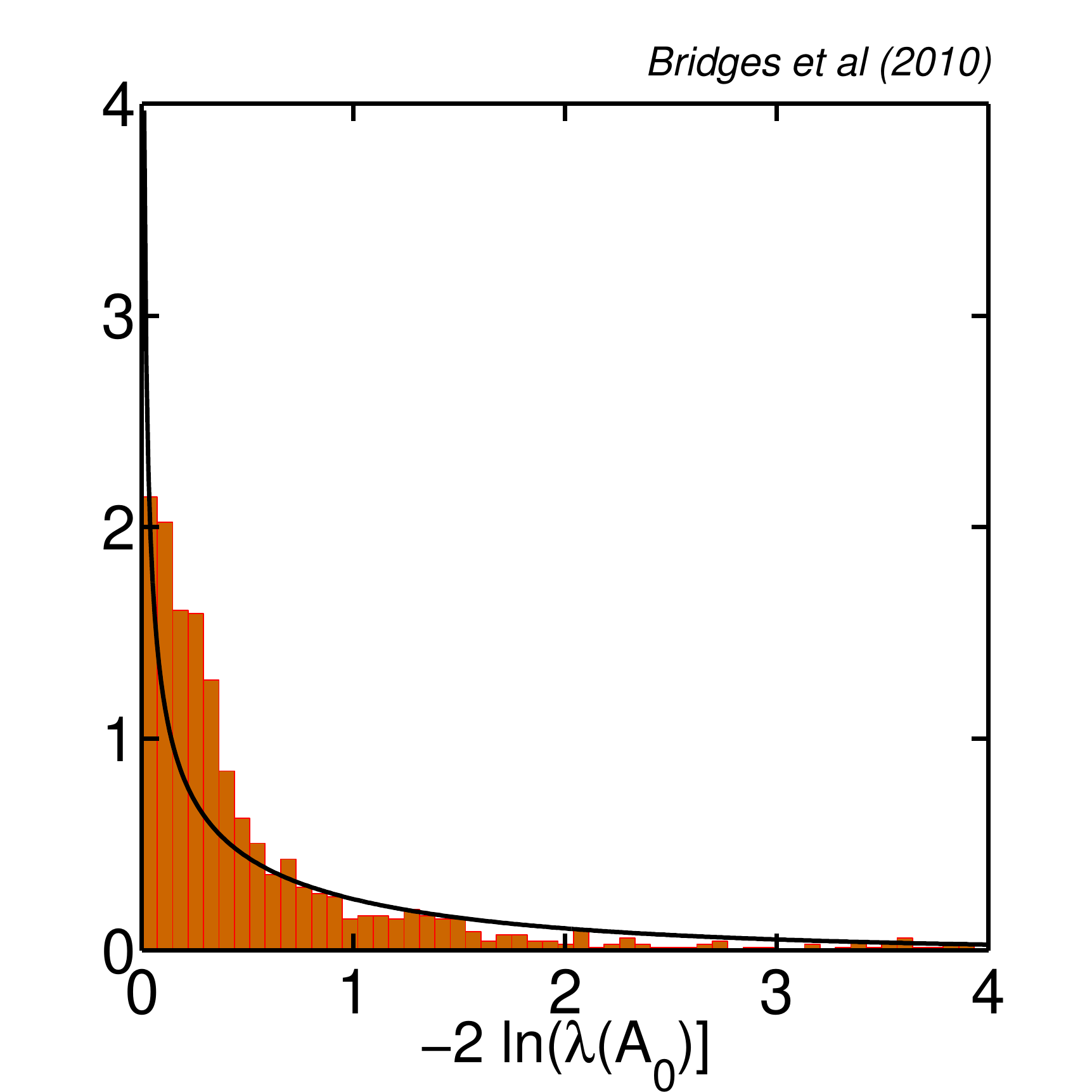}
\includegraphics[width=.24\linewidth]{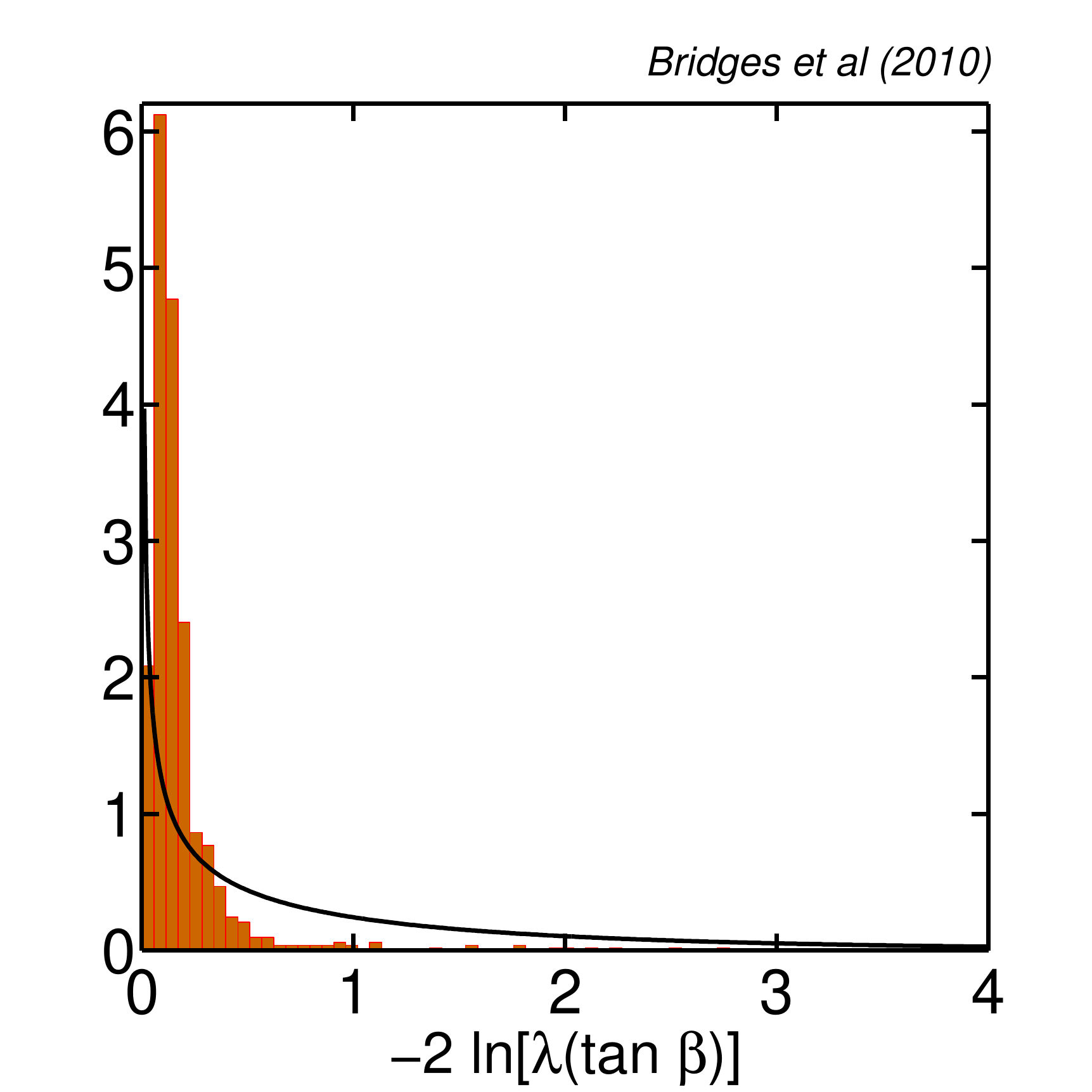} 
\caption[test]{The distribution of $-2\ln\lambda$ for the scans in $\bfm{m}$ (top row) and $\Theta$ (bottom row) from repeated realizations of pseudo-experiments. The solid line is a $\chi^2$ distribution with 1 dof expected from Wilks' theorem.  The distributions from scans in the weak-scale masses (top row) follow quite closely the $\chi^2$ distribution, while this is not the case for the scans using the CMSSM parameters as fundamental quantities. }
\label{fig:chisquare} 
\end{center}
\end{figure}

Recall that the profile likelihood ratio is defined as $\lambda(\theta) = L(\theta, \hat{\hat{\psi}})/L(\hat\theta, \hat{{\psi}})$, where $\hat{\hat{\psi}}$ is the conditional maximum likelihood estimate (MLE) of $\psi$ with $\theta$ fixed, and $\hat{\theta}$, $\hat{\psi}$ are the unconditional MLEs.  When the fit is performed in the space of the weak-scale masses, there are no boundary effects, and the distribution of $-2\ln \lambda(\bfm{m})$ (when $\bfm{m}$ is true) is distributed as a chi-square with a number of degrees of freedom given by the dimensionality of $\bfm{m}$.  Since the likelihood is invariant under reparametrizations, we expect $-2\ln \lambda(\theta)$ to also be distributed as a chi-square.    If the boundary is such that $\bfm{m}(\hat{\theta},\hat{\psi}) \ne \hat{\bfm{m}}$ or $\bfm{m}(\theta,\hat{\hat{\psi}}) \ne \hat{\hat{\bfm{m}}}$, then the resulting interval will modified.  More importantly, one expects the denominator $L(\hat\theta, \hat{\psi}) < L(\hat{\bfm{m}} )$ since $\bfm{m}$ is unconstrained, which will lead to $-2\ln \lambda(\theta) < -2\ln \lambda(\bfm{m})$.  In turn, this means more parameter points being included in any given contour, which leads to over-coverage.  Evidence for this effect can be seen in Fig.~\ref{fig:chisquare}, which shows the value of $-2\ln\lambda$ for the scans in $\bfm{m}$ and $\Theta$. 

The impact of the boundary on the distribution of the profile likelihood ratio is not insurmountable.  It is not fundamentally different than several common examples in high-energy physics where an unconstrained MLE would like outside of the physical parameter space.  Examples include downward fluctuations in event-counting experiments when the signal rate is bounded to be non-negative.  Another common example is the measurement of sines and cosines of mixing angles that are physically bounded between $[-1,1]$, though an unphysical MLE may lie outside this region.  The size of this effect is related to the probability that the MLE is pushed to a physical boundary.  If this probability can be estimated, it is possible to estimate a corrected threshold on $-2\ln\lambda$.  For a precise threshold with guaranteed coverage, one must resort to a fully frequentist Neyman Construction.

\section{Summary and conclusions}
\label{sec:conclusions}

We have presented a coverage study of state of the art techniques used for supersymmetric parameter estimation based on realistic toy data from an ATLAS sensitivity study.  The techniques employ both MCMC and nested sampling to explore the parameter space and the posterior samples can be used to provide both Bayesian credible intervals and profile likelihood-based frequentist confidence intervals.  We have tested our coverage analysis pipeline in a simple Gaussian toy model with the same dimensionality as the supersymmetric parameter space under consideration, and have confirmed the reliability of both the Bayesian posterior and profile likelihood intervals in this setting obtained with nested sampling and MCMC.

The likelihood function used in these studies is a simplified representation of the actual likelihood based on the information in the ATLAS publication.  We note that in order to perform coverage studies, one not only needs the ability to evaluate the likelihood function based on the pseudo-experiment, but also the ability to generate the pseudo-experiment itself.  Both the representation of the likelihood function and the ability to generate pseudo-experiment are now possible with the workspace technology in RooFit/RooStats~\cite{RooStats}. We encourage future experiments to publish their likelihoods using this technology.  Furthermore, we point out that these likelihood functions should include the nuisance parameters associated with experimental systematics, since they may be correlated across the different measurements entering a global fit.

We observe good coverage properties for intervals based directly on the weak-scale sparticle masses,  confirming the reliability of both the Bayesian credible intervals and profile likelihood intervals obtained with nested sampling and MCMC in this setting.   In contrast, we observe significant over-coverage in the fundamental CMSSM parameters, which we attribute to boundaries in the parameter space imposed by requirements of physically realizable theories. Our findings are specific to the benchmark point considered in our analysis, and coverage properties of benchmark points in different regions of the CMSSM parameter space might differ substantially. Currently, there is no known way to generalize the results, thus coverage requires a case-by-case analysis. The technique presented in this paper is, however, readily applicable to any benchmark point, provided a suitable likelihood function for LHC data is available, and that the neural network training is extended appropriately.

Coverage studies are computationally expensive to perform, and the most intensive step for an individual likelihood evaluation in this context is the evaluation of the RGEs in the spectrum calculators.  Thus, we have introduced a method to approximate the spectrum calculators with neural networks.  We observe consistency in the obtained intervals with speed-ups of $10^4$ by utilizing neural network approximations. In principle, the same techniques can be applied to accelerate the computation of other observables, such as the relic density. 

In order to assess the actual coverage properties of both Frequentist and Bayesian intervals obtained from future data, it will be unavoidable to perform detailed studies using a large number of pseudo-experiments. This paper demonstrated for the first time how neural network techniques can be employed to achieve a very significant speed-up with respect to conventional spectrum calculators. Such methods are necessary in order to cope with the very large number of likelihood evaluations required  in coverage studies. Our work is but a first step towards making accelerated inference techniques more widespread and more easily accessible to the community.

\subsection*{Acknowledgements} 
The authors wish to thank Louis Lyons for many useful discussions and suggestions, the authors of Ref~\cite{Yashar} for discussing their work with us prior to publication, Yashar Akrami, Jan Conrad and Pat Scott for helpful comments on a draft of this work and an anonymous referee for useful comments.  We  would like to thank the PROSPECTS workshop (Stockholm, Sept 2010) organizers, the BIRS Discovery Challenges workshop 2010 (Banff, July 2010) organizers and the CosmoStats09 conference (Ascona, July 2009) organizers  for stimulating meetings which offered the opportunity to discuss several of the issues presented in this paper. 
K.C. is supported  by the US National Science Foundation grants PHY-0854724 and PHY-0955626. F.F. is supported by Trinity Hall, Cambridge. R.RDA is supported by the project PROMETEO
(PROMETEO/2008/069) of the Generalitat Valenciana and the Spanish MICINN's
 Consolider-Ingenio 2010 Programme under the grant MULTIDARK
 CSD2209-00064. R.T. would like to thank the
Galileo Galilei Institute for Theoretical Physics for hospitality
and the INFN and the EU FP6 Marie Curie Research and Training Network
``UniverseNet'' (MRTN-CT-2006-035863) for partial support. The use of Imperial College London High Performance Computing service is gratefully acknowledged. Part of this work was performed using the Darwin Supercomputer of the
University of Cambridge High Performance Computing Service
(http://www.hpc.cam.ac.uk/), provided by Dell Inc. using Strategic
Research Infrastructure Funding from the Higher Education Funding
Council for England.

\bibliographystyle{JHEP}
\bibliography{bfhkrt1_v2}

\end{document}